\documentclass[fleqn,usenatbib]{mnras}
\usepackage{subcaption}

\usepackage{newtxtext,newtxmath}
\usepackage[normalem]{ulem}

\usepackage[T1]{fontenc}
\usepackage{anyfontsize}

\usepackage{graphicx}   
\usepackage{amsmath}    
\usepackage{mathtools}
\usepackage{comment}
\usepackage{epsfig}
\usepackage[dvipsnames]{xcolor}
\usepackage{xargs}
\usepackage{tabularx}
\usepackage{longtable}
\usepackage{threeparttable}
\usepackage{hyperref}
\hypersetup{colorlinks=true, citecolor=blue, linkcolor=blue}
\usepackage{xspace}
\let\oldAA\AA
\renewcommand{\AA}{\text{\oldAA}\xspace}

\makeatletter 
  \patchcmd{\NAT@citex}
    {\@citea\NAT@hyper@{%
      \NAT@nmfmt{\NAT@nm}%
      \hyper@natlinkbreak{\NAT@aysep\NAT@spacechar}{\@citeb\@extra@b@citeb}%
      \NAT@date}}
    {\@citea\NAT@nmfmt{\NAT@nm}%
    \NAT@aysep\NAT@spacechar\NAT@hyper@{\NAT@date}}{}{}

  \patchcmd{\NAT@citex}
    {\@citea\NAT@hyper@{%
      \NAT@nmfmt{\NAT@nm}%
      \hyper@natlinkbreak{\NAT@spacechar\NAT@@open\if*#1*\else#1\NAT@spacechar\fi}%
        {\@citeb\@extra@b@citeb}%
      \NAT@date}}
    {\@citea\NAT@nmfmt{\NAT@nm}%
    \NAT@spacechar\NAT@@open\if*#1*\else#1\NAT@spacechar\fi\NAT@hyper@{\NAT@date}}
    {}{}
\makeatother

\newcommand{\redtxt}[1]{\textcolor{red}{#1}}

\newcommand{\hi}{\text{H\,{\sc i}}\xspace} 
\newcommand{\hii}{\text{H\,{\sc ii}}\xspace}

\newcommand{\heii}{He\,{\sc ii}\xspace}

\newcommand{\oiii}{[O\,{\sc iii}]}

\newcommand{\oip}{O\,{\sc i}\xspace}
\newcommand{\ciii}{C\,{\sc iii}]\xspace}
\newcommand{\civ}{C\,{\sc iv}\xspace}

\newcommand{\feii}{Fe\,{\sc ii}\xspace}

\newcommand{\niv}{N\,{\sc iv}]}

\newcommand{\lya}{Ly\ensuremath{\alpha}\xspace}
\newcommand{\lyb}{Ly\ensuremath{\beta}\xspace}
\newcommand{\ha}{H\ensuremath{\alpha}\xspace}
\newcommand{\hb}{H$\beta$\xspace}

\newcommand{\Nhi}{\ensuremath{N_{\hi}}\xspace}
\newcommand{\meanxhi}{\ensuremath{\langle x_{\hi} \rangle}\xspace}
\newcommand{\jwst}{\textit{JWST}\xspace}

\newcommand{\ergs}{$\rm erg~s^{-1}$}
\newcommand{\ergscm}{$\rm erg~s^{-1}~cm^{-2}$}
\newcommand{\kms}{\ensuremath{\rm km~s^{-1}}\xspace}
\newcommandx{\fluxdcgs}[1][1=-20]{$\times 10^{[#1]}$~erg~s$^{-1}$~cm$^{-2}$~\AA$^{-1}$\xspace}
\newcommandx{\fluxcgs}[2][1=-20,2=\ensuremath{\times}]{${#2}10^{#1}$~erg~s$^{-1}$~cm$^{-2}$\xspace}

\newcommand{\target}{Abell2744-QSO1\xspace}

\newcommand{\cloudy}{\textsc{Cloudy}\xspace}
\newcommand{\pyneb}{\textsc{PyNeb}\xspace}

\title[FUV AGN signatures in an LRD]{\centering Holes in the BH$^\star$? AGN signatures in the FUV spectrum \\of a black-hole dominated Little Red Dot at $z=7.04$
}

\author[Ji et al.]{Xihan Ji,$^{1,2}$\thanks{E-mail: \href{mailto:xj274@cam.ac.uk}{xj274@cam.ac.uk}}
Gabriele Pezzulli,$^{3}$
Francesco D'Eugenio,$^{1,2}$
Roberto Maiolino,$^{1,2,4}$
Stefano Carniani,$^{5}$
\newauthor
Sandro Tacchella,$^{1,2}$
Gareth Jones,$^{1,2}$
Aaron Smith,$^{6}$
Joris Witstok,$^{7,8}$
Andrew C. Fabian,$^{9}$
Sophia Geris,$^{1,2}$
\newauthor
Anishya Harshan,$^{1,2}$
Yuki Isobe,$^{1,2}$
Lucy R. Ivey,$^{1,2}$
Ignas Juodžbalis,$^{1,2}$
Robert Pascalau,$^{1,2}$
Jan Scholtz$^{1,2}$,
\newauthor
Callum Witten$^{10}$
\\
\\
$^{1}$Kavli Institute for Cosmology, University of Cambridge, Madingley Road, Cambridge, CB3 0HA, UK\\
$^{2}$Cavendish Laboratory, University of Cambridge, 19 JJ Thomson Avenue, Cambridge, CB3 0HE, UK\\
$^{3}$Kapteyn Astronomical Institute, University of Groningen, Landleven 12, NL-9747 AD Groningen, the Netherlands\\
$^{4}$Department of Physics and Astronomy, University College London, Gower Street, London WC1E 6BT, UK\\
$^{5}$Scuola Normale Superiore, Piazza dei Cavalieri 7, I-56126 Pisa, Italy\\
$^{6}$Department of Physics, The University of Texas at Dallas, Richardson, Texas 75080, USA\\
$^{7}$Cosmic Dawn Center (DAWN), Copenhagen, Denmark\\
$^{8}$Niels Bohr Institute, University of Copenhagen, Jagtvej 128, DK-2200, Copenhagen, Denmark\\
$^{9}$Institute of Astronomy, University of Cambridge, Madingley Road, Cambridge CB3 0HA, UK\\
$^{10}$Department of Astronomy, University of Geneva, Chemin Pegasi 51, 1290 Versoix, Switzerland
}

\begin{document} 
\label{firstpage}
\pagerange{\pageref{firstpage}--\pageref{lastpage}}
\maketitle
 
\begin{abstract}
  It has been suggested that ``Little Red Dots'' (LRDs) might be accreting black holes enshrouded by dense gas in a nearly closed geometry, which completely covers the central black hole, leading to an atmosphere-like structure known as the ``black-hole star'' ($\rm BH^\star$).
  We test this scenario by analysing new \jwst spectroscopy in the far ultraviolet (FUV, rest-frame) of the prototypical LRD Abell2744-QSO1, at $z=7.04$.
  We found the presence of broad \lya emission with an FWHM of $\sim 1000$ \kms, and detections of \oip, \civ, and/or \feii emission lines.
  The NIRCam imaging and NIRSpec slit images indicate that the low-velocity component ($v\lesssim 200$ \kms) of \lya is likely spatially extended, but the high-velocity component ($v\gtrsim 200$ \kms) of \lya remains unresolved.
  Based on the multi-component kinematics and flux of \lya relative to Balmer lines, we conclude that the observed line profile is unlikely to be broadened by subsequent resonant scattering through the interstellar medium.
  This suggests that the high-velocity component of \lya originates in the broad-line region, although resonant scattering in the dense gas likely makes \lya broader than \ha as observed.
  The nebular features of this LRD indicate that there is at least one relatively optically thin direction where \lya can escape from the broad-line region (BLR).
  We also found indications that photons from the BLR are powering fluorescence of \feii and \oip on a larger physical scale.
  The FUV features thus challenge the fully-covered geometry interpretation and suggest that there are ``holes'' in the $\rm BH^\star$, or the absorbing medium is simply clumpy.
  
\end{abstract}

%

\section{Introduction}

Since the advent of the \textit{James Webb Space Telescope} \citep[\jwst,][]{jwst0,jwst1}, the early Universe has been open for population studies of accreting supermassive black holes.
Among the active galactic nuclei (AGN) uncovered by \jwst, a specific population, now usually referred to as ``Little Red Dots'' (LRDs) due to their compact morphologies and red colors in the rest-frame optical \citep{harikane2023,greene2024,matthee2024,kocevski2023,Kocevski_lrd_2024,Kokorev_lrd_2024}, has caught wide interest due to their peculiar properties.
Despite the frequent detections of prominent broad Balmer lines in LRDs, these objects differ from the general AGN population at low redshift in many aspects, including a peculiar ``V''-shaped spectral energy distribution (SED) with the turnover point always around 3000\,--\,4000 \AA \citep{setton_lrd_2024,Hviding_2025}, high occurrence rates of deep Balmer absorptions \citep[$>20\%$, e.g.,][]{matthee2024,juodzbalis_rosetta_2024,ji_lrdbreak_2025,deugenio_qso1_2025,deugenio_lrdoutflow_2025,irony,naidu_lrd_2025}, presence of rarely seen forbidden and permitted \feii emission lines and low-ionization metal absorptions \citep{ji_lord_2025,linxiaojing_locallrd_2025,irony,Torralba_feii_2025}, lack of hard X-ray emission and potential lack of radio emission \citep[][although not limited to LRDs but applied to the general high-$z$ AGN population discovered by \jwst]{ananna_2024,yue_lrd_2024,Maiolino2024_Xrays,Mazzolari2024radio}, and dust emission weaker than in normal AGN (\citealp{akins_lrd_2024,Casey_2025,Setton_lrddust_2025}; although significant hot dust emission is still present, e.g., \citealp{juodzbalis_rosetta_2024,Delvecchio2025,Brazzini_2026,Perez-Gonzalez2026}).
Many LRDs also show little variability \citep{kokubo_harikane2024,zhang_var_2024}, but significant variability has been confirmed in two gravitationally lensed LRDs \citep{ji_lrdbreak_2025,Furtak_qso1var_2025,zhang_crosslrd_2025} as well as some local LRD analogues (\citealp{Lin2026_variab}; although see also \citealp{burke_var_2025}).

The above observational evidence has led to the speculation that LRDs
have significant amounts of neutral, dense \citep{juodzbalis_rosetta_2024,Inayoshi_maiolino_2025}, and possibly turbulent gas \citep{ji_lrdbreak_2025} in front of their accretion discs that blocks our line of sight (LOS), filtering the observed SEDs and producing the peculiar features.
This is in clear contrast to typical Type 1 AGN, where the accretion discs and broad-line regions (BLRs) are unobscured along the LOS.
This simple model has been shown to successfully reproduce the spectral shapes and Balmer absorptions observed in LRDs \citep{ji_lrdbreak_2025,degraaff_lrd_2025,naidu_lrd_2025,Taylor_lrd_2025}.
More recently, it has been speculated that the geometry of the blocking gas could be spherical and similar to a stellar atmosphere, making LRDs ``Black Hole Stars'' \citep[$\rm BH^\star$,][]{naidu_lrd_2025,Begelman_2025}.
In one of the currently most frequently discussed pictures, LRDs produce a blackbody-like optical continuum and broad emission lines through a layered structure in its ``atmosphere'' \citep{Liu_speddlrd_2025,liu_tlusty_2026,chang_2025,Kido_2025,Inayoshi_gasremoval_2025,degraaff_2025,linxiaojing_locallrd_2025,Sun2026}.
Further indications potentially supporting the thick gas envelopes come from the broad-line profiles.
According to \citet{Rusakov_escattering_2025}, \citet{Sneppen_2026}, and \citet{matthee2026engineflowslittlered}, many broad-line AGN discovered by \jwst (including LRDs) have broad \ha lines with exponential-like wings, which might be explained with electron scattering in Compton-thick conditions with $\tau_{\rm e}\approx 1$ (although see \citealp{Laor_2006} for the scenario of optically-thin electron scattering).
However, the dominant mechanism of line broadening in LRDs has remained debated \citep[see e.g.,][]{juodzbalis_jadesagn_2025,juodzbalis_specast_2025,Brazzini_2025,scholtz2026littleredbluedots}.

Despite many observational and theoretical efforts, the origin and fate of the gas envelopes around LRDs remain unclear.
It has been speculated that the envelopes could be the gas from which the BHs were born through direct collapses \citep{Begelman_2025,Pacucci2026}.
\citet{inayoshi_firstacc_2025} and \citet{Inayoshi_gasremoval_2025} suggest that LRDs could represent the first accretion event of AGN, and the gas envelopes will later be removed by star-forming (SF) activities.
Through modeling of gas kinematics, \citet{juodzbalis_specast_2025} found that for an LRD at $z=7.04$, the kinematic mass of the whole system is dominated by the BH itself, suggesting that the BH in this LRD could have grown much earlier than the stellar evolution.
Such a picture is further supported by the extreme metal deficit in this LRD \citep[][and possibly in some other LRDs; Ivey et al. in preparation; Harshan et al. in preparation]{Maiolino_metalpoorlrd_2025}, and primordial BHs provide a possible channel for reproducing the observations \citep{Zhangsaiyang_pmbh_2025}.

Another related, key question that remains poorly understood is the actual structure of the hypothesized gas envelope, which has been assumed to be spherically symmetric in many works for simplicity \citep[see, however,][]{pm_2026}.
In addition, it remains debated how the different spectral components seen in observations are ordered spatially in the gas envelope \citep[see][]{degraaff_2025,Sneppen_2026}.
Thus far, most observational studies on LRDs have focused on the rest-frame optical regime, where broad Balmer lines can be measured.
The rest-frame ultraviolet (UV) remains a less explored but promising regime for deciphering the gaseous structure.
Recently, \citet{Tang_nv_2025} reported tentative identifications of high-ionization lines in an LRD, suggesting that ionizing radiation might leak from the gas envelope.
\citet{tripodi2025deepdivebroadlineregion}, on the other hand, found \oip$\lambda 1304$ emission in an LRD at $z=5.3$, which supports strong fluorescent excitation in the gas envelope.
Both of these works suggest that UV emission directly excited by the accreting BHs is visible in at least some LRDs.
In contrast, \citet{torralba+2025} found weak, narrow, and spatially-offset \lya emission in an LRD at $z=4.5$, and suggest that the broad \lya must have been extinguished by a gas envelope with a near unity covering factor.
The same conclusion was drawn by \citet{asada_2026} by comparing the UV luminosities of LRDs with those of SF galaxies.
Clearly, whether the emission in the UV has an AGN or SF origin depends on the geometry and structure of the gas envelope, and more observational evidence is needed.

In this work, we present analyses based on new observations of the rest-frame far ultraviolet (FUV) spectrum of an LRD at $z=7.04$. This LRD was initially discovered by \citet{furtak_abell2744_2023,furtak2023} as a triply imaged source in the lensing cluster field Abell2744 and named as \target.
This source was subsequently studied by \citet{ma_lrd_2024,ji_lrdbreak_2025,deugenio_qso1_2025,Furtak_qso1var_2025,juodzbalis_specast_2025,Maiolino_metalpoorlrd_2025} and was shown to be a benchmark that reveals the gas envelope in the LRD population and a strong candidate for heavy seeding by mechanisms including primordial BHs.

The structure of the manuscript is as follows.
In Section~\ref{sec:data}, we describe the observations and data reduction.
In Section~\ref{sec:measure}, we describe the spectral measurements we made.
In Section~\ref{sec:diagnostics}, we describe the diagnostics we performed on the nebular features identified in the FUV and discuss their implications.
In Section~\ref{sec:uv_continuum}, we discuss the origin of the UV continuum.
In Section~\ref{sec:igm}, we discuss the intergalactic medium (IGM) transmission inferred from the observations of \target.
We draw our conclusions in Section~\ref{sec:conclude}.
Throughout this work, we assume a flat $\rm \Lambda CDM$ cosmology with $h=0.674$ and $\Omega _{\rm m} = 0.315$ \citep{planck2020}.
All magnitudes are given in the AB system and wavelengths are given in vacuum.


\section{Observations and data reduction}
\label{sec:data}

\begin{figure*}
    \centering
    \includegraphics[width=0.99\linewidth]{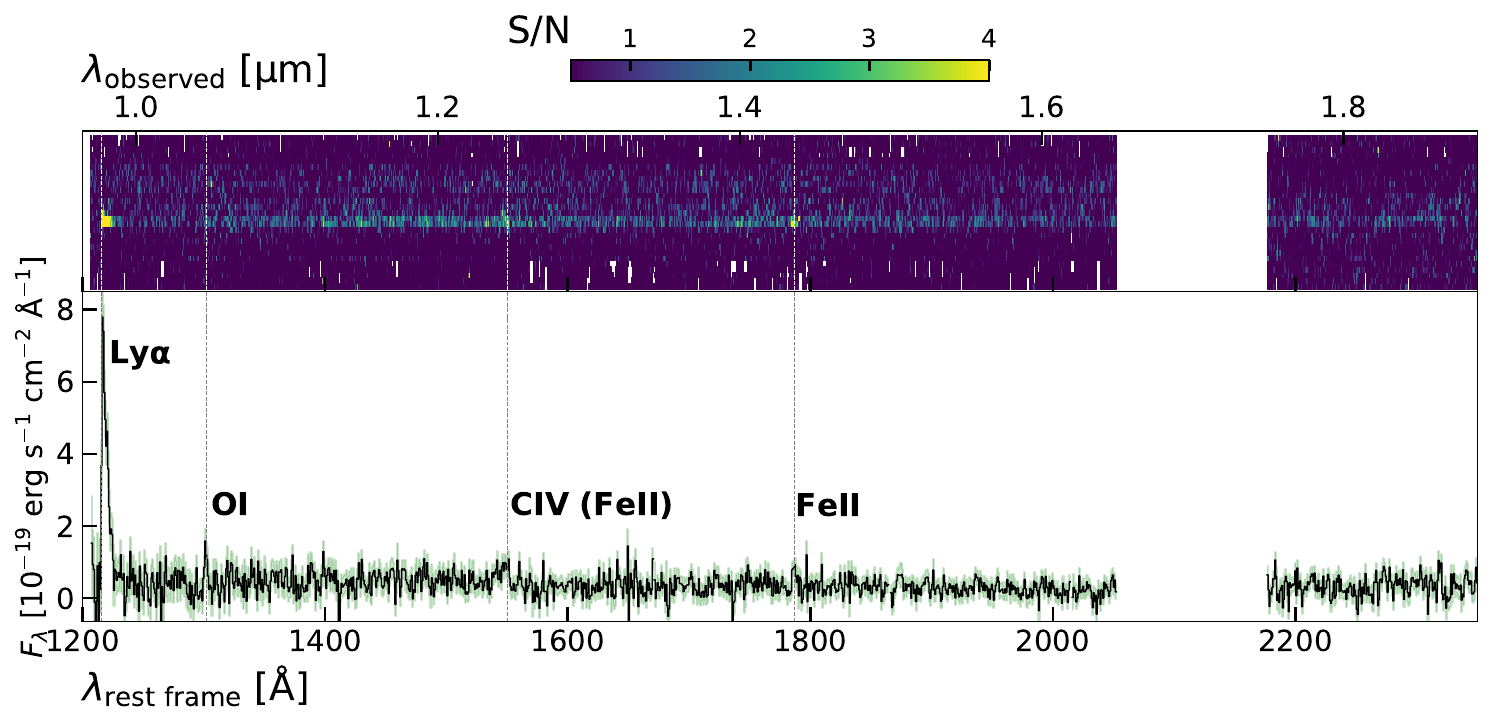}
    \caption{
    Reduced \jwst/NIRSpec G140M/F100LP spectra of \target, which cover the rest-frame FUV.
    The top panel shows the 2D signal-to-noise (S/N) map, and the bottom panel shows the 1D spectrum.
    Identified lines are marked, including \lya, \oip$\lambda 1302$, \civ$\lambda \lambda 1548,1551$ (or \feii), and \feii$\lambda 1787$.
    The shaded region in the 1D spectrum represent the $1\sigma$ uncertainty.
    }
    \label{fig:spec_2d1d}
\end{figure*}

In this work, we focus on \jwst observations of the triply imaged LRD, \target.
For the imaging, we used NIRCam data from the public data release of Ultra-deep NIRSpec and NIRCam ObserVations before the Epoch of Reionization (UNCOVER) survey \citep{uncover,uncoverdr_2024}. The raw data were obtained as part of the \jwst programmes 1324 \citep[GLASS, PI: T. Treu;][]{glass,glass_imaging}, 2561 (UNCOVER, PIs: I. Labbé, R. Bezanson), 2767 (PI: P. Kelly), 3516 \citep[ALT, PI: J. Matthee;][]{alt}, 3990 \citep[BEACON, PI: T. Morishita;][]{beacon}, and 4111 \citep[MegaScience, PI: K. Suess;][]{megascience}.

The spectroscopic data we used mainly come from new observations on image B (named by \citealp{furtak2023}) of \target from the \jwst Cycle 4 GO programme SPURS (SPectroscopic Ultra-deep Reionization-era Survey, PID: 9214; PIs: C.~Mason, D.~Stark).
The observations were carried out throughout November 6, 2025 and November 9, 2025, using the NIRSpec spectrograph \citep{jakobsen2022} and a single configuration of the Micro-Shutter Assembly \citep[MSA;][]{ferruit2022}, with three medium resolution ($R\sim 1000$) grating/filter combinations, which are G140M/F100LP, G235M/F170LP, and G395M/F290LP.
The observations in each of these three gratings used respectively 20, 13, and 16 groups per integrations, and 6, 10, and 3 integrations per exposure, all in NRSIRS2 mode \citep{Rauscher_2017}.
All observations used three nodded exposures for accurate background subtraction; this three-nod sequence was executed four times for G140M, and once for G235M and G395M.
The total integration times with G140M/F100LP, G235M/F170LP, and G395M/F290LP gratings are 29.5~h, 8.0~h, and 3.0~h, respectively.
The G140M grating covers the rest-frame far ultraviolet (FUV) regime of \target, and the G235M and G395M gratings cover the rest-frame near ultraviolet (NUV) to optical regimes.
In this work, we focus on the FUV spectrum, since
the G235M and G395M spectra are either very noisy due to the short integration time, or because previous MSA or IFS observations achieved better sensitivity at these wavelengths (see Appendix~\ref{appendix:nuv}).

The raw data of the above observations were queried from The Mikulski Archive for Space Telescopes (MAST) and reduced by the NIRSpec GTO pipeline version 5.1.
Details of the pipeline are described in \citet{scholtz_jadesdr4_2025}.
In brief, the GTO pipeline includes 1) pixel-to-pixel background subtraction directly in the count-rate images; 2) extraction of the spectral trace and assignment of wavelength and spatial coordinates; 3) wavelength correction -- this does not necessarily remove the wavelength-dependent offset for all sources (see Appendix~\ref{appendix:nuv}); 4) flat-field correction, which has been extended beyond the nominal wavelength range of the R1000 spectrum, where there is still non-negligible throughput. The extrapolation, however, is irrelevant to this work; 5) path loss correction; 6) resampling the extracted spectral traces onto an irregular 2D grid; 7) construction of the master background, which is an alternative for the pixel-by-pixel background subtraction and is irrelevant to this work; 8) extraction of the 1D spectrum with a 5-pixel-wide aperture. {Where relevant, we discuss any changes due to different data-reduction choices, including two-nod background subtraction \citep[appropriate for extended sources;][]{scholtz_jadesdr4_2025}, and three-pixel boxcar extraction (which increases the signal-to-noise ratio S/N for compact sources).}

In Figure~\ref{fig:spec_2d1d}, we plot the reduced NIRSpec G140M spectra of \target in 2D (top) and 1D (bottom).
The spectra are shifted to the systemic redshift of \target, $z=7.0367$ determined with narrow optical lines by \citet{deugenio_qso1_2025} based on data from the BlackTHUNDER programme (PID 5015; PIs: H. Übler, R. Maiolino).
{In the 2D spectrum, we note the continuum emission outside the central trace. Since \target is generally unresolved with NIRCam imaging, this signal must come from interlopers or other sources of contamination, as discussed in Appendix~\ref{appendix:background}.}
Several prominent emission lines visible in both 1D and 2D spectra are identified and marked, including \lya, \oip$\lambda 1302$, \civ$\lambda \lambda 1548,1551$ or \feii$\lambda \lambda 1548,1550$, and \feii$\lambda 1787$.
Next, we describe the spectral measurements we made for the FUV spectrum of \target.

\section{Spectral measurements}
\label{sec:measure}

We fit the G140M spectrum of \target using the Penalized PiXel-Fitting code \citep[\textsc{pPXF},][]{cappellari2004,cappellari2017} and model the continuum and emission lines simultaneously.
We use a power law with a Voigt–Hjerting optical depth distribution near \lya to model the continuum.
This is motivated by the possible damped \lya absorber (DLA) in \target \citep{ji_lrdbreak_2025}.
We also tested models with a two-photon nebular continuum as recently suggested for some high-$z$ galaxies \citep{cameron_9422}, which are discussed in Section~\ref{sec:uv_continuum}.
The DLA parameters (i.e., the gas column density, Doppler parameter, and redshift) are highly degenerate given the S/N of the spectrum.
Regardless, our focus is on emission lines, which are largely unaffected by the exact determinations of the continuum parameters as long as the continuum shape is well fitted. 
We discuss the physical origin of the continuum later in Section~\ref{sec:uv_continuum}.

We use Gaussians convolved with the instrumental line spread function (LSF) of the G140M grating from \citet{degraaff_lsf_2024} to describe lines including \oip$\lambda 1302$, \civ$\lambda \lambda 1548,1551$, and \feii$\lambda 1787$, and a line potentially associated with \feii at 1542 \AA.
While the FUV \oip is actually a resonant triplet including lines at 1302 \AA, 1304 \AA, and 1306 \AA, only a single line is identified and is closest to \oip$\lambda 1302$ given the systemic redshift of \target. 
In addition, \oip appears slightly blueshifted compared to other lines.
Therefore, when performing the line fitting, we left the kinematics of \oip free but tied the kinematics of \civ and \feii.
For diagnostic purposes, we also attempted to fit \heii$\lambda 1640$ and \ciii$\lambda 1908$, and none of the lines were detected.
After tying the kinematics to \civ and \feii, \heii is detected at $\sim 2\sigma$ but \ciii remains undetected.

\begin{figure*}
    \centering
    \includegraphics[width=0.9\linewidth]{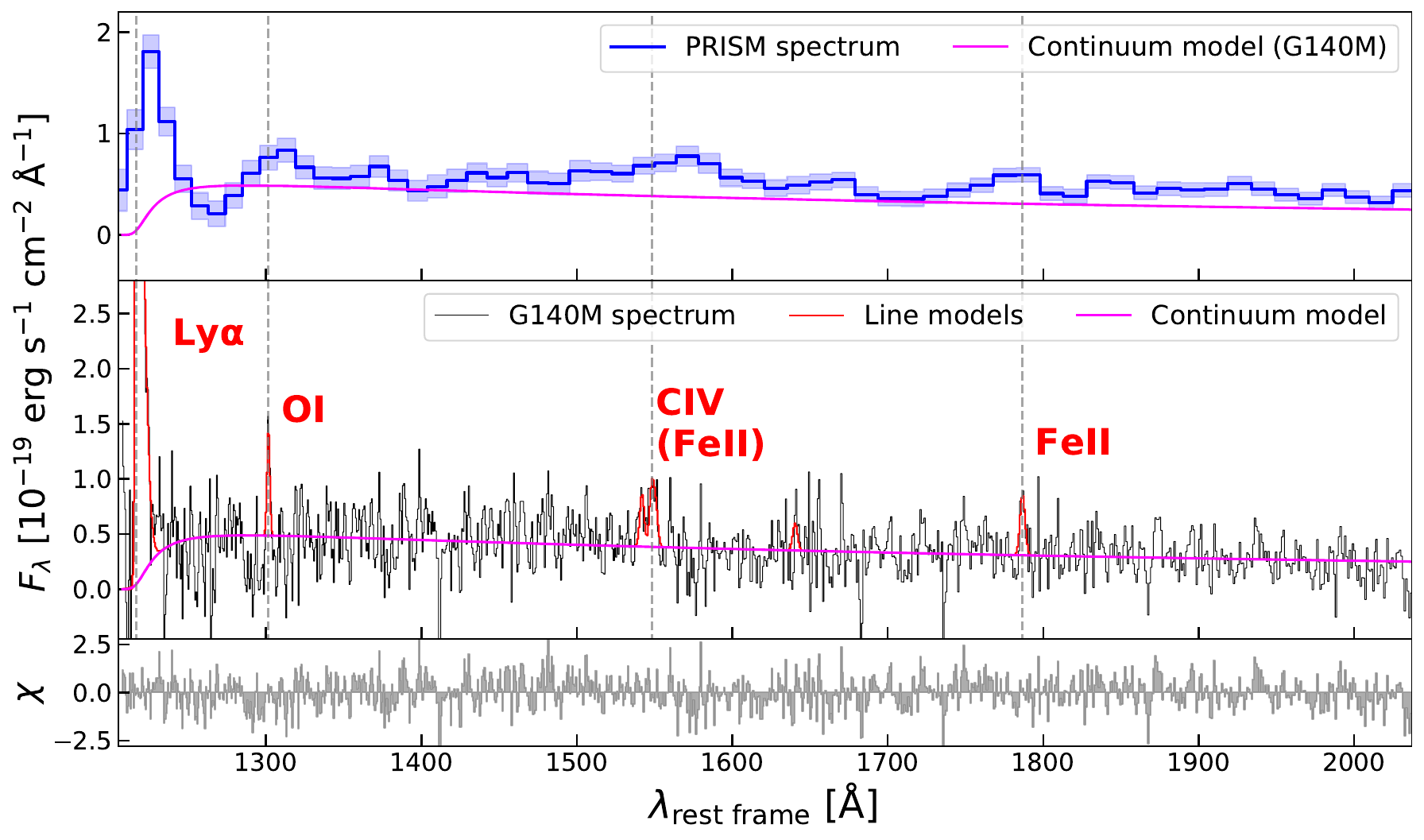}
    \includegraphics[width=0.9\linewidth]{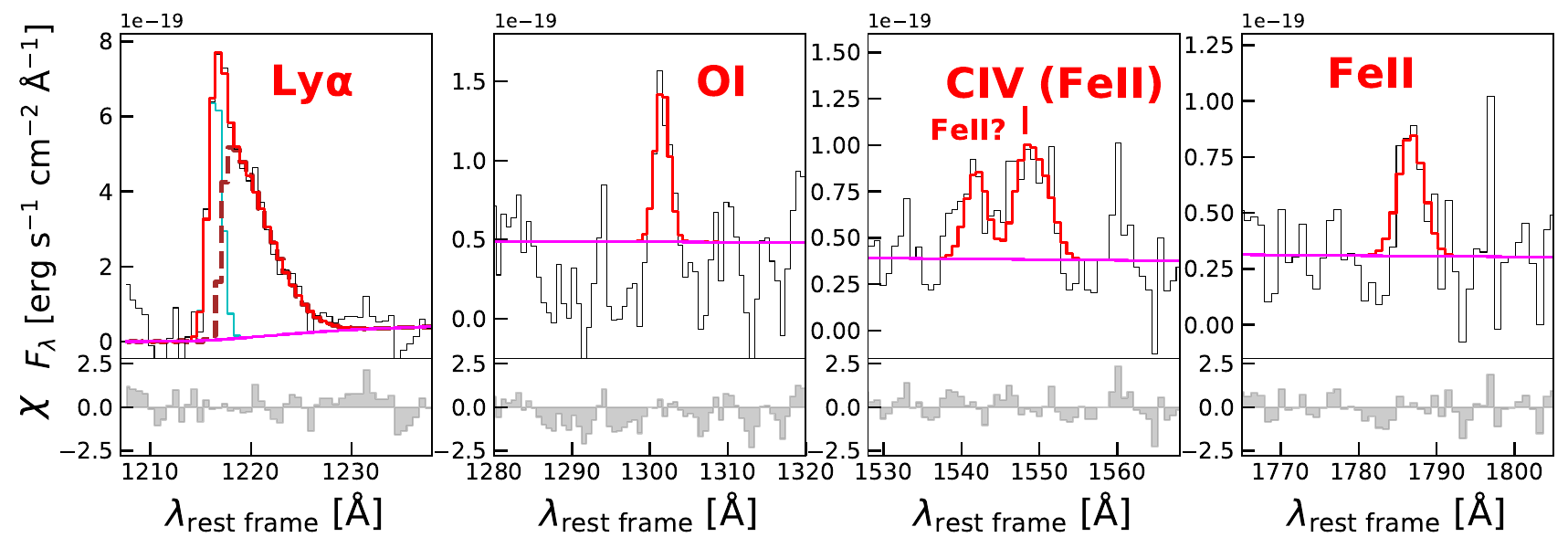}
    \caption{Best-fit spectral model for \target in the FUV (where $\rm \chi \equiv residual/\sigma$) based on the R1000 G140M spectrum. 
    Emission lines detected at $>3\sigma$ are labeled with their centres indicated by the dashed lines.
    The wavelength grids of the fitted models are sampled logarithmically by \textsc{pPXF}, and for consistency, the observed spectrum is log rebinned in the same way.
    For \lya, we also plot our decomposition into the narrow (solid cyan) and broad (dashed brown) components in the bottom left panel.
    For comparison, we plot the R100 PRISM spectrum (with $1\sigma$ uncertainties) of the same image (B) of \target in the top panel.
    Judging from the \lya peaks, there appear to be a sub-pixel scale wavelength offset between PRISM and G140M, which is a known issue for NIRSpec MSA data \citep{scholtz_jadesdr4_2025}.
    Prominent emission lines in the R1000 spectrum might have matched flux excess in the R100 spectrum, although the cross match is subject to the wavelength offset.
    }
    \label{fig:spec_fit}
\end{figure*}

For \lya emission, which is clearly asymmetric and probably affected by radiative transfer (RT) effects, 
we performed independent sets of fits under different assumptions.
Our first fit remains agnostic about the detailed RT physics, which adopts a two-component model including a symmetric Gaussian + a skewed Gaussian to describe the line profile.
This is also our fiducial fit to measure the flux and kinematics of \lya.
As we show in the next section, the two-component fit is physically motivated, as the peak and the wing of the \lya profile have different spatial extensions.
Fitting the profile as a single skewed Gaussian would not change our overall conclusion based on the flux and kinematics of the total \lya.
The skewed Gaussian is defined as
\begin{equation}
    SG(\lambda)=\frac{2A}{\sigma}G(\lambda_0, \sigma)\, \mathrm{CDF}[\alpha(\lambda_0, \sigma)] \, ,
    \label{eq:skewed_gaussian}
\end{equation}
where $A$ is the amplitude, $G(\lambda_0, \sigma)$ is a Gaussian distribution with mean $\lambda_0$ and dispersion $\sigma$, CDF is the cumulative distribution function of the Gaussian distribution, and $\alpha$ is a parameter related to the skewness (i.e., larger $\alpha$ implies a more asymmetric profile; see \citealp{skewed_gaussian}).
The kinematics of the two components of \lya are both set free and independent of other UV lines to account for the effect of resonant scattering.
The skewed Gaussian is broader and stronger than the symmetric Gaussian, although their relative strength is highly uncertain and the symmetric Gaussian component is only marginally detected at $\sim 2\sigma$.
Therefore, we also measured the properties of \lya as a single component after adding the best-fit Gaussian and skewed Gaussian.
The whole \lya is described by the total flux, peak velocity, and the velocity dispersion.
We plot our best-fit model in Figure~\ref{fig:spec_fit}, where the \lya fit is highlighted in the bottom left panel.
Finally, as a sanity check, we validate our flux measurement in the G140M/F100LP spectrum against that from the 
NIRSpec MSA PRISM ($R\sim 100$) spectrum for the same image (B) of \target.
The PRISM spectrum comes
from the UNCOVER survey and is reduced with the NIRSpec GTO pipeline.
With the same continuum model, we obtained $F_{\rm Ly\alpha,~PRISM}=(40.8\pm 3.9)\times10^{-19}$ \ergscm, consistent with $F_{\rm Ly\alpha,~G140M}=(40.1\pm 2.7)\times10^{-19}$ \ergscm, {suggesting that our measurement of \lya is not affected by its proximity to the filter edge}.
For reference, as shown by \citet{scholtz_jadesdr4_2025}, the v5.1 GTO pipeline typically produces consistent fluxes between PRISM and grating spectra within 4\% around 1 $\mu$m.

In addition to this fiducial fit, we also performed fits with physically motivated models incorporating IGM transmission and potential RT effects on the \lya line shape.
These models are described in more detail in Section~\ref{sec:igm}.

We measured the fluxes, velocity offsets, and velocity dispersions of all emission lines with \textsc{pPXF}. We extracted uncertainties 
using a Monte Carlo method, where we manually added noise to the spectrum and remeasured line properties. 
The noise was extracted from a normal distribution whose standard deviation is set to the $1\sigma$ flux uncertainty estimated by the pipeline.
The $1\sigma$ uncertainties of line measurements were computed as the standard deviations of the corresponding distributions after 1,000 trials.
We plot the best-fit model in Figure~\ref{fig:spec_fit} and summarize the emission line properties in Table~\ref{tab:lines}.
For comparison, we overlay the NIRSpec PRISM for image B.
Comparing the peaks of \lya from G140M and PRISM spectra, there is a wavelength offset on $\lesssim 1$ PRISM pixel scale ($\sim 1,000$ \kms near \lya). 
Such sub-PRISM pixel offset is a known issue for NIRSpec MSA wavelength calibration and may be caused by the different intra-shutter positions despite the correction applied by the GTO pipeline \citep[see][]{scholtz_jadesdr4_2025,Witstok_lya_2026}.
Several emission lines identified in the G140M spectrum, including \oip, \civ, and \feii, might correspond to flux excess in the PRISM spectrum.
However, the peaks of the flux excess in two spectra do not match exactly, which again might be affected by the wavelength offset.
Good wavelength agreement is found near \ha when comparing the G395M spectrum and the PRISM spectrum (see Appendix~\ref{appendix:nuv}).
Since the offset correction and the associated uncertainty is related to the native pixel size \citep{scholtz_jadesdr4_2025}, we considered the grating wavelength as the more exact solution in this work.
Next, we describe diagnostics with the measured emission lines and their implications.

\begin{table*}
    \centering
    \caption{Emission line measurements based on the NIRSpec G140M/F100LP spectrum.
    All velocity offsets, $\Delta v$, are given relative to the systemic redshift of the optical narrow lines, $z_{\rm optical}=7.0367$, from \citet{deugenio_qso1_2025}.
    All lines are fitted with Gaussian profiles, except the broad \lya, which is fitted with a narrow Gaussian + a broad skewed Gaussian (see text for details).
    In either case, $\sigma$ is the standard deviation of the line profile.
    The fluxes are not corrected for lensing magnification, which is $\mu =7.29^{+0.36}_{-2.18}$ according to \citet{furtak2023} and $\mu =9.1^{+0.9}_{-0.8}$ according to \citet{ji_lrdbreak_2025}.
    All upper limits listed are $3\sigma$-upper limits.
     }
    \begin{tabular}{l c c c c}
    \hline
      Line & $\lambda _{\rm rest}$ [\AA] & $\mu$ Flux [$\times 10^{-19}$\,\ergscm] & $\Delta v$ [\kms] & $\sigma$ [\kms]  \\
    \hline
    \lya [narrow]  & 1215.67 & $10.4\pm 5.1$ & $180\pm 200$ & $140\pm 77$ \\
    \lya [broad]$^{\rm a}$  & 1215.67 & $29.9\pm 5.0$ & $480\pm 180$ & $1030\pm 75$ \\
    \lya [total]  & 1215.67 & $40.1\pm 2.7$ & $270\pm 50$ & $670\pm 40$ \\
     & & & & ($\rm FWHM= 1300\pm 140$ \kms) \\
    \oip$^{\rm b}$ & 1302.17 & $2.2\pm 0.6$ & $-130\pm 70$ & $180\pm 60$ \\
    \feii? (tentative) & 1542.18 & $1.7\pm 0.6$ & $5\pm 70$ & $250 \pm 60$ \\
    \civ (\feii) & 1548.19, 1550.77 & $3.1\pm 0.6$ & $5\pm 70$ & $250\pm 60$ \\
    \heii & 1640.42 & $0.96\pm 0.57$ (or $<1.7$) & $5\pm 70$ & $250\pm 60$ \\
    \feii & 1786.75 & $2.2\pm 0.5$ & $5\pm 70$ & $250\pm 60$ \\
    \ciii & 1906.68, 1908.73 & $<3.1$ & $5\pm 70$ & $250\pm 60$ \\
    \hline
    \end{tabular}
    \begin{tablenotes}
        \small
        \item $\bf Notes.$
        \item $\rm ^a$ Modeled as a skewed Gaussian parameterized by the amplitude, mean, standard deviation, and the skewness parameter ($\alpha$; see Equation~\ref{eq:skewed_gaussian}).
        \item $\rm ^b$ If this line is actually \oip$\lambda 1304$, it would be blueshifted by 750 \kms with respect to the systemic redshift of \target.
    \end{tablenotes}
    \label{tab:lines}
\end{table*}

\section{Nebular diagnostics}
\label{sec:diagnostics}

The FUV spectrum of \target reveals several nebular emission lines, as summarized in Table~\ref{tab:lines}.
One immediate question is what the source of ionization and excitation for these lines is.
Thus far, the FUV nebular emission in LRDs has been speculated to originate from SF in the host galaxy (e.g., \citealp{naidu_lrd_2025,torralba+2025,degraaff_2025}), massive stars \citep{wang_heii_2025}, or a mix of SF and AGN ionization \citep[e.g.,][]{labbe_monster,rinaldi_lrd_2024,Tang_nv_2025,tripodi2025deepdivebroadlineregion,Pacucci2026,pm_2026}.
In all these scenarios, the optical continuum and broad lines are thought to arise from a massive, accreting black hole behind a high-column density of gas.
The key difference between these scenarios is whether the hypothesized gas envelopes around the black holes
are similar to stellar atmospheres, which have a near-unity covering factor and are extremely optically thick to UV emission (i.e., a low-temperature blackbody with $T_\mathrm{eff}\sim5000$~K).
In this case, no UV photons can escape from the inner regions. This would imply that no UV broad lines can be observed, and no classical narrow-line region (NLR) can exist since no extreme UV (EUV) photons from the accretion disc can leak out \citep{juodzbalis_jadesagn_2025,Inayoshi_gasremoval_2025}. In alternative scenarios where the obscuring gas is clumpy, there are lines of sight that allow the ISM to directly see the accretion disc and produces nebular emission and an NLR.
The observed UV continuum could then be the transmitted and possibly dust-reddened/attenuated accretion disc radiation or its scattered component \citep[see][]{Pacucci2026,pm_2026}.

In the FUV spectrum of \target, there are four interesting features.
\begin{enumerate}
    \item The \lya has a clear broad component with a velocity dispersion of $\sigma \gtrsim 1000$ \kms.
    \item There is a $\sim 3\sigma$ detection of \oip$\lambda 1302$, a resonant transition also observed in QSOs \citep{baldwin_qsoline_1978} and LRDs \citep{tripodi2025deepdivebroadlineregion}.
    \item There is a $\sim 5\sigma$ detection of a line near 1550 \AA, which could be \civ, a high-ionization resonant transition, or \feii, a low-ionization resonant transition, as we discuss later in this section.
    \item There is a $\sim 4\sigma$ detection of \feii$\lambda 1787$, which is part of the \feii UV 191 group frequently observed in QSO spectra \citep{vestergaard2001}.
\end{enumerate}
It is clear that the FUV spectrum of \target is dominated by permitted transitions, and their widths are all broader than the width of narrow, optical emission lines including \ha, \hb, and \oiii$\lambda 5007$, which is $\sigma_{\rm n}=22\pm 6$ \kms according to high-resolution ($R\sim 2700$) spectroscopic observations with \jwst/NIRSpec \citep{deugenio_qso1_2025}.
Next, we discuss the implications of these features.

\subsection{Origin of \texorpdfstring{Ly$\boldsymbol{\alpha}$}{LyA} and cloud geometry}
\label{subsec:lya}

\begin{figure}
    \centering
    \includegraphics[width=\columnwidth]{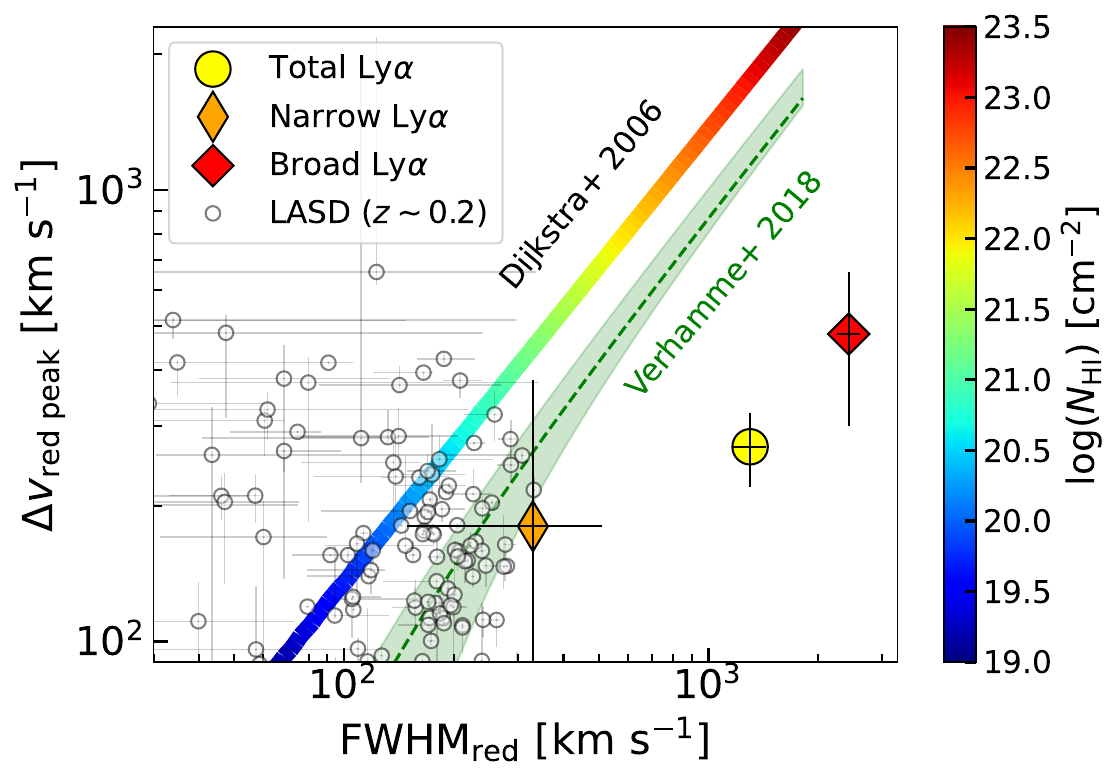}
    \caption{
    Scaling relations between the FWHM of the red peak of \lya and the velocity offset of the red peak from the analytical model of \citet{Dijkstra_lya_2006} and the empirical calibration by \citet{Verhamme_2018}.
    Both relations assume \lya is broadened by resonant scattering.
    The plotted model of \citet{Dijkstra_lya_2006} assumes the scattering medium is a static spherical cloud.
    Note that the axes are in log scales and the kinematics of the total and broad \lya deviate significantly ($>3\sigma$) from the relations.
    In addition, we plot 126 \lya emitters from LASD with systemic redshift measurements, where the median redshift is $z=0.21$.
    This comparison implies the broad component of \lya observed in \target cannot be reproduced by resonant scattering in the idealized geometry prescribed by \citet{Dijkstra_lya_2006}'s model or the physical conditions in \citet{Verhamme_2018}'s sample or LASD, although the narrow component is consistent with \lya from SF galaxies.
    }
    \label{fig:lya_relation_2}
\end{figure}

\begin{table}
        \centering
        \caption{Input parameters for \textsc{Cloudy} H\,{\sc i} models.}
        \label{tab:hi_models}
        \begin{tabular}{l c}
            \hline
            \hline
            Parameter & Values \\
            \hline
            $\log~Z/Z_\odot$ & $-2$ \\
            \hline
            $\log U$& $-2$ \\
            \hline
            $\log (n_{\rm H}/{\rm cm^{-3}})$& 10 (fiducial); 9, 11 (Appendix~\ref{appendix:tba}) \\
            \hline
            $\log (N_{\rm H}/{\rm cm^{-2}})$ & 22, 22.5, 23, 23.5, 24 \\
            \hline
            $v_{\rm turb}$/\kms & 20, 40, 60, 80, 100, 120 \\
            \hline
            Geometry & Open; plane-parallel\\
            \hline
            AGN SED & $M_{\rm BH}=10^7~M_\odot$, $\lambda _{\rm Edd}=0.1$\\ 
             & \citep{pezzulli_2017} \\
            \hline
            Dust & No dust\\
            \hline
            Atomic data & CHIANTI (v7, \citealp{chianti0};\\
             & \citealp{chianti_v7})\\
            \hline
            Solar reference & \citet{grevesse2010} \\
            \hline
        \end{tabular}
\end{table}

\begin{figure*}
    \centering
    \includegraphics[width=0.75\linewidth]{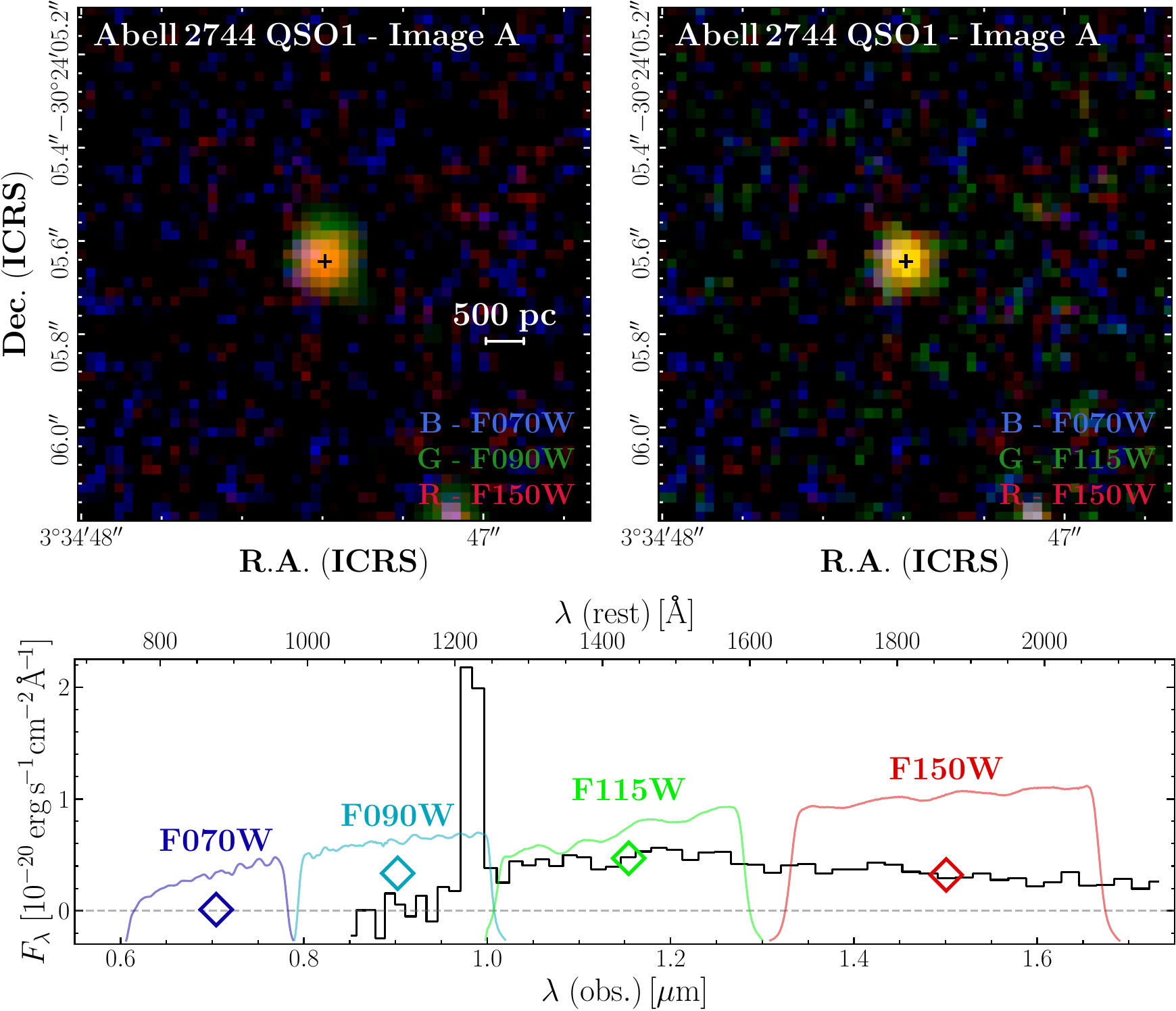}
    \caption{
    \jwst NIRCam color composite images and photometry for the image A of \target.
    \textit{Top:} Color composite images with bands covering the rest-frame UV.
    The F090W band on the left panel shows clear spatial extension in the north-south direction compared to the F115W band on the right panel.
    \textit{Bottom:} Comparison between the photometric measurements and the NIRSpec PRISM spectrum for the image A in the UV.
    While the F090W band covers the strong \lya emission, the F115W band covers only the continuum, suggesting that at least part of the \lya emission in \target is spatially extended.
    The extended \lya likely originates from scattered emission (see text).
    }
    \label{fig:lya_nircam}
\end{figure*}

\begin{figure}
    \centering
    \includegraphics[width=\columnwidth]{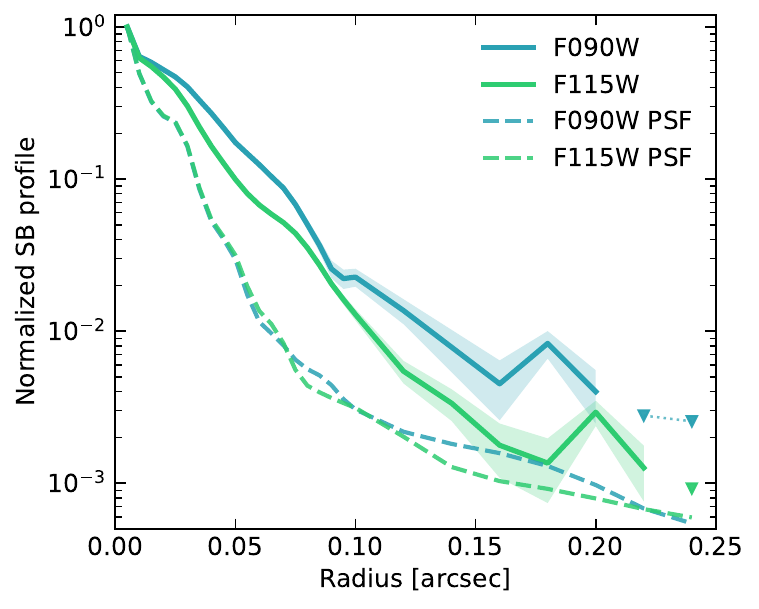}
    \caption{Radial profiles of normalized fluxes in the F090W and F115W bands for the image A of \target.
    The inverted triangles at $\rm Radius > 0.\!\!''2$ show the $2\sigma$ upper limit of the background.
    The dashed lines represent the PSF profiles in the two bands, which are similar.
    The F090W profile is more extended than the F115W profile, although the difference in $r_{\rm eff}$ is marginal (see text).
    }
    \label{fig:profile_fit}
\end{figure}

\begin{figure*}
    \centering
    \includegraphics[width=0.75\linewidth]{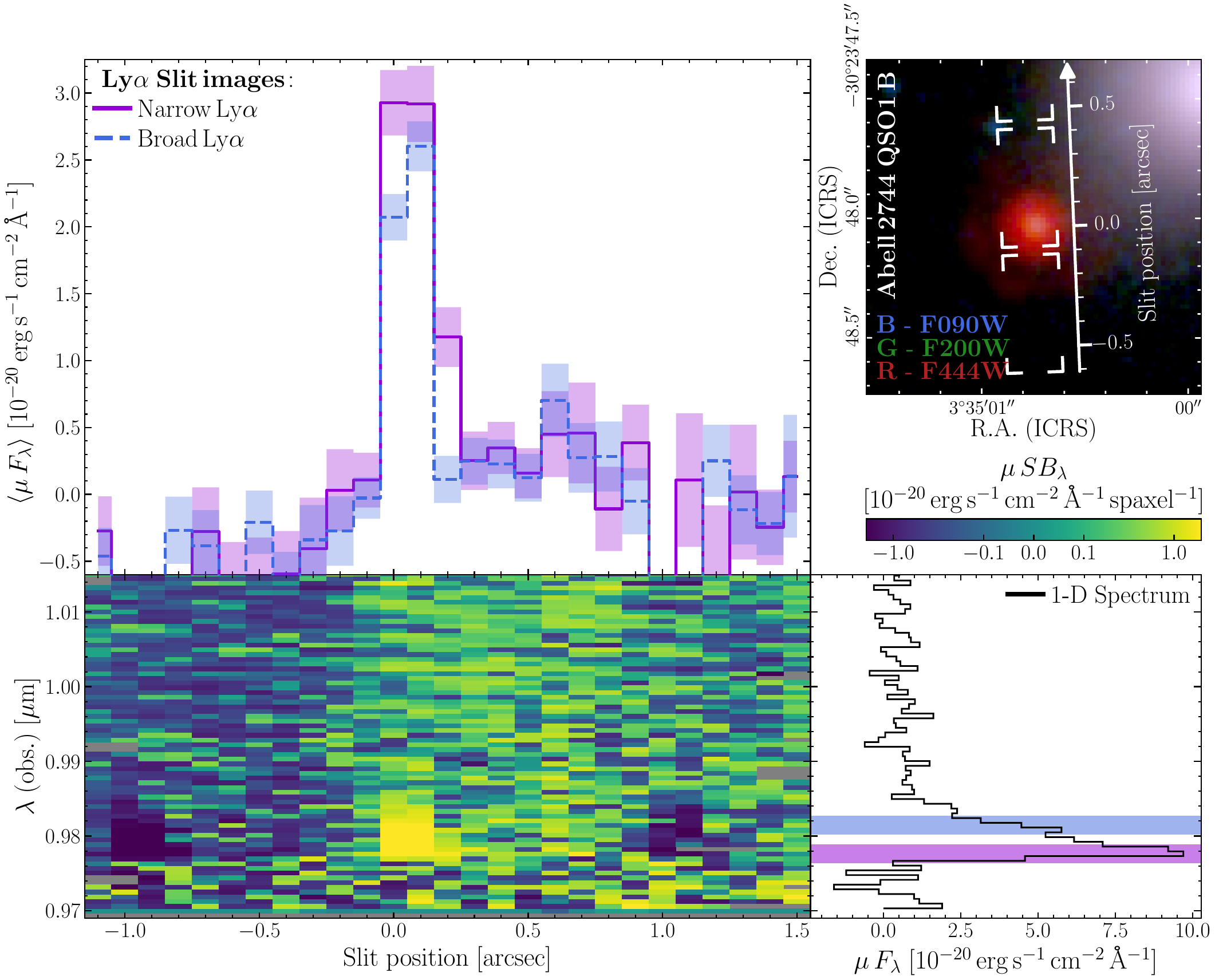}
    \caption{
    Surface brightness (SB) map for the spectral region around \lya in the G140M spectrum of the image B of \target.
    The top-right inset shows the MSA shutters from PID~9214 overlaid over a false-colour NIRCam stamp of image B;
    the position along the slit is indicated, with the origin
    assigned to the peak of image.
    The map shows the SB distribution in 2D space spanned by spectral and spatial pixels, the latter with the same
    origin as in the NIRCam image.
    The spectral core of \lya (purple band) is spatially extended compared to the red wing of \lya (blue band), where the flux distribution is extended by roughly one spatial pixel $0.\!\!''1$ (top panel), or 200\,--\,300 pc in the source plane.
    This implies that the spatial structure seen in the F090W image of Figure~\ref{fig:lya_nircam} is dominated by the narrow \lya, whereas the broad \lya is unresolved. {We note that a spurious continuum emission is visible as positive (negative) flux excess to the
    right (left) of the source (see Appendix~\ref{appendix:background} for more details), but this contaminating continuum does not affect the analysis of \lya.}
    }
    \label{fig:lya_2d}
\end{figure*}

The existence of strong Balmer absorption over broad Balmer lines in \target, as well as many other LRDs, suggests a significant population of neutral hydrogen at the first excited level ($n=2$), reaching a column density of $N_{\rm H(n=2)}\sim 10^{14-15}~{\rm cm^{-2}}$ \citep[e.g.,][]{juodzbalis_rosetta_2024,ji_lrdbreak_2025,deugenio_qso1_2025}.
This high column density of excited hydrogen can be populated with collisional excitation \citep[i.e., $n_{\rm H}\sim 10^{10}~{\rm cm^{-3}}$ in the BLR,][]{Inayoshi_maiolino_2025,ji_lrdbreak_2025} and \lya trapping that leads to pumping from the ground state \citep{ferland_netzer_1979,hall_2007,juodzbalis_rosetta_2024}, meaning that, in both cases, the \lya from the BLR should be extremely optically thick (with a line center optical depth of $\tau _{0,~\rm Ly\alpha}>10^{10}$).
In an open geometry where a slab of obscuring gas blocks our line of sight (LOS), this implies nearly all \lya photons escape in directions different from the LOS and no broad \lya is observable.
However, such a geometry does not prevent observations of broad \lya scattered back into the LOS by clouds not directly blocking the BLR.
In contrast, in a closed geometry 
with a covering factor of $C_f\approx 1$, \lya would still escape after $\sim \tau$ characteristic wing scatterings (although the extremely optically thick core scattering might end up with destruction by $2p\rightarrow 2s$; see \citealp{Nebrin2025}), provided that the gas is nearly dust-free to avoid significant absorption of \lya by dust during the escape (also assuming \lya lost by fluorescence to other lines, such as N\,{\sc v}$\lambda 1240$ and \feii, is negligible).
In the context of the $\rm BH^\star$, 
if the scattering medium becomes the 
hypothesized Compton-thick ``photosphere'' that produces the blackbody-like emission in the optical of LRDs \citep{Kido_2025,Liu_speddlrd_2025,linxiaojing_locallrd_2025,degraaff_2025},
\lya might start to become thermalized at high densities via collisional mixing of $2s$-$2p$ and blend into the continuum.
In this subsection, we investigate the origin of \lya and the geometry of gas clouds in this LRD indicated by \lya measurements.

\subsubsection{\lya with two components}

In the case of \target, we see a clear broad \lya with $\sigma = 1030\pm75$ \kms (and a Full Width at Half Maximum of $\rm FWHM\approx 1.95\sigma \approx 2000$ \kms since the $\alpha$ parameter is $\alpha \approx 14 \gg 1$ and the corresponding skewness is $\gamma _1 \approx 0.98 \sim 1$) in Figure~\ref{fig:spec_fit}, and the total \lya profile has $\rm FWHM= 1300\pm 140$ \kms.
Such broadening is unlikely to be made by resonant scattering in static gas alone, given the small shift of the \lya peak compared to the systematic redshift of \target.
To test the resonant scattering scenario, we checked the analytical model of \citet{Dijkstra_lya_2006}, where a static, spherical shell of gas with a constant temperature of $T=10^4$ K and no dust is assumed as the scattering medium, and the column density is varied in a range of $N_{\rm H}=10^{19}$\,--\,$10^{23.5}~{\rm cm^{-2}}$.
Assuming the blue peak of \lya has been extinguished by the IGM and we only see the red peak, we extracted the peak velocity and the FWHM from the models and compare them with our measurements for \target in Figure~\ref{fig:lya_relation_2}.
Both the total \lya and the broad \lya deviate significantly from the models, where their FWHM would imply $\rm N_{\rm H}>10^{22}~{\rm cm^{-2}}$ but their velocity shifts are less than the predictions from resonant scattering.
Only the kinematics of the narrow \lya could be consistent with the models at $\sim 2\sigma$.
Assuming instead a slab geometry would slightly steepen (i.e., worsen) the slope of the $\Delta v$\,--\,$\rm FWHM$ relation \citep[see][]{Neufeld_1990,Dijkstra_lya_2006} but would not change the above conclusions. However, departures from this analytic relation are possible due to complex RT effects \citep{Nebrin2025,Smith2025}.

We also compared our measurements with the empirical relation between the \lya peak velocity and FWHM measured from high-redshift \lya emitters (LAEs), where the \lya broadening is typically interpreted as due to resonant scattering.
However, we note that the details of the peak location (for a given degree of resonant scattering) are set by complex RT involving gas kinematics, geometry, and dust.
Taking the relation calibrated from 13 SF galaxies at $z=$3\,--\,6 by \citet{Verhamme_2018} and using $\rm FWHM=1300\pm 140$ \kms for the whole profile measured in QSO1, one expects the red-peak velocity to be $\Delta v_{\rm LAE}=890\pm 160$ \kms, significantly higher than $\Delta v_{\rm QSO1}=270\pm 50$ \kms we measured.
As noted by \citet{Verhamme_2018}, the empirical relation is broadly consistent with recent
RT calculations of \lya with Monte Carlo methods for a wide range of physical parameters.
If the scattering medium is expanding (e.g., due to outflows) instead of being static, the fraction of emergent \lya flux redward of the line center can change from half to nearly one depending on the expanding velocity, which facilitates \lya escape. 
In high-$z$ LAEs, more extreme outflows than typically observed in low-$z$ LAEs can be produced by certain phases of bursty SF, which also reduces the velocity shift of the red \lya peak, thereby moving the \citet{Dijkstra_lya_2006} relation down to the empirical \citet{Verhamme_2018} relation.
Still, the combination of strong broadening and low peak velocity of \lya seen in \target compared to high-$z$ LAEs implies an BLR origin and/or additional RT physics atypical of LAEs.

Notably, for the narrow \lya in QSO1, whose measured width is $\rm FWHM_{narrow}=330\pm 180$ \kms, the expected velocity shift from the \citet{Verhamme_2018} relation is $\Delta v_{\rm LAE;~narrow}=260\pm 180$ \kms, fully consistent with the measured shift of $\Delta v_{\rm QSO1;~narrow}=180\pm 200$ \kms.
This suggests that the narrow component of \lya in QSO1 could have been broadened by resonant scattering. In contrast, the inconsistent behaviour of the broad component suggests a different physical environment.
Finally, we compare \target with LAEs with systemic redshift measurements in the Lyman Alpha Spectrum Database \citep[LASD,][]{lasd}\footnote{\url{http://lasd.lyman-alpha.com}}, which includes 126 sources at a median redshift of $z=0.2$ and a maximum redshift of $z=2.2$.
While the LASD sample shows a large scatter in terms of $\Delta v_{\rm red~peak}$ and $\rm FWHM_{\rm red}$, the parameter space covered by these low-$z$ \lya emitters again better overlap with the narrow \lya rather than the broad \lya in \target.

Another piece of evidence against the picture where the bulk of \lya is intrinsically narrow but broadened by resonant scattering in the ISM comes from line fluxes.
\citet{ji_lrdbreak_2025} and \citet{deugenio_qso1_2025} reported the flux of narrow \hb for \target using R2700 grating observations of image A, and \citet{Maiolino_metalpoorlrd_2025} and \citet{deugenio_qso1_2025} reported a narrow line Balmer decrement consistent with the Case B value using the same observations.
If \lya has the same physical origin as narrow Balmer lines, one can estimate its intrinsic flux.
At $T_{\rm e}=10^4$ K and low-density limit, the Case B flux ratio between \lya and \hb is roughly 24 \citep{storey_hummer_1995}.
Directly converting the narrow \hb flux from image A to the expected Case B \lya flux for image B, we obtained $\mu _{\rm B}F_{\rm Ly\alpha;~expected}=$[(5.7\,--\,7.5)$\pm 1.0$]\,$\times10^{-18}$ \ergscm, where the range reflects the uncertainty of the lensing model.
Compared to the measured \textit{total} \lya flux of $\mu _{\rm B}F_{\rm Ly\alpha;~measured}=(4.0\pm 0.3)\times 10^{-18}$ \ergscm, there should be only modest loss of \lya after scattering if the whole \lya comes from the ISM that produces narrow \hb.
Since there has been reported time variability in the EW of \hb, we also made a direct comparison between the narrow \hb measured from the R2700 spectrum of image A and the total \lya measured from the R100 spectrum of image A \citep[see][]{ji_lrdbreak_2025,deugenio_qso1_2025}, and we found
$\mu _{\rm A}F_{\rm H\beta}=(2.0\pm 0.3)\times 10^{-19}$ \ergscm and $F_{\rm Ly \alpha}/F_{\rm H\beta}=25\pm4$ consistent with the Case B value,
meaning that there is negligible loss of \lya due to scattering should it share the origin with the narrow \hb.
The high escape fraction inferred in this scenario is in clear contrast to observations \citep[e.g.,][]{Saxena_2024} and simulations of LAEs throughout cosmic time, where \lya-to-Balmer line ratios are reduced by dust absorption \citep[might reach $\sim 50$\%,][]{smith_lya_2019} and IGM attenuation \citep[which starts to play a role at $z>5$ and reaches 10\,-\,50\% at $z\sim 7$,][]{smith_lya_2022}.
We do note that the intrinsic $F_{\rm Ly \alpha}/F_{\rm H\beta}$ can be boosted by collisional excitation \citep{Ferland_1985}, which would alleviate the issue with the high implied \lya escape in this case.
However, \citet{Maiolino_metalpoorlrd_2025} inferred an ISM density of $n_{\rm e}<10^{7}~{\rm cm^{-3}}$ for \target based on narrow optical lines, which would limit $F_{\rm Ly \alpha}/F_{\rm H\beta}<34$, again pointing towards strong \lya escape if it is of ISM origin.

The kinematical and flux measurements of \lya indicate that at least part of \lya is not from the ISM.
We further explore this idea from imaging in the following.

\subsubsection{Spatial extension of \lya}
\label{subsubsec:lya_nircam}

As we have discussed above, the broad component of \lya is unlikely to be broadened by resonant scattering in the ISM. In contrast, the narrow component of \lya can originate in the ISM and undergoes resonant broadening based on its kinematics and flux.
The two-component decomposition of \lya is supported by the spatial extensions of different wavelength/velocity channels.
In Figure~\ref{fig:lya_nircam}, we show colour-composite images with \jwst Near Infrared Camera \citep[NIRCam,][]{nircam0} in the top panels.
The cutouts correspond to image A (since image B suffers from foreground contamination)\footnote{The rest-frame time difference between images A and B is roughly 4 months \citep{furtak2023,ji_lrdbreak_2025}.
Thus, we do not expect variation in observed fluxes over large scales.
}.
We used three bands per image, including F070W, F090W, F115W, and F150W for the two images.
The corresponding band widths, responses, and photometric points are shown in the bottom panel, where we also plot the NIRSpec PRISM spectrum for the image A from the UNCOVER survey and reduced by the GTO pipeline.
Comparing the two coloured images, \target is more spatially extended in the F090W band.
From the bottom panel, one can see that the main difference between the two bands is that F090W covers almost exclusively \lya on its edge, since wavelengths blueward of \lya are absorbed by the IGM. In contrast, F115W covers only the FUV continuum without strong lines.
We can rule out that the spatial extension is driven by the continuum, since one would not expect the continuum to vary over such a short wavelength range. Besides, the continuum is clearly weaker in F090W. Moreover, we can rule out that the extension is only apparent, for example, due to higher S/N in the F090W image; in fact, the F115W has higher S/N than the F090W image.
Therefore, the only remaining explanation for the spatial extension seen only in the F090W image is that (at least part of) \lya is spatially extended.
As already noted by \citet{ji_lrdbreak_2025}, previous \jwst integral field unit (IFU) observations indicate that \lya might be spatially offset from the optical emission by $\sim 100$ pc in the source plane of image A.

As a more quantitative check on the FUV morphology of \target, we used \textsc{pysersic} \citep{pysersic} to fit the radial profiles of F090W and F115W images of image A, assuming a single Sérsic profile.
The extracted radial profiles of the normalized fluxes in the two bands together with the PSF profiles are plotted in Figure~\ref{fig:profiles}, where the F090W profile shows the largest spatial extension.
Giving a tangential lensing shear of $\mu _{\rm t}=3.52$ \citep{furtak2023}, we found that the effective radii in the source plane for the two FUV bands are $r_{\rm eff,~F090W}=34\pm 5$ pc and $r_{\rm eff,~F115W}=24\pm 2$ pc, respectively.
The size derived for the F115W band is less than one spatial pixel (30 pc) and is compatible with $r_{\rm eff}<30$ pc derived with F150W band by \citet{furtak2023}.
The F090W image does appear more spatially extended, and the difference between $r_{\rm eff,~F090W}$ and $r_{\rm eff,~F115W}$ is of $2.1\sigma$ significance.
The spatial offset between the centres of F090W and F115 images is also sub-pixel with $\Delta r=24\pm 3$ pc.
While the difference in the spatial extensions of F090W and F115 images is marginal based on the single Sérsic-profile fit, \lya might have different components with different sizes.

To further investigate which part of the \lya profile is spatially extended, we analysed slit images from the G140M observations.
Figure~\ref{fig:lya_2d} shows the surface brightness profile of the G140M spectrum along the slit for the image B of \target.
We calculated the median flux density as a function of the slit position, considering only spectral pixels in two narrow windows centred on the narrow and broad \lya components (indicated by the purple and blue shaded regions in the 1D spectrum, respectively). 
The resulting slit images show that
narrow \lya (solid purple) is spatially extended, since it spans three spatial pixels ($0.\!\!''1$ per pixel, with a NIRCam-inferred point spread function of $\rm FWHM\approx 0.\!\!''03$). 
In contrast, broad \lya (dashed blue) is spatially unresolved, as it occupies only two spatial pixels.
The RGB image in Figure~\ref{fig:lya_2d} illustrates the nominal slit location with respect to the source, with the same slit coordinates as the 2D spectrum.
This implies that in the image plane, narrow \lya extends to the north relative to the continuum, consistent with the spatial extension seen in the F090W image of image A shown in Figure~\ref{fig:lya_nircam}. 
However, repeating this spectral analysis for image A is impossible, since \lya in image A is only observed with the low-resolution PRISM, where the narrow and broad components cannot be spectrally separated.
Regardless, using the cluster lens models of \citet{Furtak_2023c}, we conclude that the location of this extension in image B is consistent with the photometric extension seen in image A (Figure~\ref{fig:lya_nircam}).

\subsubsection{Profile of the broad \lya wing}
\label{subsubsec:lya_profile}

While we show that the spatially unresolved \lya is likely to be ``intrinsically'' broad, we have not determined the precise broadening mechanism.
On the one hand, the broad component can simply come from virial broadening.
This is supported by the agreement between the virial BH mass for \target based on the broad \ha and the kinematic mass measurement based on the narrow \ha by \citet{juodzbalis_specast_2025}.
On the other hand, as recently noted by \citet{Rusakov_escattering_2025} and \citet{matthee2026engineflowslittlered}, broad Balmer lines observed in LRDs tend to show exponential-like wings.
One possible physical interpretation for such line profile is non-virial broadening by thermal electrons \citep{Laor_2006,chang_2025}.
While the above two broadening mechanisms are physically different, both virial broadening and electron scattering broadening require a BLR-like environment.
As shown by \citet{deugenio_qso1_2025}, the broad \ha profile obtained from the previous NIRSpec IFU observations of \target does show exponential-like wings, although other models such as double-Gaussian, Lorentzian, and Voigt profiles can fit the line equally well.
Additionally, even if the profile is exponential, \citet{scholtz2026littleredbluedots} recently illustrated that such a profile can also simply emerge from the virialized clouds in a stratified BLR (see also Trefoloni et al. in preparation).

\begin{figure}
    \centering
    \includegraphics[width=\columnwidth]{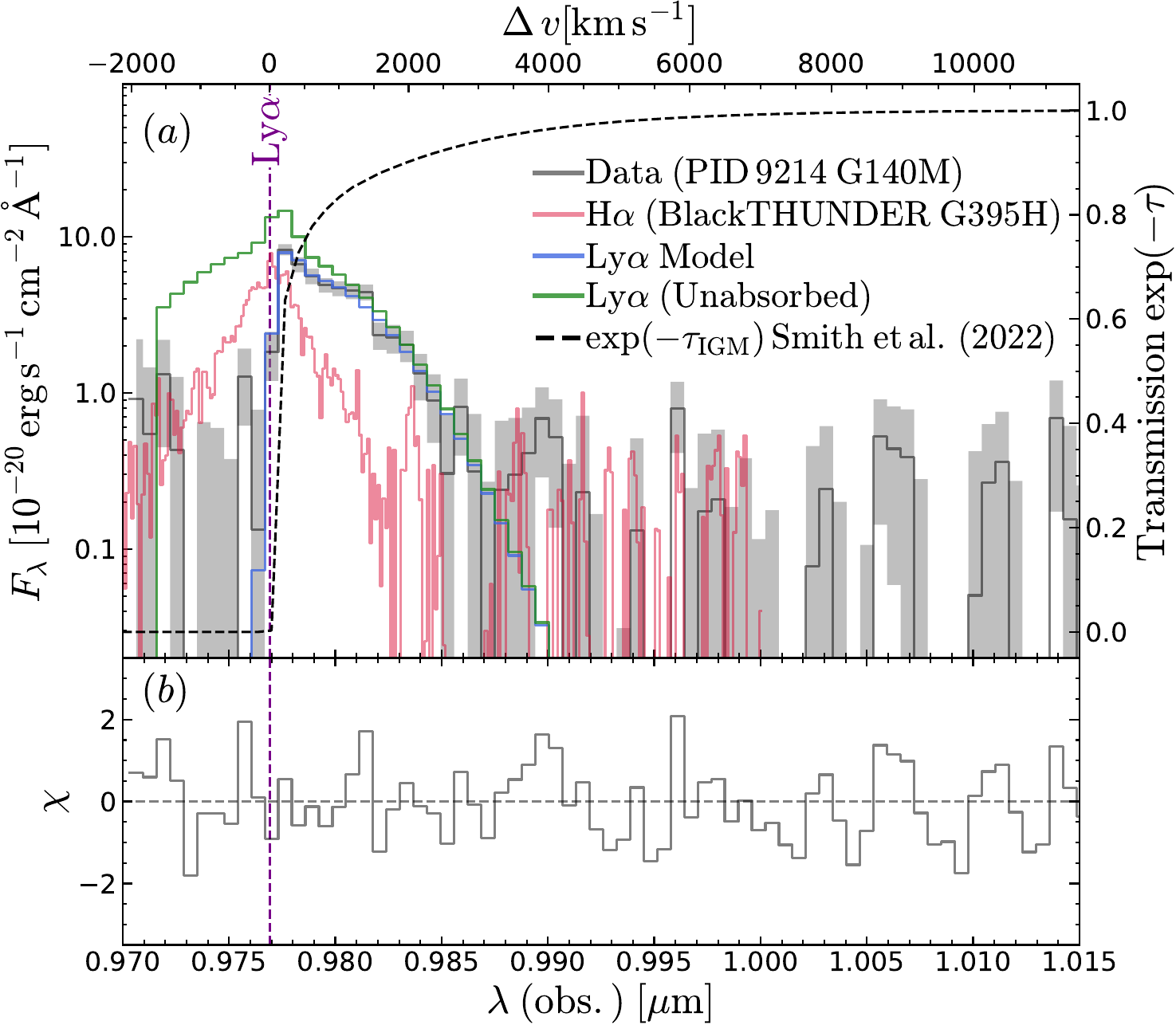}
    \includegraphics[width=\columnwidth]{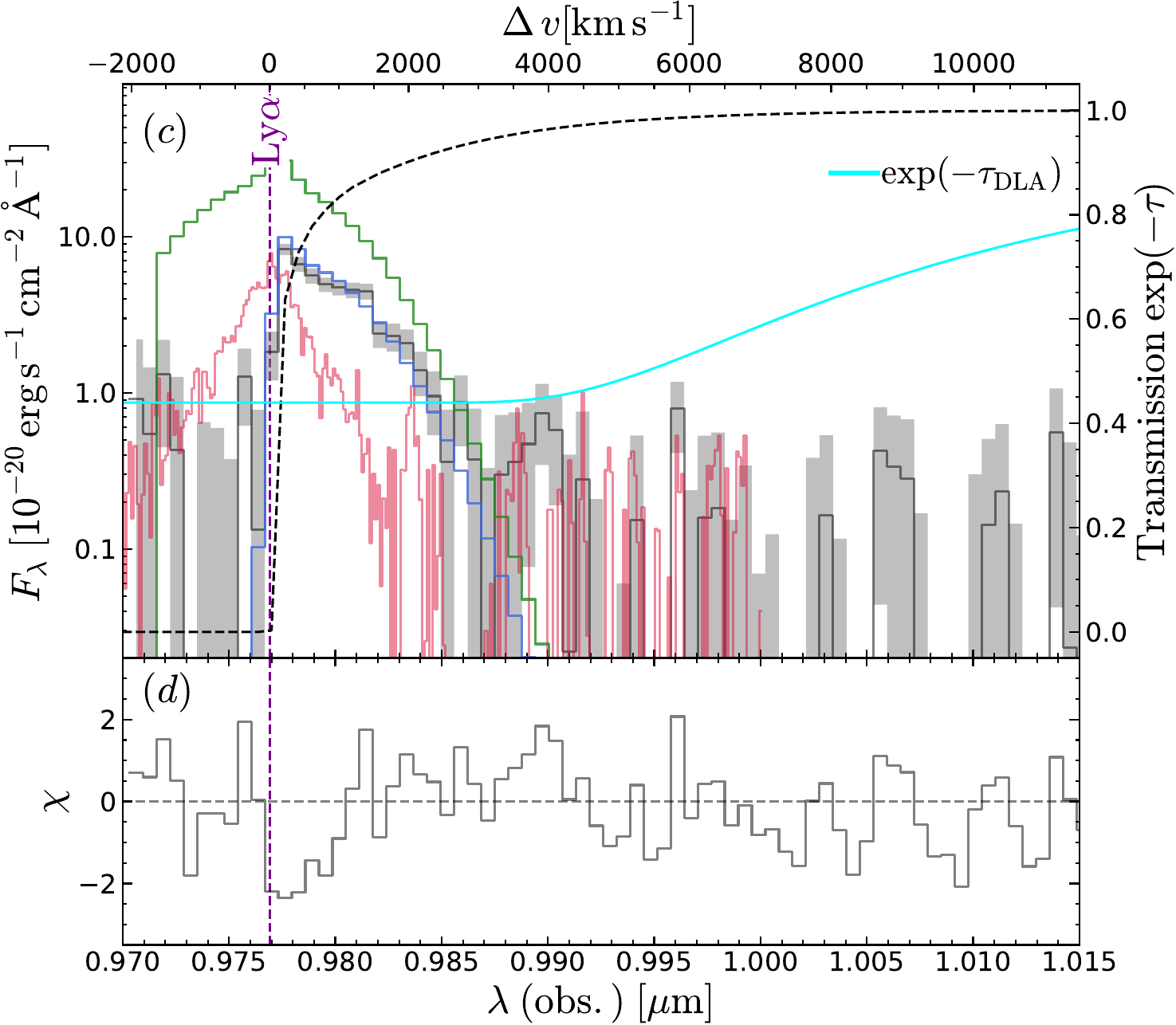}
    \caption{Comparison between the \lya and \ha profiles.
    The two lines are shown in the velocity space with the zero point set to the systemic redshift and the bottom axes are the observed wavelength space of the G140M spectrum.
    The best-fit spectral model consists of a narrow Gaussian, a broad Gaussian, and a linear continuum. Unlike in Figure~\ref{fig:spec_fit},
    here the intrinsic broad Gaussian model is symmetric (green line) but modified by the mean IGM transmission curve from \citet[][right-hand scale]{smith_lya_2022}
    to capture the observed
    asymmetry (blue line).
    The observed \ha line (red) is considerably narrower. We also considered an
    exponential-wing model for \lya (see text), which yields qualitatively similar results without
    worsening the fit. \textcolor{black}{In the bottom plot, we show a model with super-imposed DLA absorption, where the covering factor $C_f$ has been set to the value inferred by \citet{deugenio_qso1_2025}.}
    The \lya + DLA model is statistically not preferred over the non-DLA model.
    }
    \label{fig:profiles}
\end{figure}

A detailed comparison of the spectral profiles of \lya and \ha can shed light on the origin of the observed lines. In Figure~\ref{fig:profiles}, we overlay \ha from the NIRSpec R2700 spectrum \citep[see][]{deugenio_qso1_2025} onto the observed \lya profile present in this work.
There is a significant difference between the FWHMs of two lines, which is larger than the resolution difference between the R2700 and R1000 spectra and suggests that either additional broadening of \lya (e.g., due to resonant scattering of the broad line) or strong absorption near the core.
We tested the latter scenario with the mean IGM transmission curve from the \textsc{thesan-1} simulation \citep[][their Fig.~12]{smith_lya_2022}.
We started by modelling the narrow and broad \lya with Gaussians; the broad Gaussian is centred on the systemic redshift, while the narrow line has an additional velocity offset. 
We also added a linear continuum model.
The impact of the continuum model is negligible due to the relatively low continuum S/N compared to \lya, and we further discuss physical models for the continuum in Section~\ref{sec:uv_continuum}.
At each model evaluation, before convolving with the instrumental
LSF, we multiplied the model by the IGM transmission curve.
The resulting model is shown in the top panel of Figure~\ref{fig:profiles}, where the best-fit \lya profile before the IGM attenuation is $\rm FWHM=2,800\pm200$~\kms, significantly broader than that of \ha.

We also considered an alternative BLR model where the intrinsic broad-line \lya is convolved with a symmetric exponential kernel, following \citet{Rusakov_escattering_2025}. 
This fit is to test whether the \lya profile can be explained with an electron scattering-dominated model.
In this model, the scale of the exponential kernel and the scattered fraction are set by the temperature and Thomson optical depth of the scattering medium.
The probability of the model parameters is estimated with a Bayesian approach, where we use informative prior probabilities, motivated by the different spatial morphology of narrow \lya (see Section~\ref{subsubsec:lya_nircam}). 
Our priors on the narrow velocity and velocity dispersions are the posterior probabilities in the first row of Table~\ref{tab:lines}. 
We further set the
redshift prior to \ha-inferred measurement, and the narrow \lya flux is penalized against very large values (we use an \texttt{erfc} prior with mean equal to the Case B \lya flux predicted from \ha, and standard deviation 10\%).
The exponential-kernel model yields an intrinsic $\rm FWHM=2,600\pm300$~\kms very similar to that of the Gaussian model, and an exponential-kernel scale of $W = 1,000\pm300~\kms$, although the posterior distributions of these parameters are strongly anti-correlated.
The \lya line widths from both models are larger than the width of the widest Gaussian component found in \ha \citep[$\rm FWHM=2,000\pm200$~\kms assuming the broad \ha is composed of two Gaussians;][]{deugenio_qso1_2025}, and much larger than the observed width of the full broad \ha ($\rm FWHM = 680\pm80~\kms$), demonstrating that broad \lya and \ha have different emergent spectral profiles.

Since \ha displays prominent rest-frame absorption \citep[optical depth at line centre $\tau_0({\rm H\alpha})=1.2_{-0.3}^{+0.6}$, covering factor $C_f=0.56\pm0.07$;][]{deugenio_qso1_2025}, we also expect a counterpart of this absorber with neutral hydrogen at $n=1$ (i.e., the ground state). 
To assess if the profile mismatch between broad \lya and \ha can be explained by this absorption, we considered a set of models where the \lya intrinsic profile has the same shape as the double-Gaussian \ha model from \citet[][their Tab.~2; for the present test, their alternative exponential model yields the same qualitative results]{deugenio_qso1_2025}, but we added $n=1$ absorption using a damped \lya absorber\footnote{We used the publicly available implementation \textsc{lymana\_absorption} \citep[][\url{https://github.com/joriswitstok/lymana_absorption}]{witstok_z13_2025}.} (DLA), with the \lya absorption profile as described in \citet{witstok_z13_2025}.
While the intrinsic profile shape is fixed, the broad \lya flux is a free parameter. We also imposed the mean IGM absorption at $z=7$ from \citet{smith_lya_2022}.
The DLA is parametrized with a covering factor, $C_f$, and a column density of neutral hydrogen, \Nhi.
We tested models where the DLA applies to both broad \lya and the continuum, to broad \lya alone, using a variable $C_f$ \citep[with prior probability $=0.56\pm0.07$ set by the posterior probability from the \ha fit in][]{deugenio_qso1_2025}, and we considered a model with fixed $\Nhi=2.7\times 10^{21}~\mathrm{cm}^{-2}$ \citep[as published in][obtained with \textsc{Cloudy} modelling of the PRISM spectrum]{ji_lrdbreak_2025}. 
To evaluate the performance of different models statistically, we used the Bayesian Information Criterion \citep[BIC,][]{Schwarz_BIC_1978,liddle_2007} defined as
\begin{equation}
    {\rm BIC} \equiv \chi ^2 + k\,\ln n,
\end{equation}
where $k$ is the number of free parameters and $n$ is the number of data points.
We considered $\rm \Delta BIC >10$ as evidence for significant model improvement (e.g., $\rm BIC_{A}-BIC_B>10$ means that model B is significantly better).
We found that all the \lya + DLA models are statistically worse than the model without DLA, with $\Delta\,\text{BIC}$ values ranging from $-100$ to $-35$.
The reason is likely that the DLA models fail to simultaneously suppress the bright core of the \ha-like profile (which requires high \Nhi) and preserve sufficient flux at $\sim 1,000~\kms$ from the the line centre. As a result, all these fixed-profile models display visible absorption near 1,000~\kms, which is not present in the observed spectrum.
An alternative, more flexible DLA model with free broad \lya width still fails near the line core, by over-predicting the \lya flux, which is shown in the bottom panel of Figure~\ref{fig:profiles}. These tests reinforce the finding that the broad \lya is broader than \ha, most likely due to RT effects in the neutral gas within or close to the BLR.

We emphasize that rejecting models with DLA absorption of \lya does not rule out the presence of $n=1$ absorption -- which is required by the observed $n=2$ absorption and further supported by modelling the UV spectrum \citep{ji_lrdbreak_2025}. Instead, the shape of the \lya profile suggests a scenario where broad \lya arises from (or is at least affected by) a relatively unobstructed LOS, in contrast to the heavily embedded Balmer emission in the optical \citep{ji_lrdbreak_2025}.
This implies an anisotropic geometry for the gas surrounding the BLR in \target, which we discuss next.

\subsubsection{Geometry of the BLR clouds}

\begin{figure*}
    \centering
    \includegraphics[width=\columnwidth]{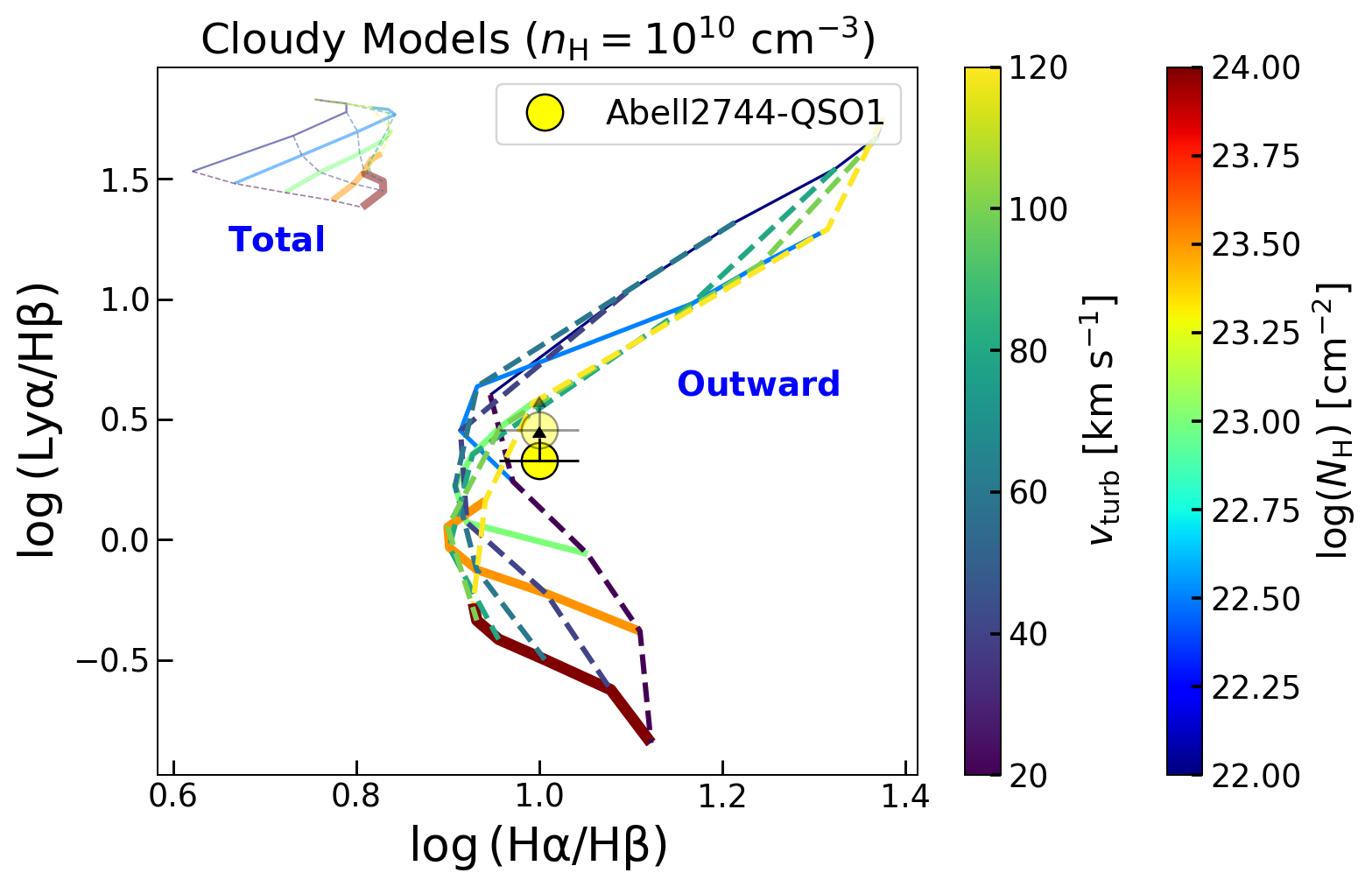}
    \includegraphics[width=.93\columnwidth]{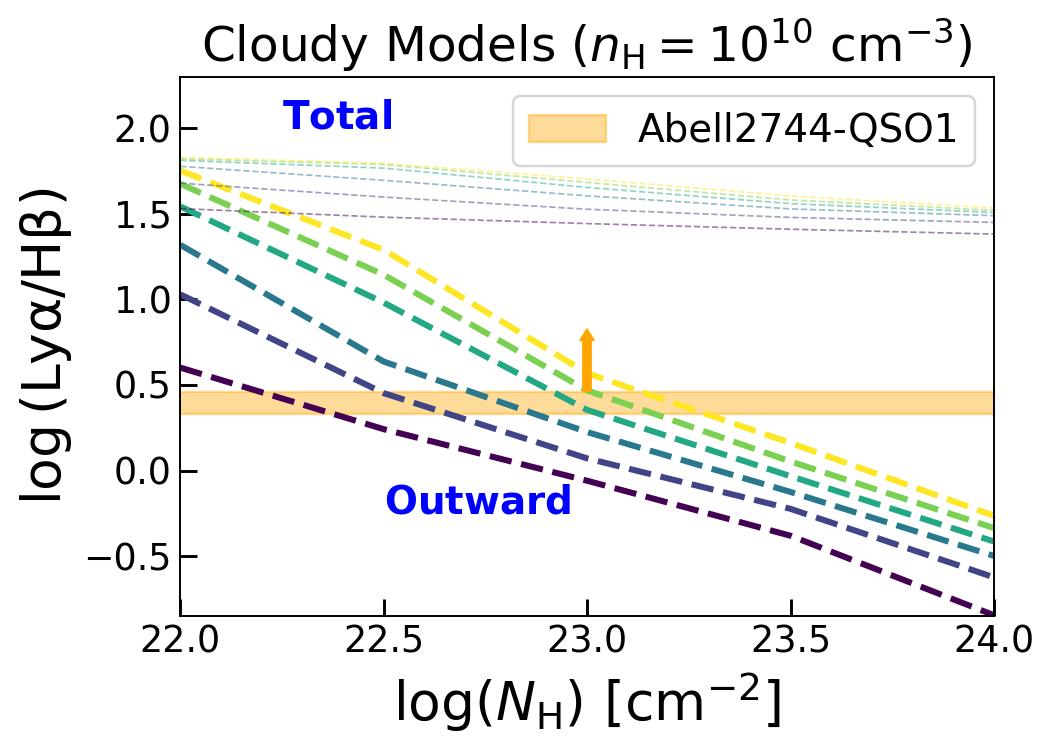}
    \caption{
    Comparison between the observed ratios of broad H\,{\sc i} lines with those predicted by \cloudy BLR models as a function of turbulence velocity and hydrogen column density at a fixed hydrogen volume density of $n_{\rm H}=10^{10}~{\rm cm^{-3}}$.
    The solid (dashed) grid lines denote constant column densities (turbulence velocities).
    The ``Total'' model grid sums emission emergent from both sides of the simulated cloud, describing approximately the case of a BLR consists of clouds equally distributed with front and back illumination and assuming that the inwardly beamed photons are not absorbed by another central source of opacity.
    In contrast, the ``Outward'' model grid considers a BLR cloud is along the LOS and the observer only sees the transmitted emission.
    The solid (transparent) symbol represents the line ratios of \target, where broad (total) \lya is used.
    The left panel shows that the line ratios of \target disfavor the case where clouds are equally distributed with back-illuminated geometry and face-illuminated geometry.
    The observed \lya-to-Balmer line ratio limit lies closer to the Outward model, and prefers $N_{\rm H}\lesssim 10^{23}~{\rm cm^{-2}}$ as shown in the right panel, lower than $N_{\rm H}\gtrsim 10^{24}~{\rm cm^{-2}}$ inferred from Balmer lines and optical continuum \citep{ji_lrdbreak_2025}.
    This indicates a distribution of column densities in the gas clouds with low-column density ``holes'', which might modulate the escape of \lya \citep[although this does not mean \lya escapes mostly through the holes, see][]{Monter_lya_2026}.
    }
    \label{fig:lyahbha_blr}
\end{figure*}

As we discussed above, the bulk of \lya seen in \target is likely of BLR origin.
This would allow us to probe the BLR conditions by combining \lya with the broad Balmer lines measured in the R2700 spectra of \target \citep{ji_lrdbreak_2025,deugenio_qso1_2025}.
Taking the flux of broad \hb measured from the NIRSpec/G395H IFU observation of the image A of \target described in \citet{ji_lrdbreak_2025} and converted it to the flux in image B, the observed broad-line ratio between \lya and \hb is $F_{\rm Ly\alpha}/F_{\rm H\beta} = 2.0$\,--\,2.6 depending on the lensing model, significantly below the Case B value of 24 \citep{storey_hummer_1995}.
Several factors can contribute to this deviation, including dust reddening, resonant scattering, and collisional excitation \citep{Ferland_1985,cloudy_2301}.
Regarding the dust reddening, it has been argued that LRDs actually have little dust obscuration, which is in part supported by the narrow-line Balmer decrements close to the Case B values \citep{deugenio_qso1_2025,ji_lord_2025,linxiaojing_locallrd_2025} and the weak dust emission in the infrared \citep[IR;][]{Setton_lrddust_2025,Casey_2025}.
For \target, the dust attenuation for narrow lines is consistent with zero after correcting for aperture losses \citep{deugenio_qso1_2025}.
Still, the Balmer decrement might not reflect the dust optical depth of \lya, since the latter can have additional path length increased by $\sim 100\times$ due to trapping.
Also, this does not exclude the possibility that the BLR is dust obscured.
The broad-line Balmer decrement for \target reaches a high value of 10, which, however, can be due to high opacities of Balmer lines in the BLRs of LRDs \citep{juodzbalis_rosetta_2024,ji_lord_2025,degraaff_2025,yanzu_lrdbal_2025}.
We leave the thorough investigation on the dust content of \target to future work as currently we lack information in the IR regime of this LRD.
In what follows, we investigate whether dust-free BLRs, as often assumed for LRDs, can reproduce the observed line ratios.

We computed photoionization models with \cloudy \citep[v17.03,][]{ferland2017} to investigate the impact of the hypothesized gas envelopes in LRDs on the broad hydrogen line ratios\footnote{We do caution that, the \cloudy computation of \lya is subject to the assumption of the redistribution function, which by default is assumed to be incomplete redistribution \citep[see e.g.,][for details]{Elitzur_1986}.
We have checked the result with complete redistribution and do not find any significant difference.
Exploring detailed partial redistribution is beyond the scope of this work and we leave it for future studies with detailed RT calculations.
}.
The model parameters are summarized in Table~\ref{tab:hi_models}.
For the fiducial model, we have assumed a metallicity of 1\% solar.
This is motivated by the measurement of \citet{Maiolino_metalpoorlrd_2025}, where the narrow-line metallicity of \target is less than 1\% solar.
We adopted an ionization parameter (defined as the ratio between the ionizing photon flux and the product of the gas density and the speed of light) of $\log U=-2$, which is a representative value for describing the BLR of some LRDs including \target \citep{ji_lrdbreak_2025}.
The exact value of $U$ has little impact on the hydrogen line ratios we investigate here.
We fixed the volume density to $n_{\rm H} =10^{10}~{\rm cm^{-3}}$, which is again a representative value to reproduce the observed spectra of LRDs \citep{Inayoshi_maiolino_2025,ji_lrdbreak_2025}.
In Appendix~\ref{appendix:tba} we show models with other densities, which poorly match the observed line ratios.
We considered a range of hydrogen column densities from $10^{22}~{\rm cm^{-2}}$ to $10^{24}~{\rm cm^{-2}}$, to test the scenario where the observed broad lines are leaked from directions with different optical depths.
We also set a range of turbulence velocity from 20 \kms to 120 \kms, which either describes microturbulence in BLRs or shear motions of BLR clouds within the line-emitting regions \citep{bottorff2000}.
As shown by \citet{ji_lrdbreak_2025} and \citet{deugenio_qso1_2025}, the highest turbulence considered here likely corresponds to gas clouds responsible for producing the Balmer-line absorption seen in \target.
For the ionizing SED, we used the AGN model SED of \citet{pezzulli_2017} at a mass of $M_{\rm BH}=10^7~\text{M}_\odot$ and an Eddington ratio of $\lambda _{\rm Edd}=0.1$ following \citet{ji_lrdbreak_2025}.
Finally, the geometry is set to plane-parallel, which is an open geometry considering gas obscuration along the LOS.
We also consider the case where both sides of the cloud are seen due to a distribution of BLR clouds, which is approximated by summing the predicted radiation from the front and the back of the cloud.
Such a case requires an exactly symmetric distribution and no obscuration in the centre of the distribution, also ignoring RT effects.
Both cases are likely not representing the actual physical geometry.
Regardless, our purpose is to estimate the possible physical conditions indicated by \lya and we leave the detailed RT calculations to future work.

In Figure~\ref{fig:lyahbha_blr}, we compare the measured broad-line ratios of \lya/\hb and \ha/\hb in \target with those predicted by \cloudy BLR models.
Note that the modelled \lya/\hb ratios should be taken as theoretical upper limits (equivalently, we plot the observed ratio as a lower limit), because our \cloudy models neglect some \lya attenuation mechanisms, such as 
dust extinction, low IGM transmission (see Section~\ref{sec:igm}), and scattering-boosted $l$-changing collisions (see \citealp{torralba+2025}. Neglecting these effects is a conservative choice for our conclusions below.
There are three noticeable features from this diagram.
First, the broad-line ratios can be well explained by BLR clouds along the LOS.
Even Balmer lines become optically thick in such clouds \citep[see e.g.,][]{yanzu_lrdbal_2025}, reproducing the large Balmer decrement and reducing the \lya-to-Balmer line ratios without the need of dust.
Second, as expected, the model shows that \lya escapes more easily with higher turbulence velocities and smaller column densities.
At moderate turbulence of $v_{\rm turb}\lesssim 100$ \kms assumed here, \lya has to escape at $N_{\rm H}\lesssim 10^{23}~{\rm cm^{-2}}$, a column density which is much smaller than the typical values of $N_{\rm H}\gtrsim 10^{24}~{\rm cm^{-2}}$ invoked to reproduce the spectral shape of LRDs \citep{Inayoshi_maiolino_2025,ji_lrdbreak_2025}, as well as to produce the hypothesized electron scattering-dominated profiles of broad Balmer lines under typical ionization conditions \citep[requiring $N_{\rm e}\sim 10^{24}~{\rm cm^{-2}}$ and $N_{\rm H}\sim 10^{25}~{\rm cm^{-2}}$;][]{Rusakov_escattering_2025,Sneppen_2026}.
While a turbulent velocity much higher than 100 \kms would allow \lya escape at higher $N_{\rm H}$, this would be in conflict with the upper limit of $v_{\rm turb}\approx 120$ \kms for hydrogen-line emitting/absorbing clouds constrained from Balmer absorptions \citep{ji_lrdbreak_2025,deugenio_qso1_2025}.
We also note that a strong microturbulence (i.e., occurring on physical scales of line production zones) would imply a large energy dissipation and heating, and thus alternative scenarios with nondissipative magnetohydrodynamic (MHD) waves and velocity shears of continuous winds have been proposed to mimic the effects of microturbulence \citep{bottorff2000}.
The simultaneous requirement to have high and low column densities suggests that \lya might have escaped from directions with less opacity, although the final \lya profile would still be at least partially shaped by optically thick directions according to \citet{Monter_lya_2026}.
Third, the ``Total'' model predicts line ratios too close to the Case B values compared to observations.
The better consistency between the observations and the ``Outward'' model suggests a back-illumination dominated geometry, although more detailed RT calculations are needed to verify.


All the above features suggest that while the observed broad lines can have a common BLR origin with no dust attenuation, the geometry of the dense gas envelope cannot be a fully closed sphere with uniform column densities. Instead, \lya needs to escape from relatively optically thin directions.
The interpretation of the geometry thus challenges the $\rm BH^\star$ scenario, where the BLR clouds are embedded in a uniform gaseous environment with a near unity covering factor \citep{naidu_lrd_2025,degraaff_2025}.
A possible geometry consistent with our analysis could be an inflated, optically thick, but relatively dust-poor ``torus'' with large scale heights compared to the BLRs of normal AGN, and \lya emission might escape from the polar directions with lower optical depths \citep[see also][]{linxiaojing_locallrd_2025,matthee_2026}.
Such a geometry would have implications on the demographics of LRDs or broad-line sources at high $z$ in general \citep[e.g.,][]{pm_2026}. 
We defer a more thorough statistical study to a separate paper (Geris et al. in preparation).
The picture of escaped \lya emission is also supported by other nebular features in the FUV of \target, which we discuss in the following subsections.

\subsection{Excitation of OI}

\subsubsection{Line strength}

\begin{figure*}
    \centering
    \includegraphics[width=\columnwidth]{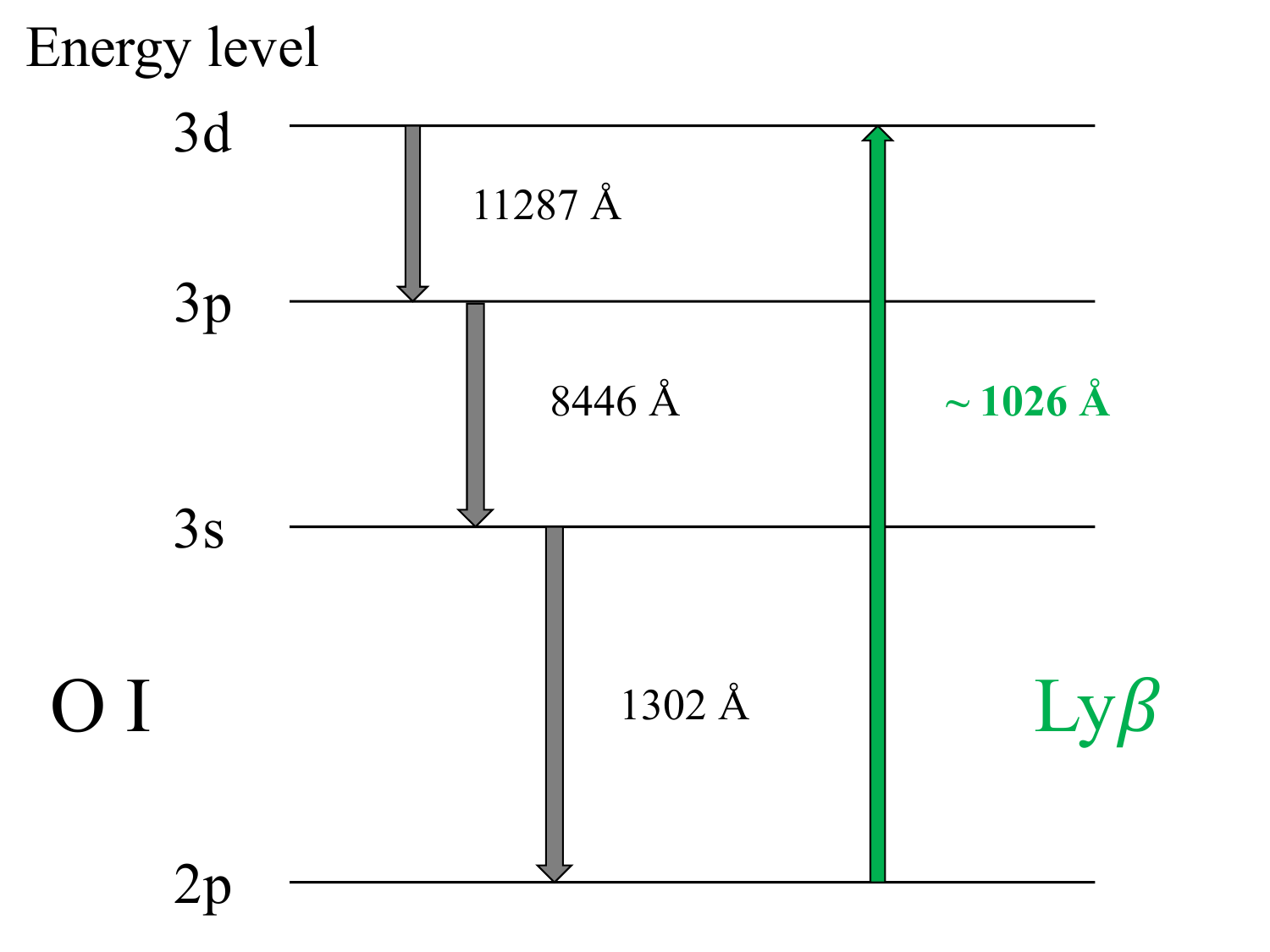}
    \includegraphics[width=\columnwidth]{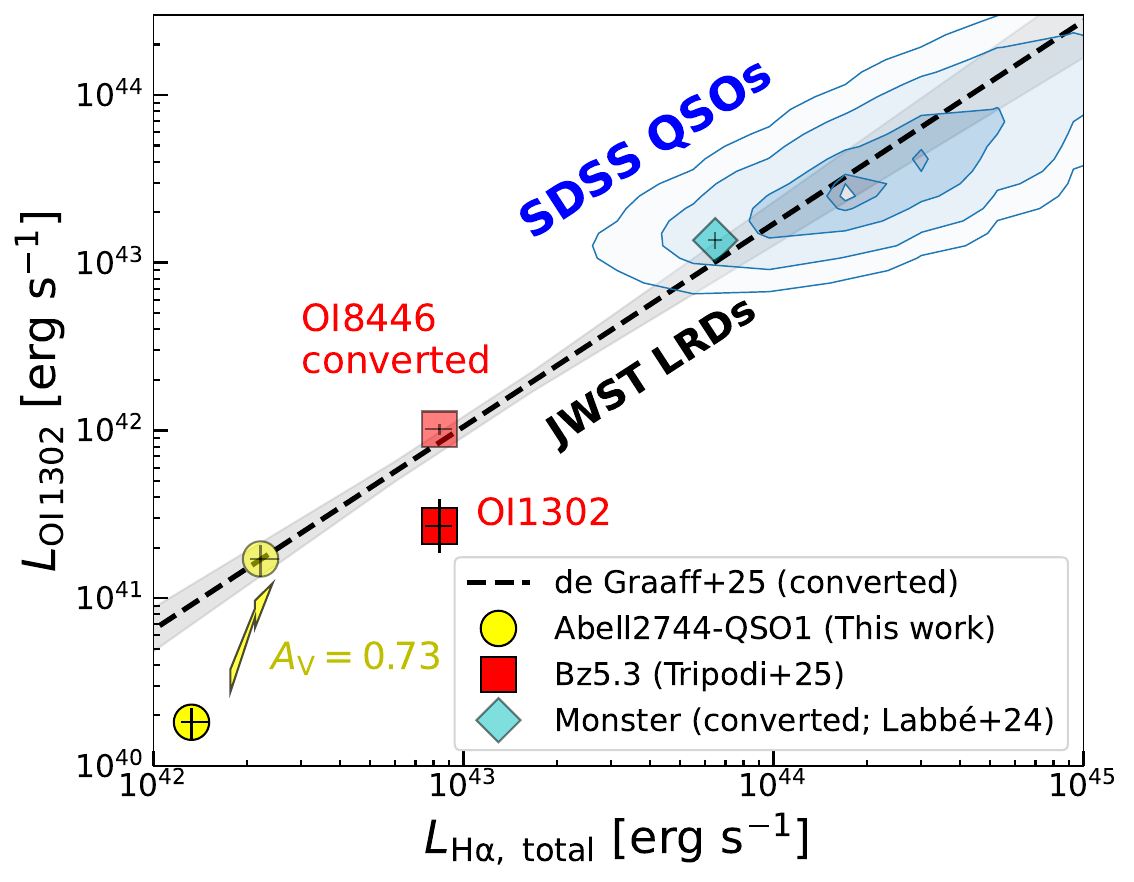}
    \caption{
    \textit{Left:} Part of the energy level diagram of \oip.
    \lyb fluorescence can pump the 3d level of \oip, leading to subsequent emission at 11287 \AA, 8446 \AA, and 1302 \AA, respectively.
    \textit{Right:}
    Comparison between \target, Bz5.3 (LRD with \oip detections reported by \citealp{tripodi2025deepdivebroadlineregion}), general LRDs at high-$z$ compiled by \citet{degraaff_2025}, and SDSS QSOs compiled by \citet{wu2022} in terms of (converted) \oip luminosities and \ha luminosities.
    The \oip$\lambda 1302$ luminosities from \citet{degraaff_2025} are converted from \oip$\lambda 8446$ assuming \lyb fluorescence as the only excitation channel.
    The dashed line is the best-fit relation and the shaded region represents the $1\sigma$ uncertainty.
    The SDSS $L_{\rm H\alpha}$ is converted from $L_{\rm bol}$, and the 5 contour levels represent 5, 16, 50, 84, 95 percentiles of the density distribution.
    We highlight an extreme LRD, A2744-45924 (also known as `Monster'), at $z=4.47$ reported by \citet{labbe_monster}, which reaches QSO luminosities.
    Both LRDs with direct \oip$\lambda 1302$ measurements (\target and Bz5.3) sit below the relation of general LRDs.
    However, if we took the \oip$\lambda 8446$ flux measured in Bz5.3 and converted it to \oip$\lambda 1302$ under the pure fluorescence case, the result (transparent square) follows the \jwst LRD relation.
    If the lower \oip$\lambda 1302$ in \target is due to dust attenuation, $A_{\rm V}\approx 0.7$ mag is required to make \target consistent with other LRDs, which is similar to Bz5.3 ($A_{\rm V}= 0.65\pm 0.20$ mag inferred by \citealp{tripodi2025deepdivebroadlineregion}).
    }
    \label{fig:o1ha_relation}
\end{figure*}

As shown in Figure~\ref{fig:spec_fit}, there is a $\sim 3.7\sigma$ detection of the line \oip$\lambda 1302$ in the FUV spectrum of \target, which is a typical resonant transition observed in QSOs and can be produced by collisional excitation, recombination, and fluorescence \citep{grandi_1980}.
Recombination is extremely unlikely, given the low metallicity of LRDs -- and \target in particular \citep{Maiolino_metalpoorlrd_2025}. In contrast, collisional excitation and fluorescent excitation from \lyb photons are both plausible, since they can be effective even in low-metallicity environments and only require high density -- a condition that is universally accepted for LRDs \citep{Inayoshi_maiolino_2025,ji_lrdbreak_2025}. 
Both collisional excitation and fluorescence can produce other \oip lines, including \oip$\lambda 8446$, which has been observed in individual LRDs \citep[e.g.,][]{juodzbalis_rosetta_2024,labbe_monster,killi_lrd_2023,wangbingjie_2024,Kokorev_feiinir_2025,linxiaojing_locallrd_2025,degraaff_2025}.
Direct observations of \oip$\lambda 1302$ (or the resonant triplet \oip$\lambda \lambda \lambda 1302,1304,1306$), on the other hand, have been reported by \citet{tripodi2025deepdivebroadlineregion} at $\sim 3\sigma$ for an LRD at $z=5.29$ (hereafter Bz5.3) in the NIRSpec/PRISM spectrum.
Our confirmation of \oip$\lambda 1302$ in \target makes it by far not only the detection at the highest redshift of $z=7.04$, but also the detection at the highest spectral resolution of $R\sim 1000$.

In the case of \lyb fluorescence, one expects a photon flux ratio of 1:1 between \oip$\lambda 1302$ and \oip$\lambda 8446$, as shown in the left panel of Figure~\ref{fig:o1ha_relation}.
The \lyb fluorescence case cannot be tested with the current data for \target due to the lack of rest-frame NIR observations to cover \oip$\lambda 8446$.
However, as reported by \citet{degraaff_2025}, there is a tight correlation between the luminosities of \oip$\lambda 8446$ and \ha.
Under the assumption that \lyb fluorescence is the only mechanism to excite \oip, and there is no dust attenuation and resonant scattering to extinguish \oip$\lambda 1302$ photons, we can convert the \oip$\lambda 8446$-\ha relation in LRDs to an \oip$\lambda 1302$-\ha relation by simply multiplying the luminosity of \oip$\lambda 8446$ by a factor of 8446/1302.
We do caution that there might be intrinsic scatters in the \oip-\ha relation, which depends on the optical depth of \ha (which enhances \lyb when trapped; \citealp{ferland_netzer_1979}), the metallicity, and whether the fluorescence is saturated or not.

In the right panel of Figure~\ref{fig:o1ha_relation}, we compare the \oip$\lambda 1302$ and \ha luminosities of \target to those converted from \oip$\lambda 8446$ and \ha luminosities of general high-$z$ LRDs shown by \citet{degraaff_2025} as well as those of SDSS QSOs compiled by \citet{wu2022}\footnote{Here we only include QSOs with S/N(\oip)\,$>5$.}.
Since \oip and \ha are not simultaneously covered in SDSS, we converted the bolometric luminosities of QSOs derived by \citet{wu2022} to the broad \ha luminosities using $L_{\rm bol}=130\,L_{\rm H\alpha}$ following \citet{sternlaor_2012}, where the intrinsic scatter in the bolometric conversion is roughly a factor of 2.
Overall, LRDs and QSOs follow a similar \oip--\ha relation, where QSOs occupy the high luminosity end.
Some extreme LRDs, such as the ``Monster'' \citep{labbe_monster}, reach the QSO regime and still follow the relation.

In Figure~\ref{fig:o1ha_relation}, we also plot the LRD, Bz5.3, which has direct detection of \oip$\lambda 1302$ as well \citep{tripodi2025deepdivebroadlineregion}.
Both \target and Bz5.3 with direct \oip$\lambda 1302$ measurements sit below the converted relation, suggesting suppression of \oip$\lambda 1302$ compared to the \lyb fluorescence case.
Since \oip$\lambda 1302$ is a resonant transition, one possibility is the loss of \oip$\lambda 1302$ due to scattering in an open geometry.
When \oip$\lambda 1302$ becomes optically thick, it might also be lost to other weak cascades from the 3s level, such as \oip$\lambda 1641$ or even \oip$\lambda 2324$ \citep{Grandi_1983}.
{Since the cross-section of \oip is lower than \hi\ with an oscillator strength ratio of roughly 0.3, and the column density ratio of $N_{\rm OI}/N_{\rm HI}\approx 0.625\,N_{\rm O}/N_{\rm H}\approx 10^{-6}$ in the ISM of \target \citep{Netzer_oiha_1976,Maiolino_metalpoorlrd_2025}, scattering of \oip$\lambda 1302$ is much less efficient compared to \lya with an optical depth ratio of $\tau_{\rm OI}/\tau_{\rm Ly\alpha} \sim 10^{-7}$\,-\,$10^{-6}$ in the case of \target \citep[see][]{Netzer_davidson_1979}.
Thus, in the same geometric configuration, if \oip$\lambda 1302$ in \target is indeed reduced by ISM scattering compared to the \jwst LRD relation by $\tau _{\rm OI}\sim 1$, \lya in the LOS should be nearly completely lost from scattering.
}

Alternatively, small amounts of dust close to the BLR might lead to the reduced \oip$\lambda 1302$.
While it has been suspected that LRDs lack the traditional dusty tori of AGN, the strength of FUV lines can still be significantly impacted at relatively low $A_{\rm V}$.
For Bz5.3, \citet{tripodi2025deepdivebroadlineregion} measured both the fluxes of \oip$\lambda 1302$ and \oip$\lambda 8446$ from the NIRSpec PRISM spectrum and concluded $A_{\rm V}=0.65\pm 0.20$ mag assuming pure fluorescence.
Interestingly, for Bz5.3, if we convert the luminosity of \oip$\lambda 8446$ to that of \oip$\lambda 1302$ in Figure~\ref{fig:o1ha_relation} as we did for other LRDs, this dust-corrected \oip$\lambda 1302$ value would be fully consistent with the \oip--\ha relation, unlike the observed \oip$\lambda 1302$.
Assuming that the \oip$\lambda 1302$ from \target suffers from dust attenuation, using the \citet{calzetti2000} dust attenuation curve, we inferred that $A_{\rm V}=0.73^{+0.10}_{-0.14}$ mag is needed to make it consistent with the \oip--\ha relation of LRDs/QSOs, similar to the case of Bz5.3.

According to \citet{Maiolino_metalpoorlrd_2025} and \citet{deugenio_qso1_2025}, the narrow-line Balmer decrement measured in the G395H spectrum of \target is consistent with the Case B value, suggesting nearly dust-free ISM.
Therefore, if \oip$\lambda 1302$ is indeed attenuated by dust, its emitting region is more dusty compared to the optical narrow-line emitting region.
If, following the interpretation of \citet{degraaff_2025}, \oip and broad \ha are co-spatial, there should be dust around the BLR.
However, given the much greater path length of \lya based on the optical depth analysis above, \lya would be nearly fully absorbed in this scenario even with moderate dust attenuation of \oip$\lambda 1302$.

Besides the loss by scattering and absorption, there are other possibilities to make \target sit below the converted \oip--\ha relation of LRDs/QSOs.
In the pure fluoresence case, the intensity ratio of \oip of \ha would depend on
\begin{equation}\label{eq:OI-Ha_theo}
    I(\mathrm{OI8446})/I(\mathrm{H}\alpha) \propto N_{\mathrm{OI}}/N_{\mathrm{HI}}\,\tau(\mathrm{Ly}\beta)\,/\epsilon (\mathrm{H}\alpha),
\end{equation}
where $\epsilon (\mathrm{H}\alpha)\propto 1/\sqrt{\ln \tau _{\mathrm{H}\alpha,0}}$ is the \ha escape probability \citep{Netzer_oiha_1976,grandi_1980}.
To reduce the ratio, one might either have i) a very low metallicity \citep{Maiolino_metalpoorlrd_2025} so that not enough $\rm O^{0}$ is present in the gas; ii) a lower \lyb (or \ha) optical depth in the \oip emitting region in \target compared to other LRDs \citep[see][]{ferland1979}.
Finally, it is possible that \lyb fluorescence is not the only excitation mechanism for \oip and collisional excitation is a ubiquitous and important mechanism in LRDs, which enhances \oip$\lambda 8446$ with respect to \oip$\lambda 1302$.
All of the above scenarios cannot be ruled out for \target.
However, for Bz5.3, since its \oip$\lambda 1302$ luminosity converted from \oip$\lambda 8446$ is in good agreement with other LRDs as shown in Figure~\ref{fig:o1ha_relation}, it cannot be explained with a low metallicity or less \lyb pumping.
In the case of pure collisional excitation, one expects a flux ratio of \oip$\lambda 7774$/\oip$\lambda 8446 \approx 0.3$, which cannot be ruled out for Bz5.3
\citep{tripodi2025deepdivebroadlineregion} but is largely ruled out for the ``Rosetta Stone'' LRD at $z=2.26$ \citep{juodzbalis_rosetta_2024}.
Clearly, more observational confirmations of \oip transitions in LRDs are needed to determine the dominant excitation mechanism.
Next, we examine the kinematics of \oip$\lambda 1302$.

\subsubsection{Kinematics}

To explain the \oip$\lambda 8446$-\ha correlation in \jwst-discovered LRDs, \citet{degraaff_2025} suggest that both lines might be produced in similar regions in the BLRs, where the production of \oip$\lambda 8446$ is dominated by \lyb fluorescence and the \ha might come from a combined effect of \lya trapping and collisional excitation.
However, in \target, it is clear that \oip$\lambda 1302$ is much narrower than \lya and other broad lines and its centroid is blueshifted.
The fact that \oip is much narrower than other broad lines is also noted by \citet{degraaff_2025}, although they mentioned that the broad \oip, even if exists, might be hard to recover in most LRDs where only NIRSpec PRISM observations are available.
There are cases where a broad component in \oip$\lambda 8446$ is confirmed in LRDs \citep[see][]{juodzbalis_rosetta_2024,Kokorev_feiinir_2025}, but a narrow component is also present in those cases and their relative contributions in general LRDs remain to be quantified.
The narrow \oip is in clear contrast to the SDSS QSOs presented by \citet{wu2022}, where over 99\% of the sources have $\rm FWHM(OI)>1000$ \kms.
The distinct kinematics of \oip in \target suggests \lyb fluorescence, if dominating the \oip production, occurs in regions separate from the sites of broad \ha production.
A possible configuration is that \oip is produced from clouds further away from the BLR, and a significant number of \lyb photons must have escaped from the BLR to pump the upper level of \oip fluorescent cascade (see Figure~\ref{fig:o1ha_relation}).
Notably, the velocity dispersion of \oip ($\sigma=180\pm60$ \kms) is higher than that of narrow \ha \citep[$\sigma=22\pm 6$ \kms,][]{deugenio_qso1_2025}, 
suggesting that it might come from a region different from the ISM.
We do caution that \oip$\lambda 1302$ is marginally resolved in G140M given an instrumental velocity dispersion of $\sigma _{\rm inst}\approx 200$ \kms at $\lambda \approx 1.05~\mu$m for a point-like source \citep{degraaff_lsf_2024}.

In the case of in-situ fluorescence, \oip$\lambda 8446$ and \ha are indirectly coupled through the ionizing flux from the accretion disc. As shown by \citet{Netzer_oiha_1976}, \ha trapping is a key element in the process (Equation~\ref{eq:OI-Ha_theo} above) and could explain the broad \oip$\lambda 8446$/\ha flux ratio seen in the nearby Seyfert NGC 4151. To trap \ha, the neutral medium needs to be warm with excited hydrogen at $n=2$. If however, as indicated by the different kinematics, \oip$\lambda 1302$ and \ha are not produced in the same spatial region, how should one interpret the \oip$\lambda 8446$-\ha relation? For fluoresence in a displaced region, one might imagine \lyb escape from the BLR and contributing to the \oip fluorescence cascade outside, but the difficulty associated with this scenario is that \lyb escape is much harder compared to \lya due to the conversion to \ha + \lya or two-photon emission when \lyb is scattered out. Alternatively, fluorescence in an \ha-thick medium may be initiated by \ha photons from the BLR. This is because \ha trapping pumps the $n=3$ level of hydrogen, which can lead to \lyb, hence \oip fluorescence.
In this scenario, a relatively low \oip$\lambda 8446$-\ha ratio could be explained by incomplete covering of thick (\oip-emitting) material around the source of \ha photons. Furthermore, in this case, \ha photons from the broad wings would be out of resonance for \lyb pumping given that the velocity dispersion of the neutral medium producing the \oip emission is low. The blueshift of \oip compared to the bulk \ha velocity further enhances the effect. This might be another explanation for the relatively low efficiency of \oip fluorescent cascade in \target compared to the median trend of QSOs.

Finally, we discuss the possibility that all FUV lines in \target come from the BLR, but are broadened differently.
In this scenario, the BLR clouds are optically thick to hydrogen lines, including \lya, due to Thomson scattering and these lines are broadened by thermal electrons \citep[see][]{Rusakov_escattering_2025,chang_2025,Sneppen_2026}, yet the clouds remain optically thin to \oip and \feii since these lines are produced in the outer BLR with much fewer free electrons.
In this scenario, the intrinsic, virial components in \lya, \oip, and \feii have similar widths, with $\sigma \sim 200$\,--\,300 \kms (see Table~\ref{tab:lines}), but the different scattering optical depths cause the different observed widths.
The line at 1550 \AA can only be associated with \feii in this case, as \civ would come from a more inner region (i.e., closer to the accretion disc) compared to hydrogen lines and be broadened by electron scattering, which is not seen.
The difficulty of this scenario for QSO1 is the unexplained intermediate velocity component in Balmer lines.
As shown by \citet{Maiolino_metalpoorlrd_2025} and \citet{juodzbalis_specast_2025}, in NIRSpec G395H IFU observations of \target, there is an intermediate broad ($\sigma \approx 200$ \kms) component in \ha and \hb that is spatially resolved on a scale of $\sim 300$ pc.
In contrast, the broad wings of Balmer lines remain spatially unresolved.
In the pure electron scattering scenario, the intermediate component would come from the unscattered BLR emission and should remain spatially unresolved.
Alternatively, one needs a $\rm BH^\star$ with an external outflow to explain the observations.
As we noted in Section~\ref{subsubsec:lya_profile}, an electron scattering-dominated, exponential-kernel model does not statistically outperform a simple double-Gaussian model in describing the \lya profile, 
although resonant scattering likely complicates the interpretation as we have discussed.
To better understand the scattering contributions to different lines, spectropolarimetry observations would be useful, as Thomson scattering should have changed the polarization of incident photons.

In summary, the detection of \oip$\lambda 1302$ in \target suggests \lyb fluorescence is likely an important process in LRDs.
However, the fact that \oip$\lambda 1302$ and \ha show distinct kinematics favours a scenario where the two lines are produced in different regions, indicating that \lyb trapping predominantly occurs in a spatial location different from where it is produced.

\subsection{High ionization versus low ionization}

Next, we discuss diagnostics related to the detection of \civ and/or \feii lines, which have very close wavelengths not separable at the resolution of NIRSpec grating spectroscopy.
Furthermore, the resonant nature of these lines makes the identifications even more difficult.
Therefore, in what follows, we discuss two different scenarios where the line found at $\sim 1550$ \AA is \civ and \feii, respectively.

\subsubsection{High-ionization case}

\begin{figure*}
    \centering
    \includegraphics[width=\columnwidth]{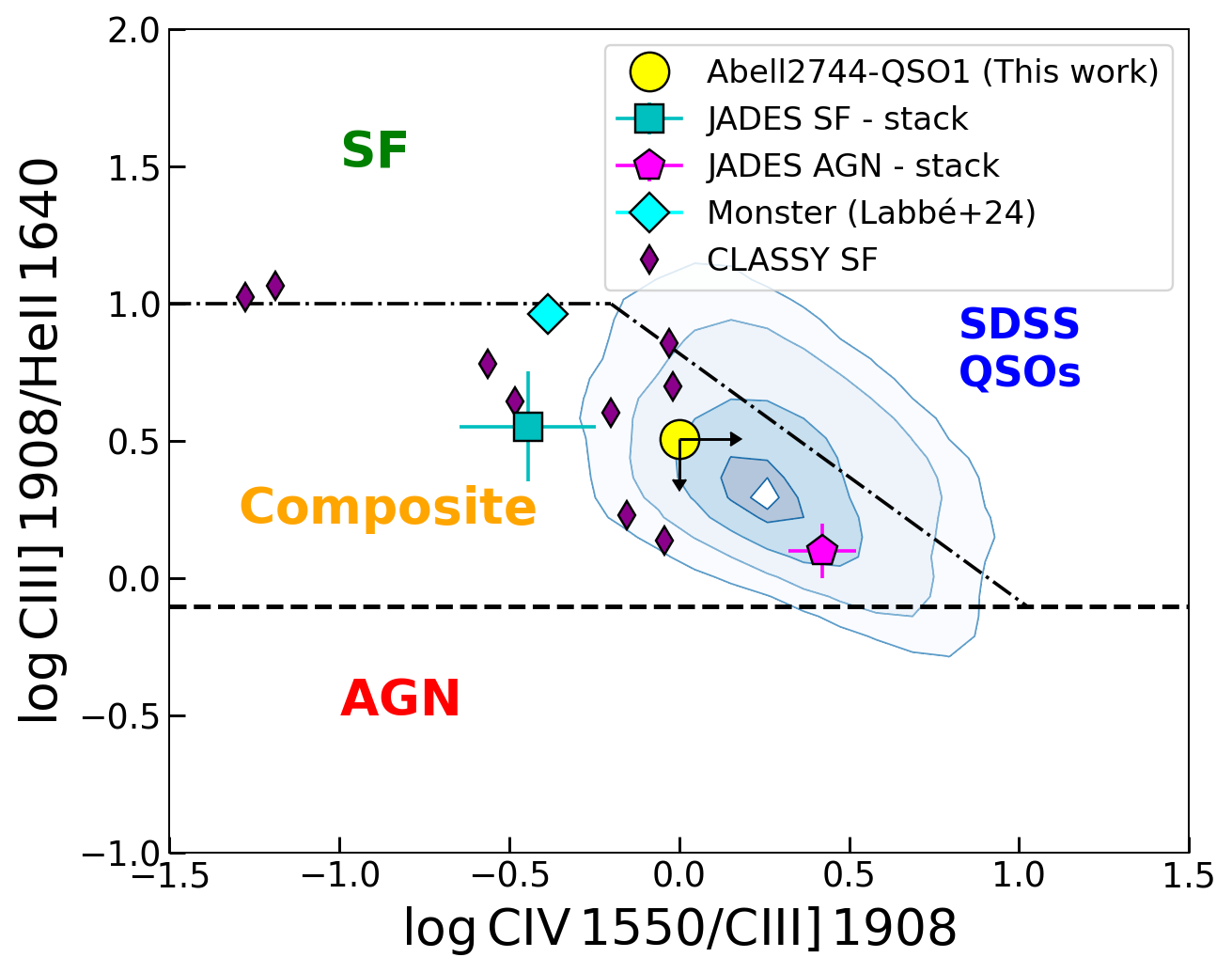}
    \includegraphics[width=\columnwidth]{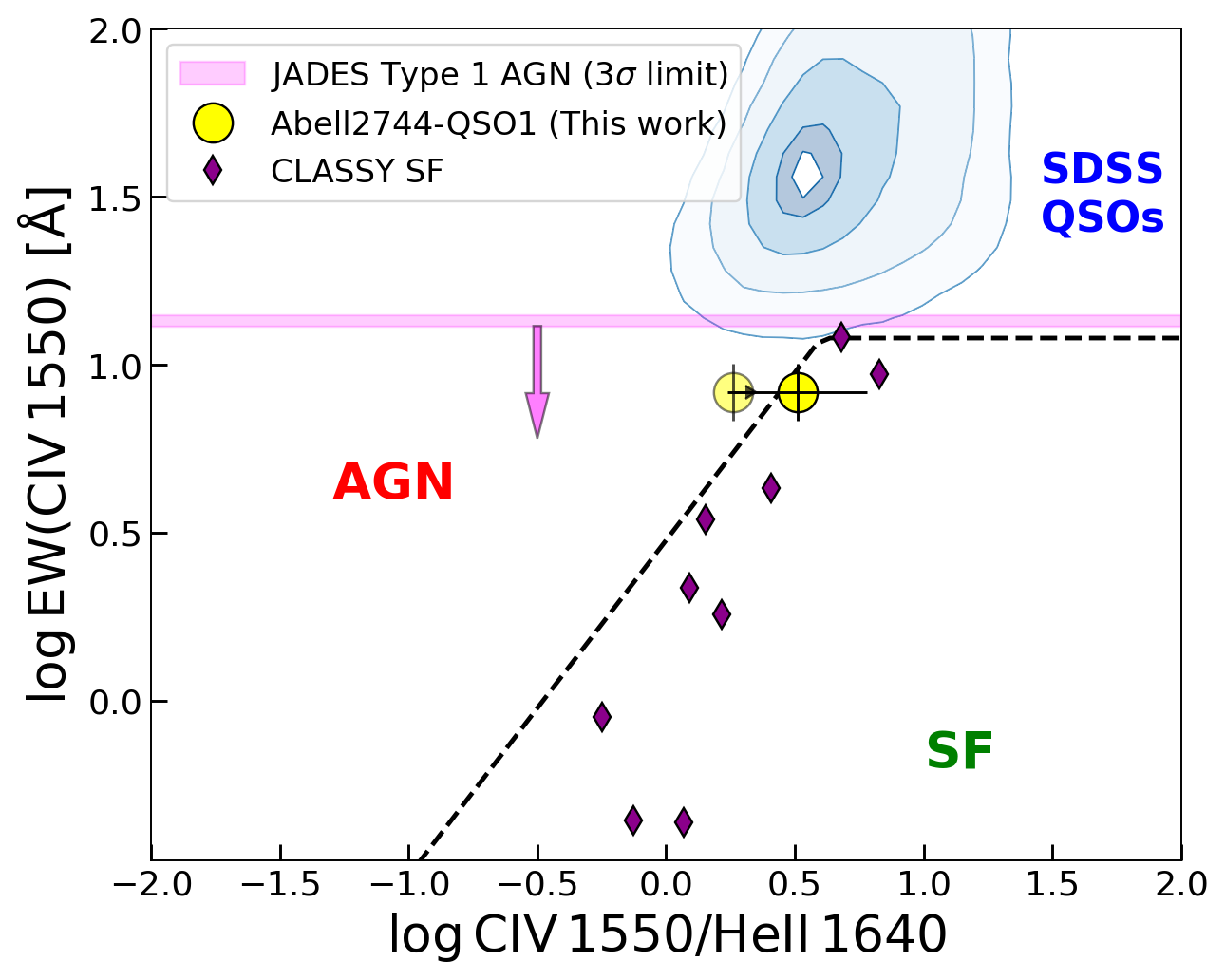}
    \caption{
    Diagnostic diagram using UV emission lines.
    The demarcation lines are from \citet{Hirschmann_2019}, and the data points are from \target (this work, assuming the 1550 \AA line is \civ), stacked spectra of SF galaxies and Type 2 AGN selected from the \jwst Advanced Deep Survey \citep[JADES,][]{rieke_jades_2023,eisenstein2023,eisenstein2023b,bunker_dr1_2024,deugenio_dr3_2024,scholtz_jadesdr4_2025} by \citet{scholtz2023}, local analogs of high-$z$ SF galaxies from the \textit{HST} COS Legacy Archive Spectroscopic SurveY \citep[CLASSY,][]{berg_2022,mingozzi_2024}, and SDSS QSOs \citep[][only sources with $\rm S/N>5$ in all line fluxes are shown]{wu2022}.
    The 5 contour levels of SDSS QSOs represent 5, 16, 50, 84, 95 percentiles of the density distribution.
    Since \civ is not decomposed into narrow and broad components in the SDSS catalog, we use the total fluxes of all emission lines for SDSS QSOs.
    Using only the narrow components of \ciii and \heii would shift the contours up in the left panel.
    In the right panel, we also plot the $3\sigma$ upper limits for EW(\civ) from the stacked spectra of Type 1 AGN in JADES at $3.5<z<5$ and $z>5$ \citep{Zucchi_2025}.
    The solid symbol for \target uses the $\sim2\sigma$ detection of \heii, whereas the transparent symbol treats \heii as non detection and uses the $3\sigma$ upper limit.
    Assuming the 1550 line in the FUV is \civ, the line ratios of \target put it in the zone between high-$z$ SF galaxies and AGN.
    }
    \label{fig:c4c3he2_diagram}
\end{figure*}

Thus far, it has been suspected that there is a general lack of high-ionization emission in the Type 1 AGN population, including LRDs, at $z\gtrsim4$ discovered by \jwst \citep{lambrides_superedd_2024,juodzbalis_jadesagn_2025,wang_heii_2025,Zucchi_2025}, where lines with ionization potentials close to or higher than that of \heii ($\sim 54$ eV) are missing.
This suggests that LRDs have intrinsically softer SEDs or mechanisms to remove EUV photons compared to QSOs that are generally featured with strong high-ionization lines, including \civ in the FUV.
Notably, the high-ionization lines might not be missing in all LRDs, as tentative identifications of N\,{\sc v}\xspace $\lambda 1240$ have been reported for an LRD at $z=6.98$ \citep{Tang_nv_2025} and both \niv$\lambda 1486$ and \civ$\lambda 1550$ are identified in an LRD at $z=4.47$ \citep{labbe_monster} -- and tentatively even at $z=8.6$ \citep{Tripodi_z8d6lrd_2025}.

If the line we measured at 1550 \AA in \target is dominated by \civ, we can perform UV diagnostic diagrams and evaluate the hardness of the ionizing radiation seen by the FUV lines.
In Figure~\ref{fig:c4c3he2_diagram}, we present two diagnostic diagrams proposed by \citet{Hirschmann_2019} involving \civ.
The left panel shows the \civ/\ciii versus \ciii/\heii diagram, where AGN tend to have enhanced \civ and \heii relative to star-formation photoionization, due to their harder ionizing SED.
If we use the $\sim 2\sigma$ detection of \heii and the $3\sigma$ upper limit on \ciii, \target is compatible with the distribution of the SDSS QSOs, despite the fact that the line ratios for the SDSS QSOs include both narrow and broad components.
Compared to high-$z$ galaxies and local analogues, the current limits on \target put it between the SF galaxies and Type 2 AGN.

The right panel of Figure~\ref{fig:c4c3he2_diagram} shows the \civ/\heii versus \civ equivalent width (EW) diagram.
Again, the measured values for \target put it on the edge of the SF-AGN demarcation.
The SDSS QSOs show systematically higher EWs in \civ, but we caution that these measurements include both narrow and broad components.
Interestingly, as shown by \citet{lambrides_superedd_2024} and \citet{Zucchi_2025}, \civ remains undetected even in the stacked spectra of Type 1 AGN selected by \jwst at $z>3.5$.
The $3\sigma$ upper limits on EW(\civ) reported by \citet{Zucchi_2025} with JADES Type 1 AGN at $3.5<z<5$ and $z>5$ are in a range of $\sim 13$\,--\,14 \AA, slightly higher than what is measured in \target, which is $8.3\pm 1.6$ \AA.

To conclude, if the 1550 \AA line is \civ, \target would lie between high-$z$ SF galaxies and AGN in terms of UV diagnostic diagrams.
The \civ detection would still be compatible with the non detection of \civ in high-$z$ Type 1 AGN selected by \jwst.
Interestingly, the existence of \civ, if truly powered by the AGN, would imply leakage of ionizing photons and formation of an NLR, suggesting the gas envelope in \target cannot have a unity covering factor and trap all the EUV photons.
Still, external, SF-dominated ionization cannot be ruled out, as shown by the CLASSY galaxies.
Next, we discuss the alternative scenario where the 1550 \AA line is not \civ, but actually \feii.

\subsubsection{Low-ionization case}

\begin{table}
        \centering
        \caption{Input parameters for \textsc{Cloudy} \feii models.}
        \label{tab:feii_models}
        \begin{tabular}{l c}
            \hline
            \hline
            Parameter & Values \\
            \hline
            $\rm [Fe/H]$ & $-2$, $-1$, 0\\
            \hline
            $\log U$& $-2$ \\
            \hline
            $\log (n_{\rm H}/{\rm cm^{-3}})$& 10 \\
            \hline
            $\log (N_{\rm H}/{\rm cm^{-2}})$ & 23, 23.5, 24, 24.5, 25 \\
            \hline
            $v_{\rm turb}$/\kms & 10, 60, 110, 160, 210 \\
            \hline
            Geometry & Open; plane-parallel\\
            \hline
            AGN SED & $M_{\rm BH}=10^7~M_\odot$, $\lambda _{\rm Edd}=0.1$\\ 
             & \citep{pezzulli_2017} \\
            \hline
            Dust & No dust\\
            \hline
            Atomic data & CHIANTI (v7, \citealp{chianti0};\\
             & \citealp{chianti_v7})\\
            \hline
            \feii data (UV) & \citet{Smyth_2019} \\
            \hline
        \end{tabular}
\end{table}

\begin{figure*}
    \centering
    \includegraphics[width=0.95\linewidth]{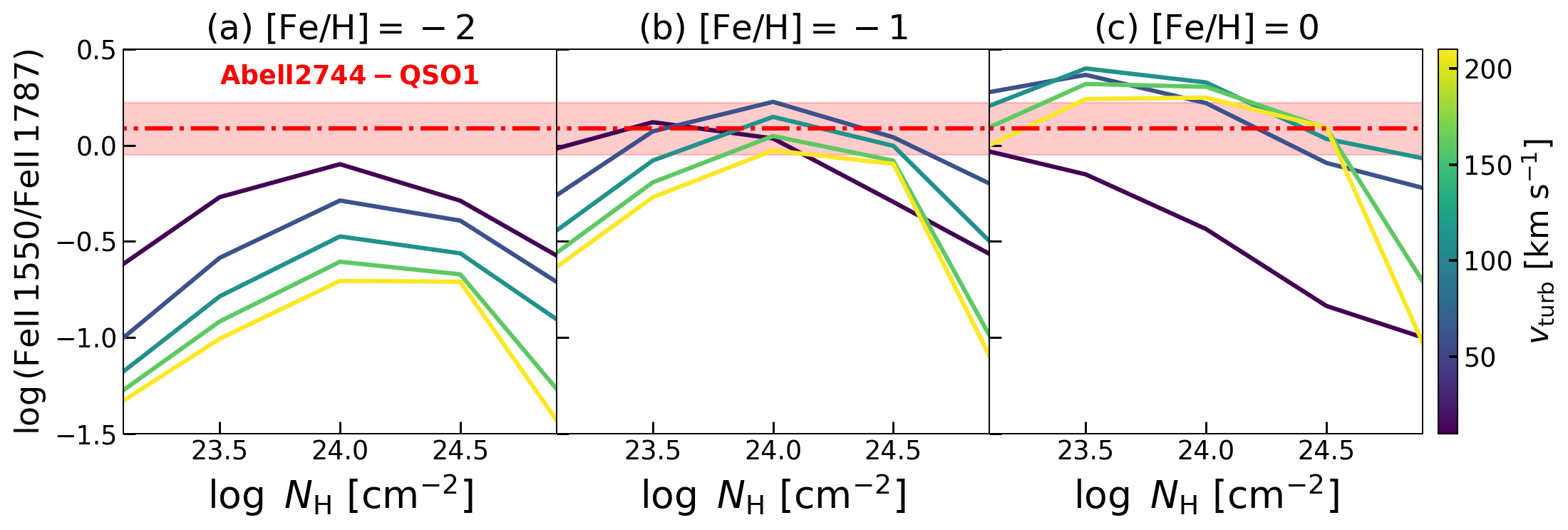}
    \caption{Comparison between the measured flux ratio between \feii$\lambda 1550$ (assuming no contribution from \civ) and \feii$\lambda 1787$ in \target and those predicted by \cloudy models.
    The strength of \feii$\lambda 1550$ is more sensitive to the iron abundance, suggesting that $\rm Fe/H$ needs to be higher than 1\% solar in the scenario where \civ is absent to match the measurement in \target. 
    }
    \label{fig:feii_r}
\end{figure*}

\begin{figure*}
    \centering
    \includegraphics[width=0.9\linewidth]{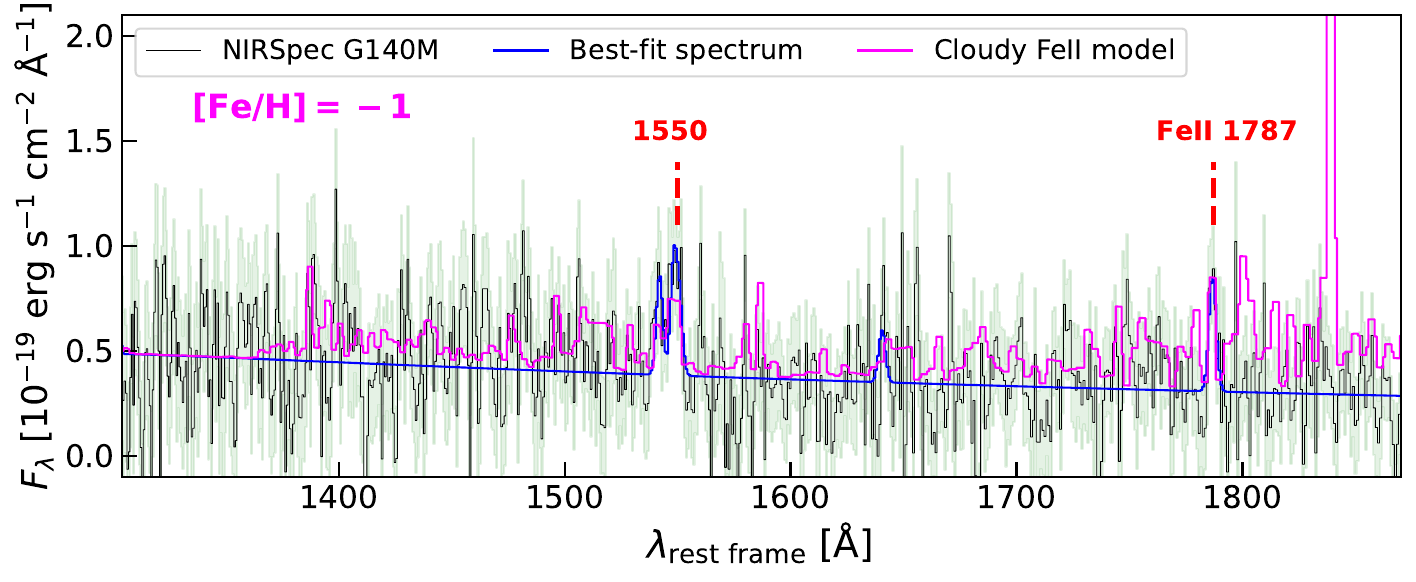}
    \includegraphics[width=0.9\linewidth]{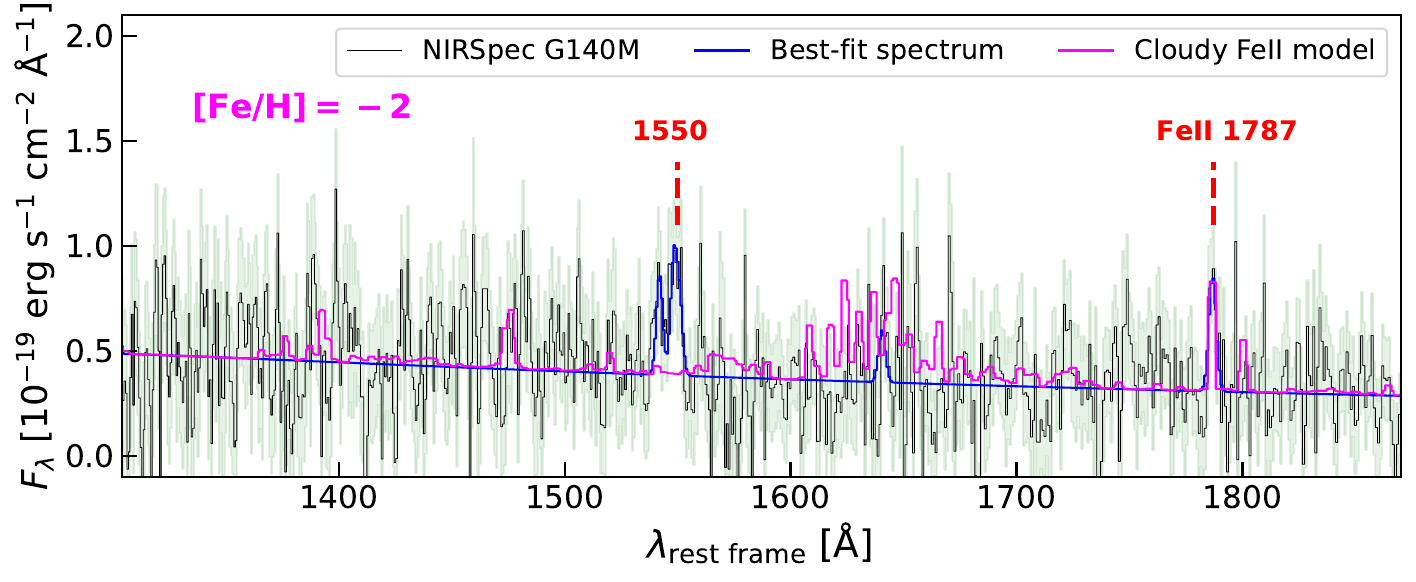}
    \caption{Comparison between the observed FUV spectrum of \target and the \feii model computed with \cloudy.
    The \feii model is convolved to the LSF of the G140M spectrum, normalized to the peak of the line observed at {1787} \AA, and added to the best-fit continuum model.
    \textit{Top:} Model assuming the iron abundance, Fe/H, is 10\% solar.
    While both 1550 \AA and 1787 \AA lines are predicted by the model, the model predicts more detectable transitions of \feii compared to the observation.
    \textit{Bottom:} Model assuming the iron abundance, Fe/H, is 1\% solar.
    In this cases, only the 1787 line remains strong as it is mainly pumped by \lya and is less sensitive to the abundance.
    }
    \label{fig:feii_eg}
\end{figure*}

Given the presence of the line at the rest-frame 1787 \AA, which is presumably associated with the \feii UV 191 group, we explored the second scenario where the line at 1550 \AA is also \feii.
Indeed, there is another known \feii group at 1550 \AA and both 1550 and 1787 groups have been reproduced in photoionization models of BLRs \citep{baldwin2004}.
A key difference, however, between \feii lines from BLRs of normal AGN and \feii lines identified in \target, is the spectral profiles.
The virial motions of BLR clouds tend to broaden \feii, and the \feii lines come with \textit{groups} rather than individual lines \citep[e.g.,][]{vestergaard2001}.
The fact that only individual \feii lines narrower than the broad hydrogen lines are observed in \target suggests that they probably have an origin in the outskirt of the BLR.
It has been noted that the \feii 1787 group can be produced by \lya fluorescence since its upper level can be pumped by 
$3d^64s~a^6D_{9/2}\rightarrow 3d^64p~^6P^\circ$ \citep{baldwin2004}.
Since Fe is observed to be strongly depleted onto dust in the ISM \citep[see e.g.,][]{jenkins2009}, \feii$\lambda$1787 is likely produced in dust-poor regions, which also facilitate fluorescence rather than dust absorption of \lya.
In comparison, the production of the \feii 1550 group is more complex due to the multiple upper-level configurations involved.

To understand the excitation of \feii, we ran another suite of \cloudy simulations as summarized in Table~\ref{tab:feii_models}, where we turned on all available levels for $\rm Fe^+$ during the calculation.
The atomic data for the UV \feii transitions mainly come from \citet{Smyth_2019}, as described in \citet{sarkar2021}.
The model grid spans a range of values in $\rm [Fe/H]$, $N_{\rm H}$, and $v_{\rm turb}$.
We allowed $v_{\rm turb}>120$ \kms (i.e., higher than the \lya calculation) as a test for stronger turbulence in the \feii emitting clouds, but our conclusions are not affected by this choice.
To obtain the model predicted flux ratio of \feii lines at 1550 \AA and 1787 \AA, we defined two narrow bands centered at these wavelengths, with band widths of 6 \AA, and calculated summed fluxes of \feii emission within the bands.

The results are plotted in Figure~\ref{fig:feii_r}.
Since the \feii UV 191 group can be produced by \lya pumping \citep{baldwin2004}, its strength can become less sensitive to the iron abundance compared to the \feii 1550 group when the fluorescence is saturated.
Indeed, from Figure~\ref{fig:feii_r}, it is clear that the relative abundance of Fe/H is a key driver of the line ratio.
To reproduce the observed line ratio, the iron abundance needs to be higher than 1\% solar.
Meanwhile, according to \citet{Maiolino_metalpoorlrd_2025}, the narrow-line oxygen abundance of \target is $\sim 0.4$\% solar.
Therefore, in the scenario where \civ is not present in the FUV, \target would need to have a near-to-supersolar Fe/O in certain regions.
Type II supernovae tend to produce $\rm [Fe/O]\sim -1.8$ \citep[as inferred from Galactic low-metallicity dwarf stars, e.g.,][]{Amarsi_2019}.
To produce a high Fe/O, regional enrichment by early Type Ia supernovae or pair-instability supernovae would be needed \citep[see discussions by, e.g.,][]{ji_gnz11_2024,Nakane_fe_2025,Isobe_sife_2025}.

Notably, it has long been suggested that bright QSOs at $z>5$ already have supersolar Fe abundances \citep[e.g.,][]{Schindler_2020}, yet the general population of Type 1 AGN discovered by \jwst is Fe-poor in their BLRs \citep{trefoloni_feii_2024}.
If LRDs do represent an early evolutionary stage of AGN, one should be able to track the Fe enrichment in this population.
Indeed, many works have reported the identifications of UV \feii features in LRDs \citep{labbe_monster,tripodi2025deepdivebroadlineregion,irony,Torralba_feii_2025}, which might indicate some level of Fe enrichment, but subject to the complex mechanisms of excitation \citep{sarkar2021,Huang_agndisc_2023}.
We will present a systematic analysis of the UV \feii features in LRDs in a different work.

Finally, we note the caveat associated with the \feii interpretation of the 1550 \AA-line and the above abundance estimation, which is the apparent absence of other \feii transitions in the FUV.
Indeed, in observations, UV \feii emission from BLRs is typically characterized by highly blended groups rather than individual, strong transitions \citep{tsuzuki2006}.
In Figure~\ref{fig:feii_eg}, we plot one of the \feii models predicted by \cloudy with $\rm [Fe/H]=-1$, $N_{\rm H}=10^{25}~{\rm cm^{-2}}$, and $v_{\rm turb}=100$ \kms and normalize it to the observed line at 1550 \AA.
In addition to the well detected lines at 1550 \AA and 1787 \AA, the tentative line at 1542 \AA might also come from \feii emission.
However, the \feii model predicts more \feii transitions that should be detectable, especially near the NUV regime.
The lack of significant detection of other \feii transitions in the G140M spectrum indicates that either the model fails to capture the actual physical processes that produce the observed \feii lines, or the 1550-\AA transition is mostly due to \civ.
Indeed, as shown in Figure~\ref{fig:feii_r}, at low Fe/H and high turbulence, \feii$\lambda 1550$ is predicted to be much weaker compared to \feii$\lambda 1787$ by \cloudy.
One of such examples is shown in the bottom panel of Figure~\ref{fig:feii_eg}, where we selected the model with $\rm [Fe/H]=-2$, $N_{\rm H}=10^{23}~{\rm cm^{-2}}$, and $v_{\rm turb}=100$ \kms and normalize it to the observed \feii$\lambda 1787$.
Such a model predicts negligible \feii emission around 1550 \AA and the \feii FUV emission is fluorescence dominated.
The low-Fe/H model would alleviate the tension between the FUV metallicity and the optical metallicity inferred by \citet{Maiolino_metalpoorlrd_2025}, which is $\rm [O/H]\approx -2.4$.
In this case, the 1550 \AA line is again dominated by \civ.

Either case, the significant detection of \feii$\lambda 1787$ in the FUV confirms again the ubiquity of the presence of \feii in LRDs, possibly tracing the gas in the outskirt of BLRs.
To perform detailed diagnostics of the gas properties, more \feii transitions or low ionization transitions from the same region, such as Mg\,{\sc ii} and Ca\,{\sc ii}, need to be observed.
While many \feii transitions and Mg\,{\sc ii} transitions are covered by the NIRSpec G235M spectrum of \target, the observing depth is too shallow to detect any transitions, as shown in Appendix~\ref{appendix:nuv}.
Deeper UV observations of LRDs with medium-to-high spectral resolutions will be key to precisely determining the gaseous conditions.

\section{Origin of the continuum}
\label{sec:uv_continuum}

\begin{figure}
    \centering
    \includegraphics[width=\columnwidth]{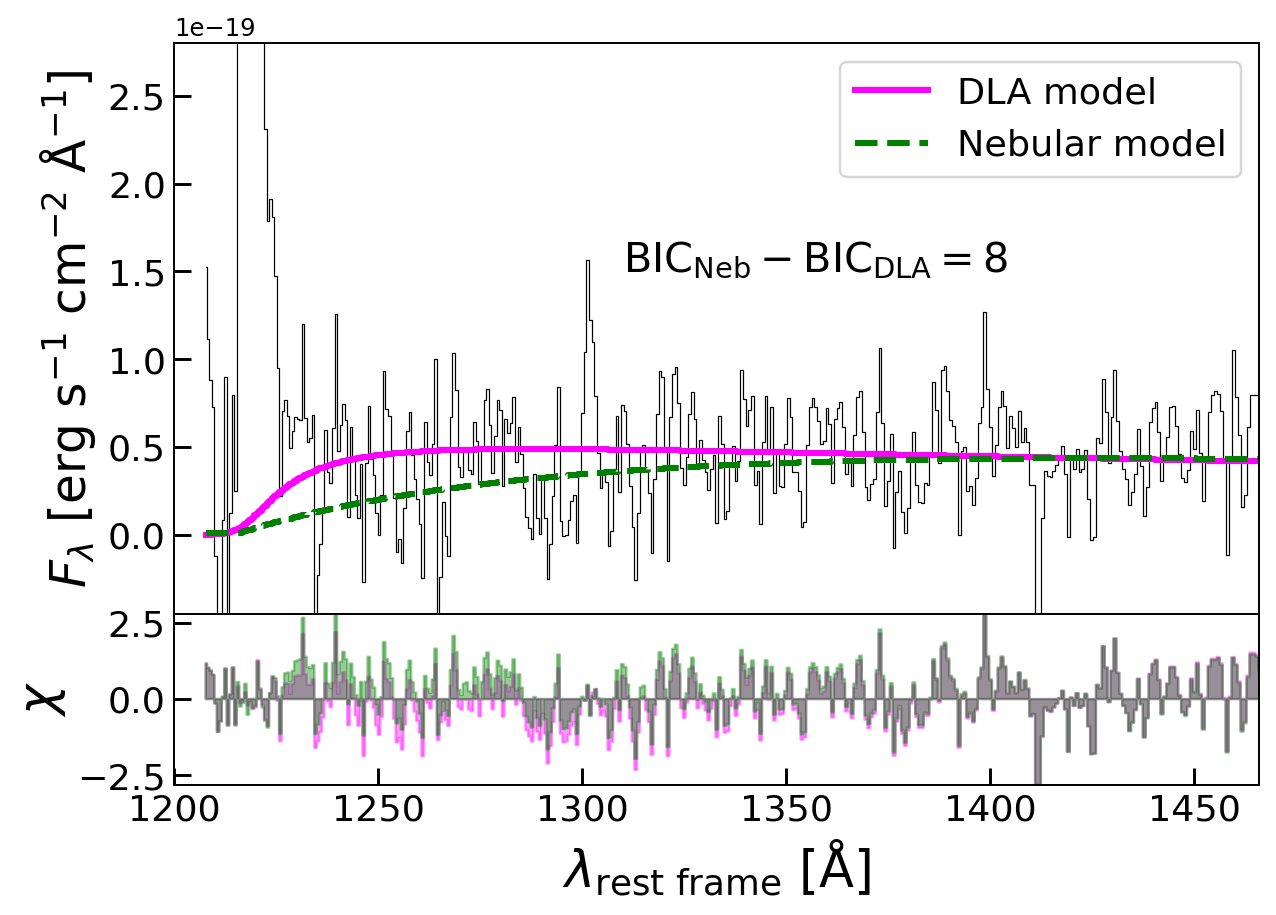}
    \caption{
    Comparison between DLA and nebular continuum models for the FUV continuum of \target.
    For clarity, we only show the continuum models, but the normalized residuals ($\chi$) are calculated with the whole continuum + emission line models.
    The main difference between the two models arise near \lya, and the DLA model is preferred over the nebular continuum-only model despite having more free parameters.
    }
    \label{fig:continuum_com}
\end{figure}

In Section~\ref{sec:measure}, we fit the FUV spectrum of \target assuming the continuum is a power law damped by a \lya absorber.
This corresponds to a physical scenario where the continuum is either starlight from young stellar populations attenuated by neutral gas, or AGN accretion-disc emission filtered through a dense gas envelope \citep[see][]{ji_lrdbreak_2025,juodzbalis_specast_2025}.
The first scenario is supported by the observations of some LRDs, where the FUV is either well resolved in NIRCam images \citep[thus ruling out an accretion disc;][]{rinaldi_lrd_2024} or is spatially offset compared to emission from longer wavelengths \citep{torralba+2025}.
Still, another possibility for the spatial extension of the FUV emission in some LRDs is a nebular continuum from a diffuse, gaseous halo, which would be dominated by a hydrogen two-photon continuum in the FUV regime (\citealp{bottorff2006}; see e.g., \citealp{Rauch_2016,tacchella2024} for discussions of the two-photon continuum halos in galaxies).
While the damped stellar/AGN continuum and the nebular continuum would not be mutually exclusive, investigating whether one of these scenarios dominates could provide insight into whether we are seeing directly the ionizing sources in the UV (power law + DLA) or actually the reprocessed emission (nebular continuum), which could help to infer the physical properties of LRDs.

In the case of \target, the FUV continuum remains unresolved in the NIRCam image \citep{furtak2023}.
Through analyses of the PRISM spectra of \target, \citet{ji_lrdbreak_2025} concluded the DLA model is preferred over the nebular continuum model, but the nebular continuum model cannot be completely ruled out due to the low resolution of the PRISM.
With the new R1000 spectrum, we tested the nebular continuum model again.
In Figure~\ref{fig:continuum_com}, spectral fitting results from a DLA continuum model and a nebular continuum model are compared.
The nebular continuum model was computed with \pyneb \citep{luridiana2015}, where we set the electron temperature, density, and normalization of the continuum as free parameters.
The free normalization is not entirely physical \citep{ji_gnz11_2024}, but we would like to test the plausibility of the nebular continuum model with enough flexibility.
From Figure~\ref{fig:continuum_com}, the main difference between the two models lies close to \lya, where the nebular continuum model predicts systematically lower fluxes due to the fixed peak of the two-photon continuum at 1430 \AA \citep{gaskell1980}.

To evaluate the statistical performance of the two models, we again used the BIC.
Compared to the nebular continuum model, the DLA model has a total of five parameters, including the normalization, power-law slope, hydrogen column density, velocity of the DLA, and the Doppler parameter of the DLA.
We obtained $\rm BIC_{Neb}-BIC_{DLA}=8$, suggesting that the fit with the DLA model, despite having more free parameters, is indeed improved compared to the nebular continuum model.
In practice, $\rm \Delta BIC>6$\,--\,10 have been used as a criterion for noticeable improvement.
Therefore, while with the R1000 spectrum, we arrive at the same conclusion as \citet{ji_lrdbreak_2025} that the DLA model is preferred, we would need higher S/N to firmly support this interpretation and rule out the nebular origin for the continuum.

\section{IGM transmission}
\label{sec:igm}

Finally, at the spectral resolution of G140M we can obtain valuable insights into the conditions of the IGM surrounding \target. Until now, we have relied on the mean transmission curves from \textsc{thesan-1} (end of Section~\ref{subsec:lya}). In this section, we attempt to fully model the IGM, using again the \lya absorption profile from \citet{witstok_z13_2025}. We fix the mean neutral fraction \meanxhi to 0.5 \citep[see e.g.,][]{naidu_2020,Jones_lya_2025}, consider an ionized bubble with radius $R_{\hii}$, and set the residual neutral fraction within the bubble to $10^{-8}$ \citep{Witstok_lya_2026}\footnote{As in Section~\ref{subsec:lya}, we use the \textsc{lymana\_absorption} package.}.

In this model, we used again informative priors on narrow \lya based on the models from Table~\ref{tab:lines}. For $R_{\hii}$, we used flat priors in log space, which avoids biasing the solution to non-zero values. We then calculated the posterior probability distributions with the same methods as in Section~\ref{subsubsec:lya_profile}.
The resulting model is shown in Figure~\ref{fig:igm} (see Appendix~\ref{appendix:igmcorner} for the full posterior distribution). The IGM transmission we derive is smaller but comparable to the mean value from \citet{smith_lya_2022}'s simulations. The model is fully consistent with no ionized bubble, given that the lower 3$\sigma$ limit (0.13\textsuperscript{th} percentile) of the marginalized posterior probability of $R_{\hii}$ is $10^{-4}$~physical Mpc (pMpc), which coincides with the lower bound of our uniform prior.
The equivalent 3$\sigma$ upper bound is $R_{\hii}<0.8$~pMpc. 
The 50\textsuperscript{th} percentile value and the confidence interval at the 68\% level are $R_{\hii}=0.20^{+0.07}_{-0.05}$.
This result is \textit{fully degenerate} with the adopted \meanxhi and thus cannot be constrained independently. 
Specifically, the model with free \meanxhi finds no evidence of a bubble ($R_{\hii}=0.004_{-0.004}^{+0.027}$~pMpc, and $R_{\hii}<0.08$~pMpc as a $3\sigma$ upper limit).
{The resulting $\meanxhi=0.18_{-0.05}^{+0.07}$ is lower than the value adopted in the fixed \meanxhi model as well as $\meanxhi=0.41$ in \textsc{thesan-1} simulations, although we note that the environment of \target could be quite different compared to the average galaxy in the simulations\footnote{There are other factors that can also contribute to the difference between the analytical model and \textsc{thesan-1} simulations.
For example, the residual neutral fraction within the ionized bubble in \textsc{thesan-1} simulations reaches a much higher value of a few $\times 10^{-3}$ (see \citealp{smith_lya_2022} for details).
}.
}

We note that while the data show substantial flux near $v=0~\kms$ in Figure~\ref{fig:igm}, this is due to the spectral resolution of the grating. While the results may suffer from the complex background, we tested this model against different data reduction and found qualitatively consistent results (see Appendix~\ref{appendix:igmcorner}).

Despite considerable uncertainties, the above finding of an insignificant ionized bubble is largely consistent with the expectations from the ionizing flux output derived from the narrow component of \ha in \target.
To estimate the size of the ionized bubble, we follow the approximation of \citet{mg_2020} given by 
\begin{equation}
    R_{\hii} \approx \left[\frac{3f_{\rm esc}\dot{N}_{\rm ion}t_{\rm ion}}{4\pi \,n_{\rm H,~IGM}(z=7)}\right]^{1/3},
\end{equation}
where $f_{\rm esc}$ is the escape fraction, $\dot{N}_{\rm ion}$ is the ionizing photon rate, $t_{\rm ion}$ is the duration of ionizing photon production, and $n_{\rm H,~IGM}\approx 1.88\times10^{-7}(1+z)^3~{\rm cm^{-3}}$ is the hydrogen density of the IGM.
Assuming the ionizing source that powers the narrow emission lines (no matter being the AGN or SF activities in the host galaxy) is also ionizing the IGM, $(1-f_{\rm esc})\dot{N}_{\rm ion}$ ionizing photons are absorbed by the ISM and $\alpha _{\rm eff,~H\alpha}/\alpha_{\rm B}\approx 1.17/2.54 \approx 0.46$ is turned into \ha under the Case B approximation \citep{draine2011}.
With a lensing corrected $L_{\rm H\alpha,~narrow}\approx 4\times 10^{40}$ \ergs from \citet{deugenio_qso1_2025}, and assuming a typical $f_{\rm esc}=0.05$ and $t_{\rm ion}=50$ Myr typically assumed for SF ionization, we obtained $R_{\hii} \approx 0.06$ pMpc.

The derived size of the ionized bubble is smaller compared to the median value of $R_{\hii}=0.20^{+0.07}_{-0.05}$ pMpc at $\meanxhi = 0.5$, but fully compatible with $R_{\hii} < 0.08$ pMpc when \meanxhi is left free.
In addition, as we have noted, in the fixed $\meanxhi = 0.5$ case, the bubble size would still be compatible with $10^{-4}$ pMpc within $3\sigma$ (see Appendix~\ref{appendix:igmcorner} for the posterior distribution), indicating that the ionized bubble is insignificant regardless of the treatment of \meanxhi.
The bubble size derived under $\meanxhi = 0.5$ (free \meanxhi) would imply an escape fraction of $f_{\rm esc}=0.68^{+0.16}_{-0.21}$ ($f_{\rm esc}<0.12$).
The above comparison indicates that the negligible bubble size from our fit can be compatible with the ionization inferred from narrow lines, although the relatively large uncertainty in the fitted bubble size cannot rule out additional ionization (e.g., from the AGN) over a relatively short time scale.
Indeed, our interpretation of broad \lya escape from the BLR would point towards ionizing photon leakage from the BLR at least during the epoch of observation.
Given that the broad \ha of \target is roughly 20 times brighter than the narrow \ha, for an ionization period of 1 Myr from the AGN \citep[which might be the same order of the ``LRD phase'' according to some models; e.g.,][]{inayoshi_firstacc_2025}, 
and assuming that the AGN dominates the ionization of the bubble,
we found $f_{\rm esc}=0.84^{+0.09}_{-0.15}$ ($f_{\rm esc}<0.25$) for $\meanxhi = 0.5$ (free \meanxhi).
Certainly, in the BLR the Case B approximation no longer applies, but \lya trapping + collisional processes would possibly indicate more efficient \ha production, and thus the estimated $f_{\rm esc}$ values are likely upper limits.

Finally, one remaining question is whether there might be any nearby galaxies around \target contributing to the growth of an ionized bubble.
As noted by \citet{furtak_abell2744_2023}, there is an overdensity in the Abell2744 field at $z=7.88$ \citep{Morishita_overdense_2023}, which was thought to be close to \target before the spectroscopic redshift was available.
Now, given $z_{\rm spec}=7.04$ for \target, this LRD does not overlap with this known overdensity.
Within the UNCOVER survey, there is one spectroscopic source at $z=7.04$ ($\rm RA=3.56959743$ deg, $\rm Dec=-30.37322304$ deg; also a spectroscopic LRD) with a projected distance (without correcting for lensing since this source is sufficiently offset from the main cluster; \citealp{Furtak_2023c}) of $\sim4.6$ cMpc (or $\sim 0.57$ pMpc) to \target.
Taking these two sources within 4.6 cMpc and comparing them with the galaxy number density at $z\sim 7$ from \citet{Bouwens_2021} integrated to $M_{\rm UV}=-16.81$ mag (i.e., the lensing corrected value for \target), there is either an underdensity ($1+\delta \sim 0.4$) or a moderate overdensity ($1+\delta \sim 3$) depending on whether the radius or the diameter of the volume is set to 4.6 cMpc.
Thus, the environment of \target remains unclear due to the lack of statistics.

In contrast, we note that another LRD discovered in the Abell2744 field at $z=4.47$, UNCOVER-45924 \citep[or the Monster;][]{greene2024,labbe_monster}, lies close to a large galaxy overdensity traced by both \ha and \lya \citep{matthee_bhenvir_2024,torralba+2025}.
\citet{torralba+2025} show that there is a weak \lya halo around UNCOVER-45924.
The \lya line profile observed in UNCOVER-45924 is much narrower ($\rm FWHM=270 \pm 50$ \kms) and fainter (relative to \ha) than what we see in \target, and more consistent with SF LAEs at high $z$.
The two LRDs also lie in two extreme luminosity ends of the high-$z$ LRD population as shown in Figure~\ref{fig:o1ha_relation}, with UNCOVER-45924 approaching the typical QSO luminosity.
The clear contrast between \target and UNCOVER-45924 in terms of environments and \lya line profiles suggest potentially different evolutionary paths and gas geometries within the LRD population.
Regarding the environments of LRDs as a population, \citet{Arita_lrdhalo_2025,Pizzati_lrdhalo_2025} have shown that LRDs appear to reside in lower-mass halos compared to QSOs.
As recently argued by \citet{Carranza-Escudero_lrdenvironment_2025}, clustering studies of high-$z$ LRDs in observations suggest that they might actually lie in less dense environments compared to galaxies at similar redshifts.
Further studies on the diversity of LRDs' environments would be informative for understanding their evolution.

\begin{figure}
    \centering
    \includegraphics[width=\linewidth]{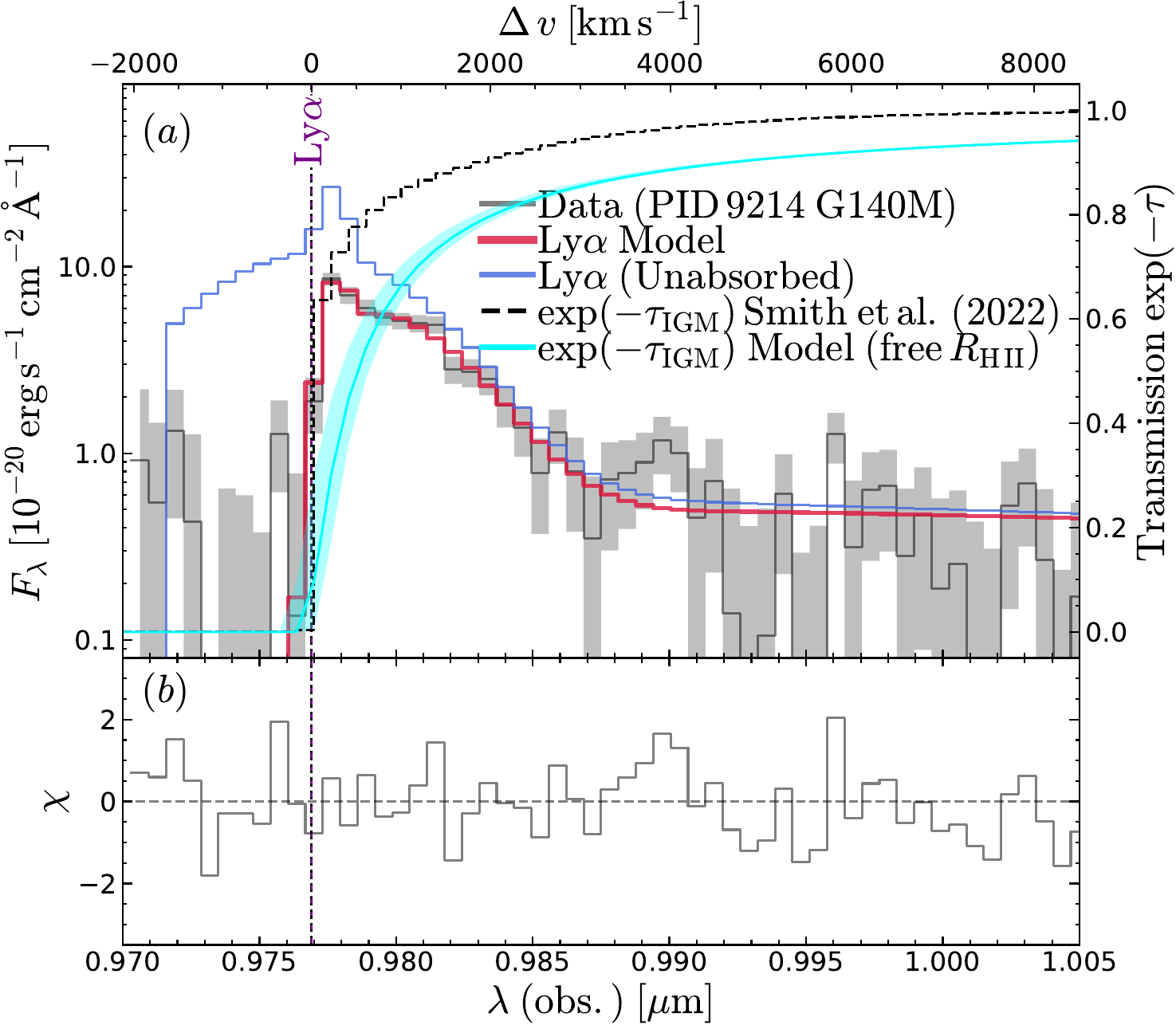}
    \caption{Best-fit IGM-absorption model, using an ionized bubble with variable radius $R_{\hii}$. While this model requires a weak ionized bubble ($R_{\hii}=0.20\pm0.05$~pMpc), this preference depends on the assumption of fixed $\meanxhi=0.5$. The performance of the IGM-absorption model is the same as the fiducial model shown in Figure~\ref{fig:profiles}, with $\Delta\,\text{BIC}=-0.3$ (i.e., no improved performance).
    The best fit IGM transmission (cyan, right axis) is compared to the mean simulated IGM transmission at $z\sim 7$ from \citet[][dashed black line]{smith_lya_2022}. 
    }\label{fig:igm}
\end{figure}

\section{Conclusions}
\label{sec:conclude}

\begin{figure}
    \centering
    \includegraphics[width=\linewidth]{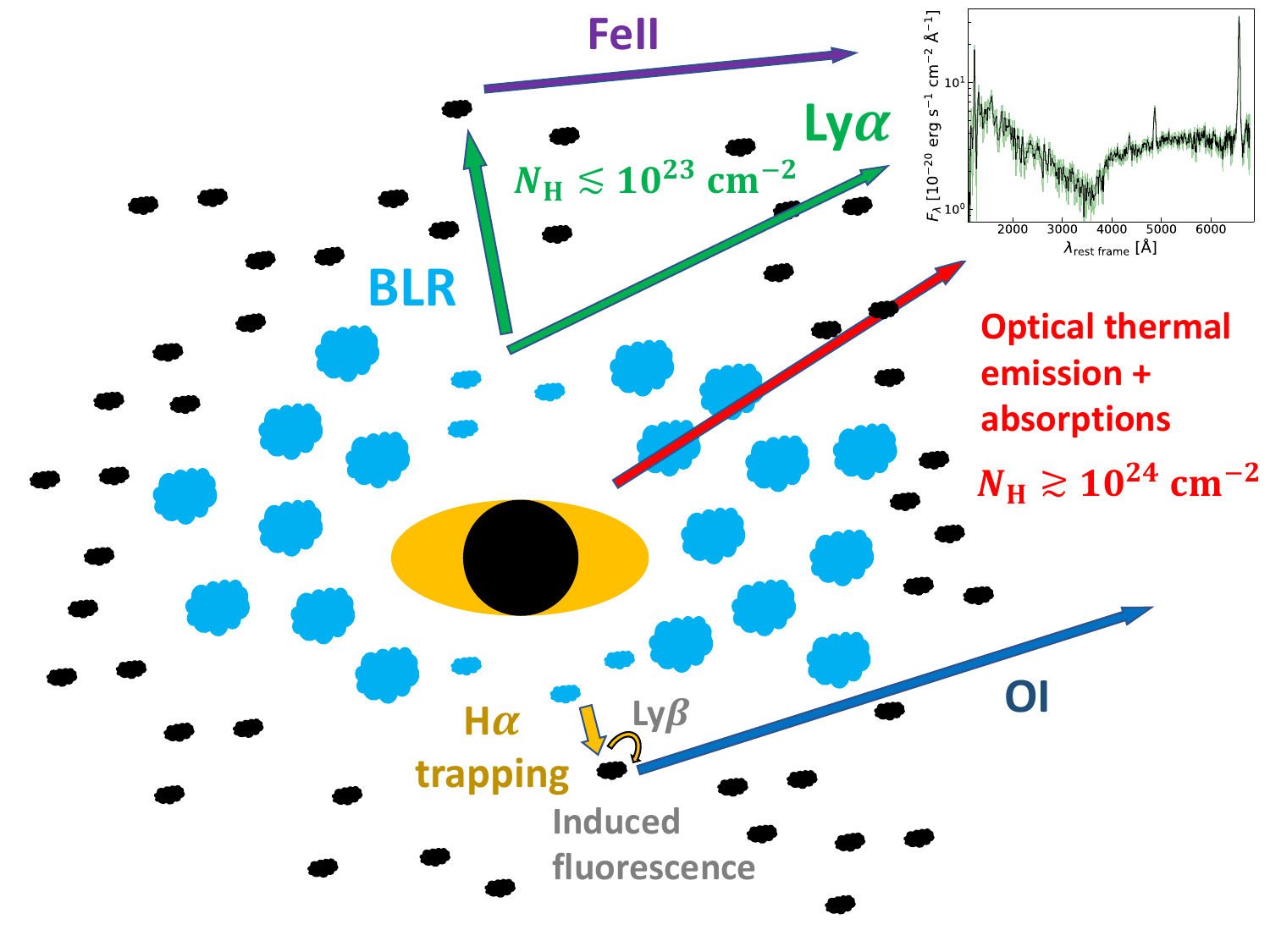}
    \caption{Schematic plot of the gaseous structure in \target.
    The accretion disc of the black hole is surrounded by clumpy, dense gas clouds (i.e., part of the BLR clouds or clouds close to the BLR), which do not have the same column densities in all directions.
    Along the LOS comes the attenuated optical continuum and broad Balmer emission characterized by large optical depths requiring $N_{\rm H} \gtrsim 10^{24}~{\rm cm^{-2}}$.
    In contrast, the observed broad \lya emission likely originates in less optically thick directions with $N_{\rm H} \lesssim 10^{23}~{\rm cm^{-2}}$.
    The escape of \lya further enables fluorescent cascades of \feii emission on a larger physical scale.
    While \lyb is difficult to escape from the BLR, 
    fluorescence cascade of \oip can occur in clouds optically thick to \ha, which is initiated by local \lyb photons pumped by \ha photons from the BLR.
    }
    \label{fig:schematic}
\end{figure}

In this work, we investigated the geometry of the line emitting clouds in the multiply imaged LRD, \target, at $z=7.04$, leveraging new observations in the rest-frame FUV with \jwst NIRSpec/G140M.
Several emission line features are identified in the FUV, including \lya, \oip$\lambda 1302$, \civ$\lambda 1550$ (or \feii), and \feii$\lambda 1787$.
Our inferred structure for the gas around the accreting black hole in \target is shown in Figure~\ref{fig:schematic}, and we summarize our findings in the following.
\begin{enumerate}
    \item There is prominent \lya emission detected in the FUV, which can be further decomposed into a narrow component ($\sigma =140\pm 77$ \kms) that dominates the core and a broad component ($\sigma = 1030\pm 75$ \kms) that dominates the wing of the line profile.
    Based on the NIRCam images and NIRSpec G140M slit images, the narrow \lya is spatially extended on a physical scale of 200\,--\,300 pc on the source plane and the broad \lya remains spatially unresolved.
    This implies different origins for the two components of \lya.
    \item The kinematics of the broad \lya are inconsistent with resonant-scattering dominated broadening, suggesting that the broad component of \lya is likely intrinsically broad.
    Only the narrow component of \lya shows consistent kinematics with conventionally resonant-scattering shaped spectra reminiscent of LAEs.
    The broad wing of \lya is significantly broader than the broad \ha in the optical observed previously with NIRSpec R2700 IFU. The difference suggests a mix of broadening mechanisms for \lya, including virial broadening, resonant scattering, and possibly electron scattering, although the electron scattering model is not statistically preferred based on our fit of \lya.
    \item By comparing the observed strengths of broad \lya and Balmer emissions to those from theoretical photoionization models, we found that the broad \lya is likely escaping from BLR clouds with total hydrogen column densities of $N_{\rm H}\lesssim 10^{23}~{\rm cm^{-2}}$, in contrast to $N_{\rm H}\gtrsim 10^{24}~{\rm cm^{-2}}$ invoked by previous work to explain the optical continuum and broad lines of LRDs including \target. This implies inhomogeneous cloud distributions in the BLR of \target, where \lya can be affected by relatively optically thin directions as shown in Figure~\ref{fig:schematic}.
    \item The \oip$\lambda 1302$ detected in \target might come from fluorescent cascade pumped by \lyb.
    In this scenario, the narrowness ($\sigma =180\pm 60$ \kms) and the blueshift ($\Delta v=-130\pm 70~\kms$) of \oip indicate that it comes from outside the BLR, possibly produced by \lyb fluorescence induced by trapped BLR \ha photons in thick clouds with low velocity dispersions.
    The strength of the observed \oip$\lambda 1302$ appears lower than the expectation converted from the scaling relation between the luminosities of \oip$\lambda 8446$ and \ha observed in other LRDs.
    Possible explanations include \oip trapping (and fluorescence), dust attenuation, a lower metallicity, a lower \lyb optical depth, or ubiquitous collisional excitation of \oip in LRDs, which will need more observations of UV \oip in LRDs to verify.
    \item If the emission line observed at the rest-frame 1550 \AA in the spectrum of \target is \civ, it suggests that ionizing radiation from the accretion disc might have leaked from the gas envelope in this LRD, although an external ionization by young stellar populations cannot be ruled out.
    If this emission line is actually associated with \feii, given the presence of \feii$\lambda 1787$ that can be pumped by \lya, the relative strength of \feii lines suggest a supersolar abundance ratio of Fe/O.
    However, as shown by \cloudy models, a high Fe abundance tends to overpredict FUV \feii lines that are not detected.
    Therefore, the \civ explanation is more plausible despite the presence of \feii$\lambda 1787$.
    The narrowness of \feii$\lambda 1787$ ($\sigma = 250\pm 60$ \kms) suggests that it comes from outside the BLR and might be pumped by leaked \lya (see Figure~\ref{fig:schematic}).
    \item Regardless of the detailed identification of the 1550 \AA line, we note that the various metal lines revealed in the UV spectrum cannot be reliably used to accurately constrain the metallicity or chemical enrichment in \target. This is due to the resonant nature and/or excitation mechanism (e.g., fluorescence) of these lines, which makes their observed intensity strongly dependent on RT effects.
    \item The UV continuum is better described by a power law attenuated by a DLA compared to a pure nebular continuum. Still, the statistical difference between the two models given the current data is not very significant ($\rm \Delta BIC=8$) and the origin of the UV continuum requires deeper observations to verify.
    \item There is no evidence for a significant ionized bubble surrounding \target. By fitting the \lya profile, we obtained a bubble size of $R_{\hii}=0.20^{+0.07}_{-0.05}$ pMpc ($R_{\hii}<0.08$ pMc) for a fixed $\meanxhi = 0.5$ (free \meanxhi), compatible with the ionization from the host galaxy as well as leaking ionizing photons from the BLR.
\end{enumerate}

To conclude, the new FUV observations for \target provide key insights into the geometry of the dense gas in the central region of this LRD.
The broad \lya and fluorescent transition of \feii indicate that the hypothesized gas envelope in \target cannot be homogeneous in all directions and needs to have ``holes'' with lower optical depths to leak \lya. This possibly suggests that the gas envelope is actually (part of) the BLR composed of clumpy medium surrounding the accreting BH.
Future medium-to-high resolution follow-up of LRDs in the rest-frame UV regime will establish a census of the gas geometry for this population.

\section*{Acknowledgements}
We thank Harley Katz for insightful and constructive discussions related to this work.
We thank Hannah Übler for comments on a draft of this manuscript.
XJ, FDE, RM, GJ, SG, AH, LI, IJ, RP, and JS acknowledge ERC Advanced Grant 695671 ``QUENCH'' and support by the Science and Technology Facilities Council (STFC) and by the UKRI Frontier Research grant RISEandFALL.
RM acknowledges funding from a research professorship from the Royal Society.

{This work used observations made with the NASA/ESA/CSA James Webb Space Telescope. The data are available at the Mikulski Archive for Space Telescopes (MAST) at the Space Telescope Science Institute, which is operated by the Association of Universities for Research in Astronomy, Inc., under NASA contract NAS 5-03127 for JWST.}

\section*{Data Availability}

All \jwst\ observations used in this paper are available through the MAST portal.
Analysis results presented in this manuscript will be shared on reasonable request to the corresponding author.

%
   \bibliographystyle{mnras} 
   \bibliography{ref} 
%

\appendix

\section{NUV-to-optical spectrum of \target}
\label{appendix:nuv}

In this Appendix, we present reduced spectra from G235M/F170LP and G395M/F290LP gratings for \target.
The observations were extracted from the same programme as the G140M/F100LP data (PID: 9214; PIs: C. Mason, D. Stark).
See Section~\ref{sec:data} for descriptions of observations and data reduction.

\begin{figure*}
    \centering
    \includegraphics[width=0.95\linewidth]{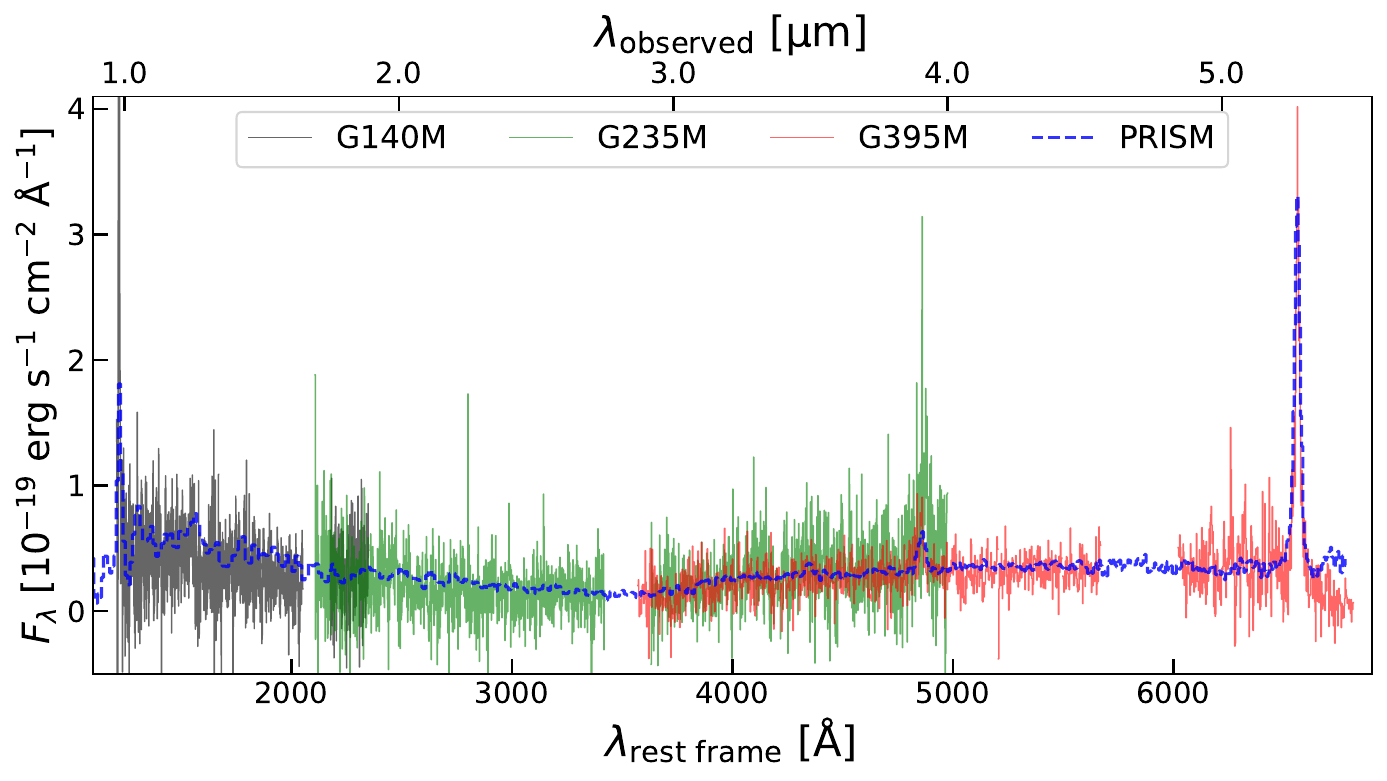}
    \caption{
    Reduced 1D spectra from all grating configurations for the image B of \target, including G140M, G235M, and G395M.
    For comparison, we show the MSA PRISM spectrum of the image B reduced by the same pipeline.
    The wavelength calibrations of the spectra show good consistency in the optical, where \ha has negligible offset on the pixel scale.
    The rest-frame NUV and optical grating spectra of \target show relatively large flux uncertainties and generally do not provide additional constraints on spectral features compared to the archival PRISM spectrum.
    Therefore, in this work, we only focus on the G140M spectrum that reveals new features in the FUV regime.
    }
    \label{fig:all_gratings}
\end{figure*}

In Figure~\ref{fig:all_gratings}, we plot the full rest-frame FUV to optical spectrum of \target covered by all three gratings as well as the MSA PRISM of the same image.
The grating and PRISM spectra are overall agreed on the flux level and show good consistency in their wavelengths in the optical as traced by the reddest strong emission line, \ha.
Besides FUV emission lines, only \ha and \hb are significantly detected in the gratings.
In the NUV-to-optical regime, the gratings generally do not provide new constraints on the spectral features of \target.
In the FUV, as we have shown in Figure~\ref{fig:spec_fit}, there appears to be a wavelength offset between the G140M spectrum and the PRISM spectrum.
This wavelength calibration issue has been noted for \jwst NIRSpec observations, which shows increasing scatter towards shorter wavelengths \citep{scholtz_jadesdr4_2025}.

\section{Issue with the background subtraction}
\label{appendix:background}

\begin{figure}
    \centering
    \includegraphics[width=\linewidth]{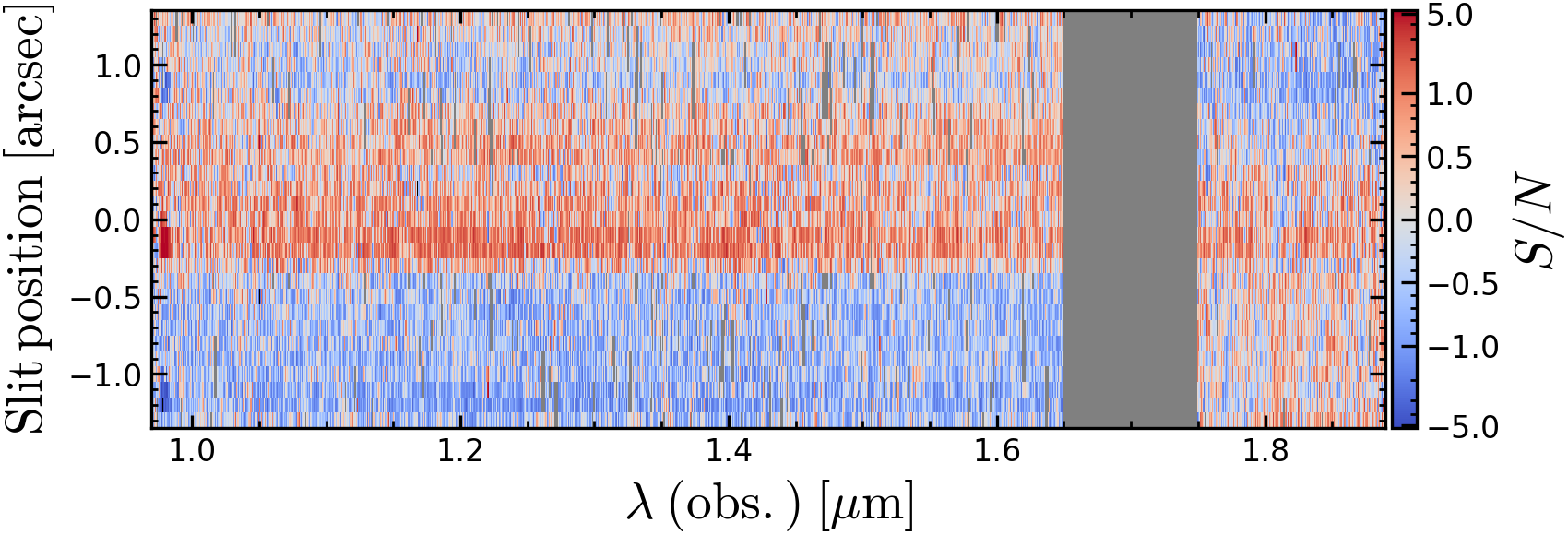}
    \caption{$\rm S/N$ map of the two-nod background-subtracted spectrum, showing the object trace (central shutter, near slit position $-0.\!\!''15$), and residual background contamination, appearing as positive (positive slit position) and negative (negative slit positions). Notice the pattern swapping going in NRS2, just after the detector gap.
    }
    \label{fig:background}
\end{figure}

As shown in Figures~\ref{fig:spec_2d1d} and \ref{fig:lya_2d}, there is a background subtraction issue in the 2D spectrum, where continuum contamination is visible.
In this Appendix, we further characterize the contamination.
In Figure~\ref{fig:background} we show the $\rm S/N$ map of the two-nod subtracted, rectified and aligned spectrum. \target is visible near $-0.\!\!''15$ along the slitlet, corresponding to the bottom edge of the central microshutter (Figure~\ref{fig:lya_2d}). The \lya emission is particularly clear, as are the negative \lya traces at the two most distant nod positions, near $-1.\!\!''2$ and $+0.\!\!''9$.
In absence of contamination, the rest of the 2D spectrum should scatter around 0, while here positive (negative) slit positions display weak but clearly detected positive (negative) signal.

There are three possible explanations. The first and most obvious is stellar light from the outskirts of a foreground cluster galaxy. Due to the placement and position angle of the slit, positive slit positions are closer to the centre of the interloper, hence are bound to accept more light, which then gives rise to the positive/negative pattern during subtraction.
To some extent, this effect is also present in the UNCOVER PRISM observations, although in that case its magnitude is remarkably smaller, most likely due to the fortunate position angle of the slit, which happens to be almost orthogonal to the imaginary line connecting the image B of \target to the centre of the foreground galaxy.

A closer look at Figure~\ref{fig:lya_2d} also reveals a compact blue source between the central and top shutters, slightly outside the shutter region. We cannot exclude that this source may also participate in the contamination; this could happen due to PSF effects scattering light inside the shutter, but also due to the finite precision of the NIRSpec/MSA MSATA acquisition process.
The hypothesis of point-source contamination seems supported by a visible gap between the positive contamination and \target (Figure~\ref{fig:background}).

Finally, we notice that the positive/negative pattern swaps order going from NRS1 to NRS2. This cannot happen if the contamination originated from the same shutter as the one being rectified in this reduction. We inspected the mask and confirm there are no other assigned sources along the same row, hence the contamination cannot be due to spectral overlap (regardless of their spectral order).
However, we do rule out contribution from either a disobedient shutter, or from a bright enough spoiler.

Regardless, our results are unchanged when changing the details of the data reduction. Future observations of the more isolated image A of \target could be used to validate our findings.

\section{Alternative models for the BLR}
\label{appendix:tba}

In this Appendix, we present additional photoionization models run with \cloudy.
When investigating the ratios of H\,{\sc i} emission lines from the BLR in \target in Section~\ref{subsec:lya}, we fixed the hydrogen density to $n_{\rm H}=10^{10}~{\rm cm^{-3}}$.
However, the BLR can in principle have a broad range of densities at $n_{\rm H}>10^{8}~{\rm cm^{-3}}$ \citep{netzer_1990}, where high density clouds might be able to reach $n_{\rm H}\sim 10^{12-13}~{\rm cm^{-3}}$ \citep{temple2021}.
Here we investigate the effects of lowering and increasing the density, respectively.

\begin{figure*}
    \centering
    \includegraphics[width=0.9\linewidth]{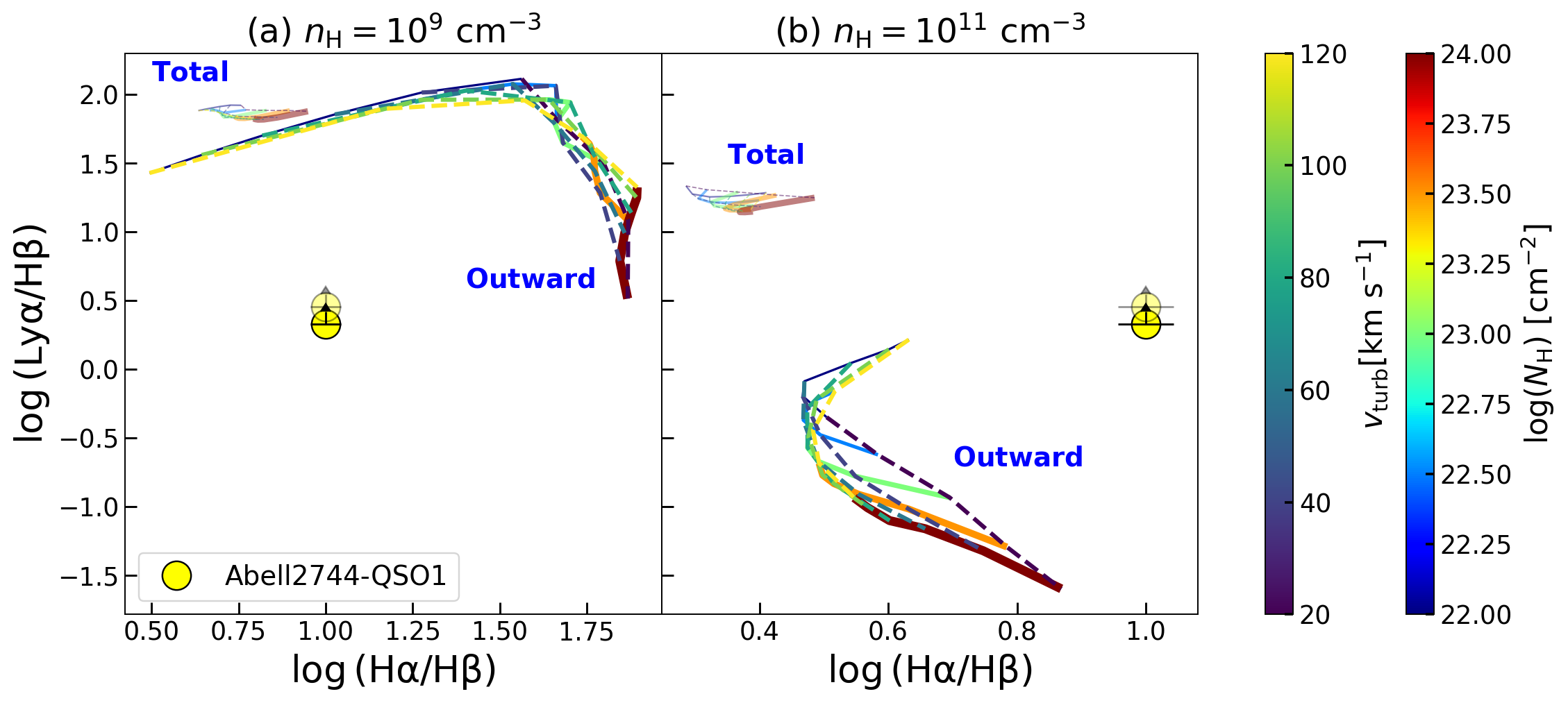}
    \caption{
    Same as the left panel of Figure~\ref{fig:lyahbha_blr} but with \cloudy models at different densities.
    Besides the density, the rest of the model parameters are the same as those listed in Table~\ref{tab:hi_models}.
    These models generally match the observational constraints worse compared to the model with $n_{\rm H}=10^{10}~{\rm cm^{-3}}$.
    }
    \label{fig:lyahbha_othermodels}
\end{figure*}

Figure~\ref{fig:lyahbha_othermodels} shows two additional model grids similar to the one shown in Figure~\ref{fig:lyahbha_blr}, but with $n_{\rm H}=10^{9}~{\rm cm^{-3}}$ and $n_{\rm H}=10^{11}~{\rm cm^{-3}}$, respectively.
At lower density, the predicted flux ratio of \lya/\hb is much closer to the Case B for $N_{\rm H}<10^{24}~{\rm cm^{-2}}$.
This suggests significantly lower \lya optical depth compared to the case of $n_{\rm H}=10^{10}~{\rm cm^{-3}}$, likely due to the less neutral gas fraction in the cloud.
In contrast, at $n_{\rm H}=10^{11}~{\rm cm^{-3}}$, \lya trapping becomes more significant and \lya/\hb is further reduced compared to the cases of $n_{\rm H}=10^{10}~{\rm cm^{-3}}$. Interestingly, the flux ratio of \ha/\hb seen in the outward direction is reduced and becomes closer to the Case B value.
This is likely caused by the fact that Pa$\alpha$ becomes optically thick at high densities, making the ``fluorescent'' production of \ha by \hb through ``\hb absorption $\rightarrow$ Pa$\alpha$ emission $\rightarrow$ \ha emission'' inefficient \citep[see][]{chang_2025,yanzu_lrdbal_2025}.
In summary, these tests show that $n_{\rm H}\sim 10^{10}~{\rm cm^{-3}}$ (in combination with other fiducial model parameters we set) best describes the observed line ratios in \target.

\section{Posterior probability distributions for the IGM transmission model}\label{appendix:igmcorner}

\begin{figure*}
    \centering
    \includegraphics[width=0.9\linewidth]{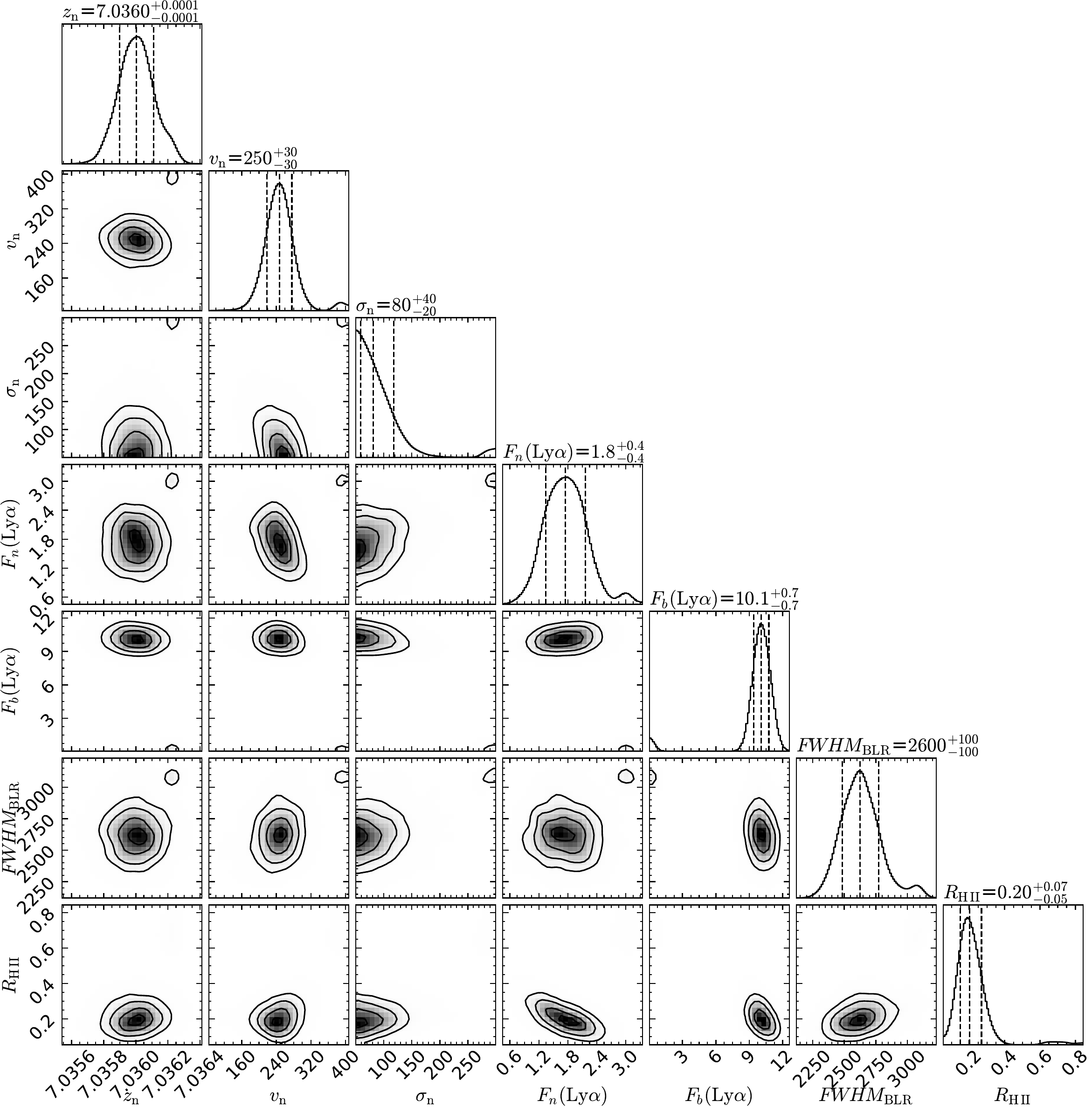}
    \caption{
    Marginalized 1- and 2-D probability distribution for the physical parameters in the fiducial IGM spectral model of Section~\ref{sec:igm} (fixed $\meanxhi=0.5$). The velocity and line widths are given in \kms, line fluxes are parametrized before IGM absorption, and are given in $10^{-18}~\mathrm{erg\,s^{-1}\,cm^{-2}}$, and $R_{\hii}$ is in pMpc.
    }\label{fig:igmcornerfixed}
\end{figure*}

\begin{figure*}
    \centering
    \includegraphics[width=0.9\linewidth]{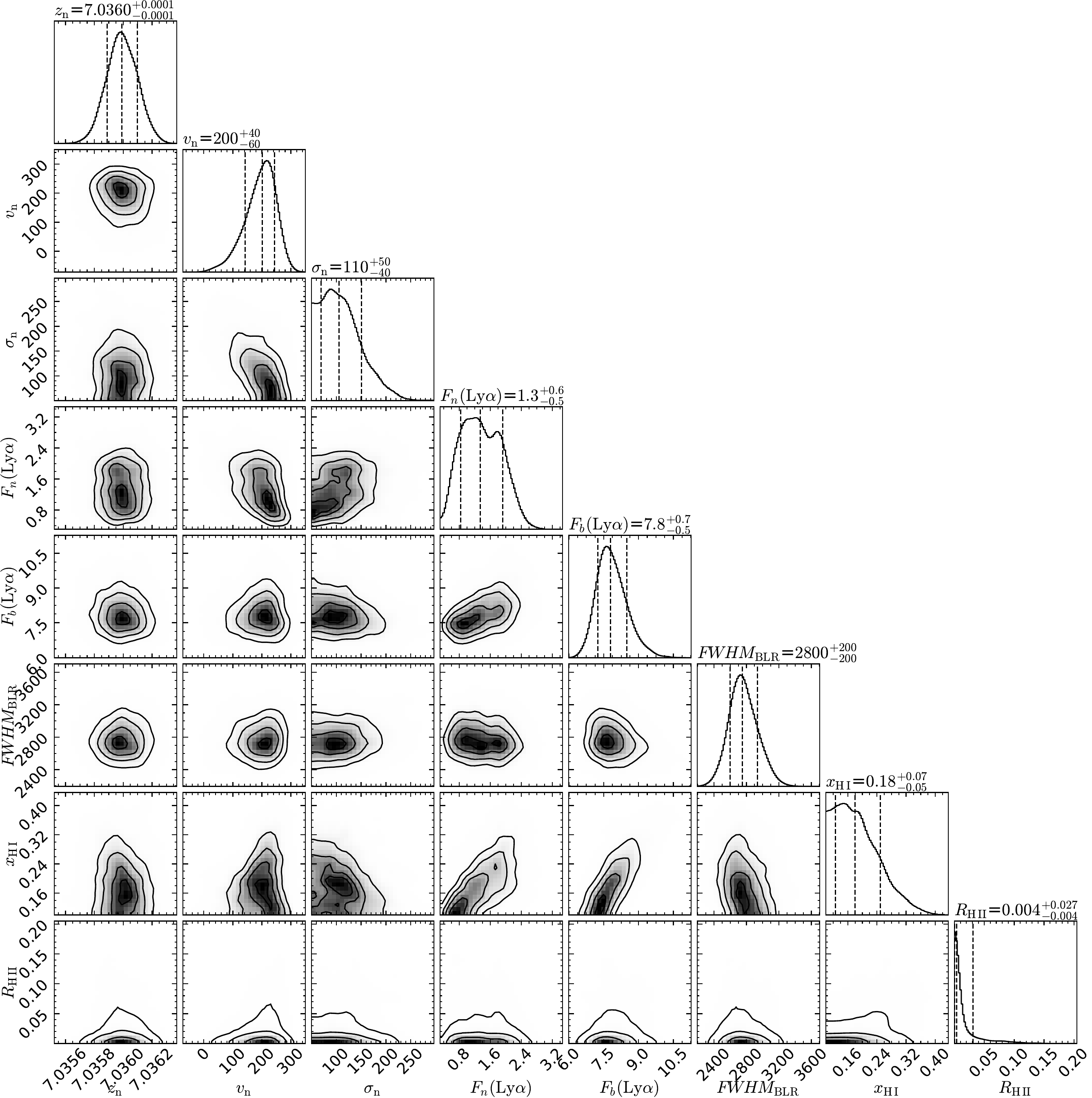}
    \caption{
    Marginalized 1- and 2-D probability distribution for the physical parameters in the alternative IGM spectral model with free \meanxhi (Section~\ref{sec:igm}). The labels and units are the same as in Figure~\ref{fig:igmcornerfixed}.
    }\label{fig:igmcornerfree}
\end{figure*}

This section summarizes the robustness of the IGM inference to physical assumptions and background subtraction. The ionized bubble radius $R_{\hii}$ changes between the fiducial model (fixed $\meanxhi=0.5$; Figure~\ref{fig:igmcornerfixed}) and the model with free \meanxhi (Figure~\ref{fig:igmcornerfree}), but it remains consistent with non detection (i.e., within a 99.73\% confidence interval) in both cases (note that the 16\textsuperscript{th}--84\textsuperscript{th} inter-percentile range for the $\meanxhi=0.5$ model does not reflect the heavy tail toward small $R_{\hii}$).

\begin{figure*}
    \centering
    \includegraphics[width=0.9\linewidth]{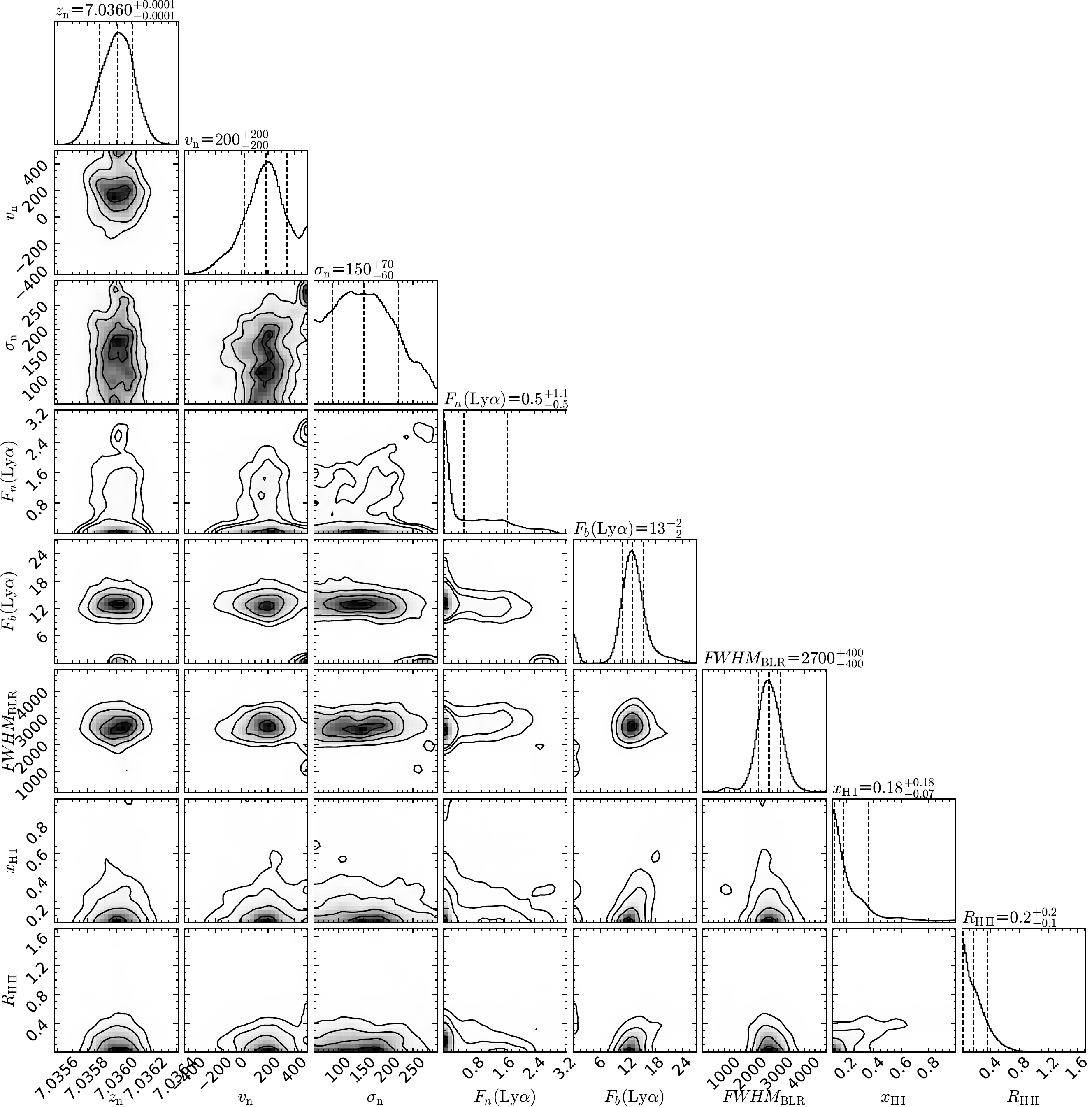}
    \caption{
    Same as Figure~\ref{fig:igmcornerfree}, but using the DJA data reduction to test the robustness against a different data-reduction pipeline and the background contamination. The labels and units are the same as in Figure~\ref{fig:igmcornerfixed}.
    }\label{fig:igmcornerdja}
\end{figure*}

We also note that the background subtraction strategy does not affect our results dramatically. Using the publicly available DJA spectrum (independently reduced compared to the GTO-reduced spectrum) to constrain the data we obtain similar results (cf., the fiducial model of Figure~\ref{fig:igmcornerfixed} to the model of
Figure~\ref{fig:igmcornerdja}).

\bsp	
\label{lastpage}
\end{document}